\documentclass{aastex631}

\usepackage{booktabs}
\usepackage{amsmath}
\mathchardef\mhyphen="2D
\usepackage{rotating}
\usepackage{longtable}
\usepackage{afterpage}
\usepackage{scalefnt}
\usepackage[toc,title,page]{appendix}
\usepackage{float}
\graphicspath{{./}{figures/}}

\begin{document}

\author[0000-0003-4442-8546]{M. G. Dainotti}
\affiliation{National Astronomical Observatory of Japan, 2 Chome-21-1 Osawa, Mitaka, Tokyo 181-8588, Japan}
\affiliation{The Graduate University for Advanced Studies, SOKENDAI, Shonankokusaimura, Hayama, Miura District, Kanagawa 240-0193, Japan}
\affiliation{Space Science Institute, Boulder, CO, USA}

\author[0000-0001-5083-6461]{B. De Simone}
\affiliation{Department of Physics "E. R. Caianiello" - Università degli Studi di Salerno - Via Giovanni Paolo II, 132, 84084 Fisciano SA}
\affiliation{INFN Gruppo Collegato di Salerno - Sezione di Napoli - Via Giovanni Paolo II, 132, 84084 Fisciano SA}

\author[0000-0002-5781-1563]{M. I. Khadir}
\affiliation{Department of Physics, University of Constantine 1 - RN79, Constantine, Algeria}

\author[0000-0002-7751-6115]{K. Kawaguchi}
\affiliation{Faculty of Science, Kumamoto University, 2-39-1 Kurokami, Chuo Ward, Kumamoto, Kumamoto Prefecture 860-0862, Japan}

\author[0000-0003-1169-1954]{T. J. Moriya}
\affiliation{National Astronomical Observatory of Japan, 2 Chome-21-1 Osawa, Mitaka, Tokyo 181-8588, Japan}
\affiliation{The Graduate University for Advanced Studies, SOKENDAI, Shonankokusaimura, Hayama, Miura District, Kanagawa 240-0193, Japan}
\affiliation{School of Physics and Astronomy, Faculty of Science, Monash University, Clayton, Victoria 3800, Australia}

\author[0000-0003-0304-9283]{T. Takiwaki}
\affiliation{National Astronomical Observatory of Japan, 2 Chome-21-1 Osawa, Mitaka, Tokyo 181-8588, Japan}
\affiliation{The Graduate University for Advanced Studies, SOKENDAI, Shonankokusaimura, Hayama, Miura District, Kanagawa 240-0193, Japan}

\author[0000-0001-8537-3153]{N. Tominaga}
\affiliation{National Astronomical Observatory of Japan, 2 Chome-21-1 Osawa, Mitaka, Tokyo 181-8588, Japan}
\affiliation{The Graduate University for Advanced Studies, SOKENDAI, Shonankokusaimura, Hayama, Miura District, Kanagawa 240-0193, Japan}
\affiliation{Kavli Institute for the Physics and Mathematics of the Universe (WPI), The University of Tokyo Institutes for Advanced Study, The
University of Tokyo, 5-1-5 Kashiwanoha, Kashiwa, Chiba 277-8583, Japan}
\affiliation{Department of Physics, Faculty of Science and Engineering, Konan University, 8-9-1 Okamoto, Kobe, Hyogo 658-8501, Japan}

\author[0000-0002-3884-5637]{A. Gangopadhyay}
\affiliation{Hiroshima Astrophysical Science Center, Hiroshima University, Higashi-Hiroshima, Japan}

\title{The quest for new correlations in the realm of the Gamma-Ray Burst - Supernova connection}

\begin{abstract}
Gamma-Ray Bursts (GRBs) are very energetic cosmological transients. Long GRBs are usually associated with Type Ib/c Supernovae (SNe), and we refer to them as GRB-SNe. Since the associated SN for a given GRB is observed only at low redshift, a possible selection effect exists when we consider intrinsically faint sources which cannot be observed at high redshift. Thus, it is important to explore the possible relationships between GRB and SN parameters after these have been corrected for astrophysical biases due to the instrumental selection effects and redshift evolution of the variables involved.
So far, only GRB prompt emission properties have been checked against the SNe Ib/c properties without considering the afterglow (AG). This work investigates the existence of relationships among GRB's prompt and AG and associated SN properties. We investigate 91 bidimensional correlations among the SN and GRB observables before and after their correction for selection biases and evolutionary effects. As a result of this investigation, we find hints of a new correlation with a Pearson correlation coefficient $>0.50$ and a probability of being drawn by chance $<0.05$. 
This correlation is between the luminosity at the end of the GRB optical plateau emission and the rest-frame peak time of the SN.
According to this relation, the brightest optical plateaus are accompanied by the largest peak times. 
This correlation is corrected for selection biases and redshift evolution and may provide new constraints for the astrophysical models associated with the GRB-SNe connection.
\end{abstract}

\keywords{Gamma-ray bursts (629) --- Core-collapse supernovae (304)}

\section{Introduction} \label{sec:intro}
\noindent Gamma-ray Bursts (GRBs) are bright and short $\gamma$-ray flashes. During their short durations, they emit the same energy that the Sun will have released at the end of its life starting from its birth: their isotropic energy ranges from $10^{47}$ to $10^{54}\,{\rm erg}$.

\citet{Mazets1981} and \cite{Kouveliotou1993} discovered a bimodality distribution of the duration of GRBs prompt emission in their observer frame with the analysis of the first BATSE catalog. Based on their findings, they defined two classes of GRBs: Short (SGRBs, with the prompt duration, $T_{90}$\footnote{It is the time during which a GRB emits $90\%$ of its energy (from $5\%$ to $95\%$).} shorter than 2 seconds) and Long (LGRBs, with $T_{90} > 2$ seconds). This bi-modality was proven to hold also in the rest-frame. SGRBs originate from binary mergers: the merging of a binary neutron star (NS) or an NS with a black hole (BH). The binary NS merger has given rise to the GW observed in association with the SGRB 170817A \citep{Abbott2017,Pian2017,Troja2017}. 
Instead, traditionally the LGRBs are thought to be generated from the Core-Collapse (CC) of massive stars, the Wolf-Rayet (W-R, \citealt{Biermann1993,Usov1994,Schaerer1994}). For a later discussion on this classification based on the observations of the Neil Gehrels Swift Observatory (hereafter Swift satellite, 2004-ongoing) GRBs, one can refer to \citet{Bromberg2013}.

Among all the phenomenological classes of GRBs, an interesting role is played by the GRBs associated with Type Ib or Ic Supernovae (SNe Ib/Ic). We here refer to these associations as GRB-SNe \citep{Wang1998,Ruffini2001,Woosley2006,Nomoto2010,Hjort2013,Cano2011,Rueda2012,Cano2014a,Cano2017a,Guessoum2017,AguileraDena2018,Gompertz2020,Moriya2020UL}, which are studied here. We here recall that most of the nearby LGRBs are believed to be associated with an SN, but only a fraction of them (2-7\%) are visible due to instrumental selection biases \citep{Rossi2021}. On the other hand, a fraction of SNe Ib/c smaller than $\sim 10\%$ is associated with a GRB \citep{Soderberg2006latetime}. An estimation of the ratio between GRBs and SNe Ib/Ic leads to a percentage of GRB/SNe $\sim1-9\%$ \citep{Guetta2007,Ruffini2016}.

 The debate of whether or not all the LGRBs are associated with SNe has been fomented for a long time since 2006. Indeed, \citealt{DellaValle2006,Fynbo2006,Melandri2014} discussed cases of GRBs for which the SNe should have been seen if these were associated with the underlined GRBs: in particular, these are the cases of GRB 060614 and 060505. The first GRB has also revealed a Short with Extended Emission nature \citep{Kaneko2015}, while for the latter the SN associated has been debated by various authors \citep{Ofek2007}.

In the GRB-SNe events, the connection is often established by looking at the late times of their optical light curves (LCs) where the associated SNe usually show a photometric feature that points out their presence, namely a bump in the LC \citep{Cano2011}. Usually, for a GRB-SN associated event, in the late time of the optical LC three main components emerge: (I) the GRB AG, (II) the contribution from the host galaxy, and (III) the SN identified through the bump.

Concerning the morphology of GRBs, 
their LCs show two main phases: the prompt emission, where we observe the main event in $\gamma$-rays and hard X-rays, and the subsequent afterglow (AG), which is observed in soft X-rays, optical, and sometimes in radio.

\begin{figure}[h]
    \centering
    \includegraphics[scale=0.25]{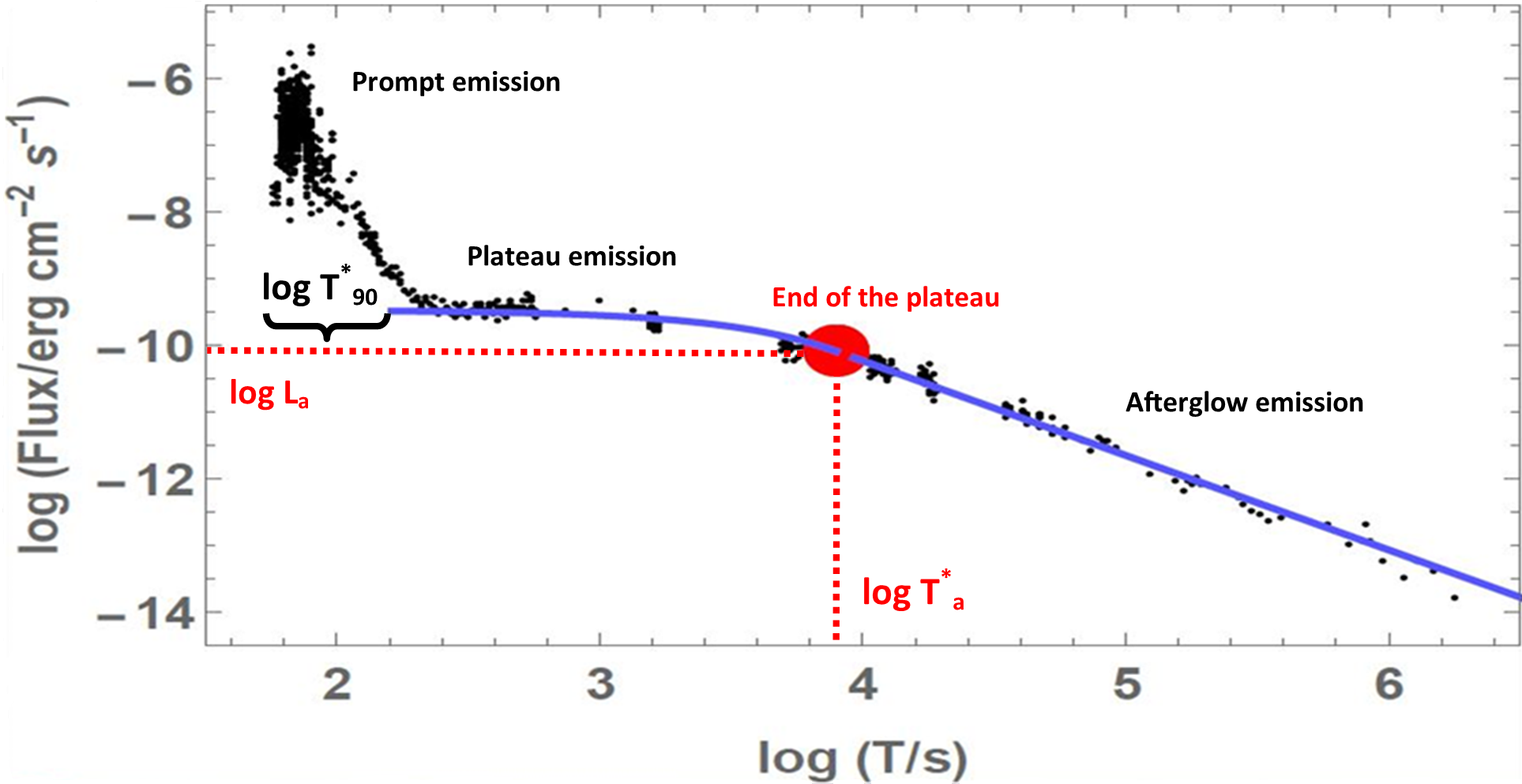}
    \includegraphics[scale=0.90]{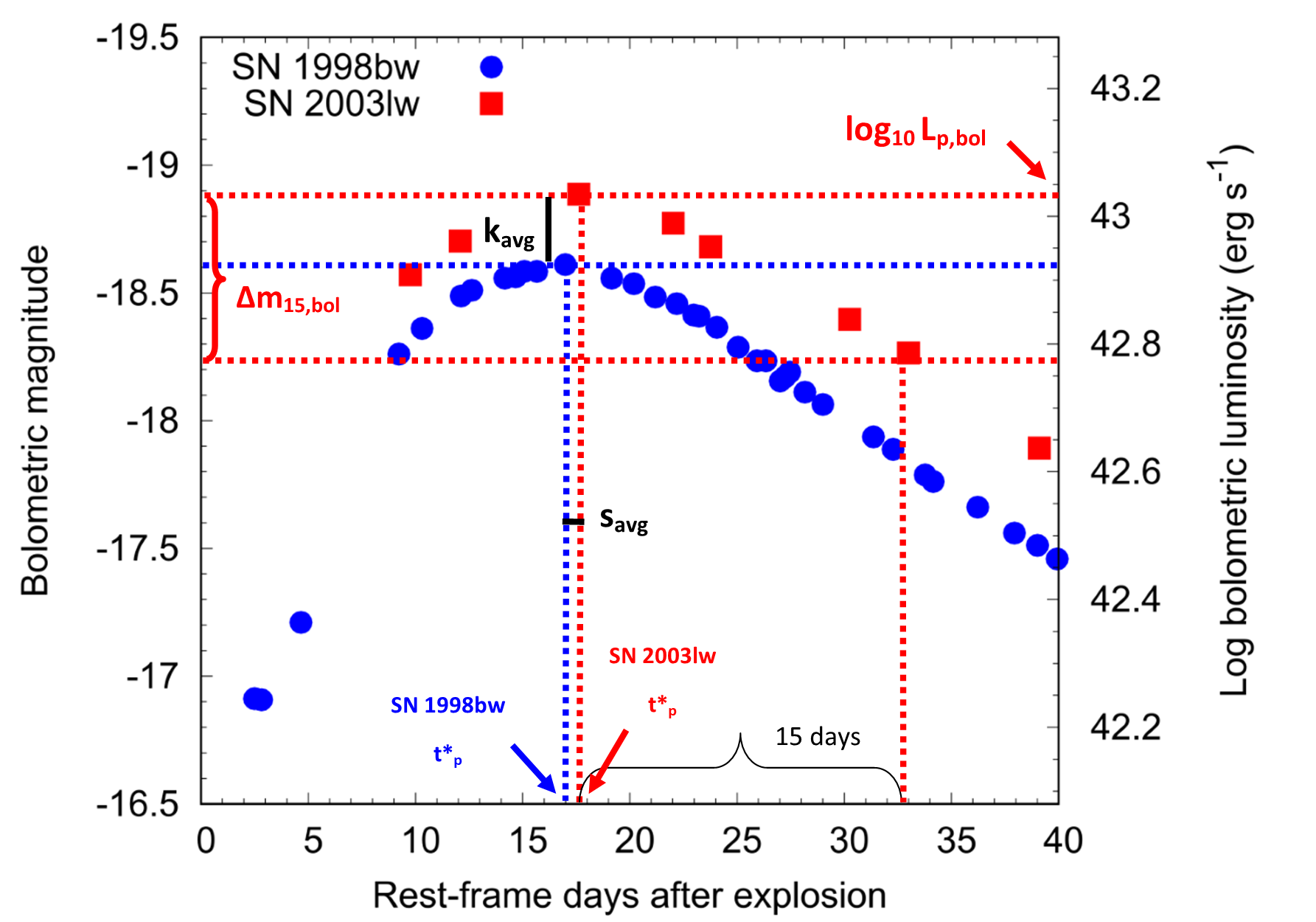}
    \caption{\textbf{Upper panel.} A schematic picture of a GRB LC in the rest frame with the main features. \textbf{Lower panel.} A plot where the properties of the SN LC in the case of association with a GRB are highlighted: we remind that the template LC is the one of SN 1998bw (colored in blue in the panel) while the examples of properties are referred to SN 2003lw.}
    \label{fig:GRBSNcartoon}
\end{figure}

The launch of the Swift satellite \citep{Gehrels2004,Sakamoto2008Swift} has uncovered a more peculiar behavior of the GRB LC than the simple power-law trend, this latter being explained within the traditional fireball model \citep{Granot1999,Piran1999,Meszaros2000,Piran2000,Meszaros2001,GranotSari2002}.
More specifically, the Swift satellite has discovered the existence of a new exciting feature in the AG phase: the plateau emission.
This flattening, which happens after the abrupt luminosity decrease of the prompt emission, is observed for more than $42\%$ of Swift X-ray LGRBs \citep{Evans2009}, see the upper panel of Figure \ref{fig:GRBSNcartoon}. Not only is the prompt emission at the center of the investigation for the search for relevant correlations, but also the plateau emission has been studied for this purpose.

The search in the AG properties is crucial as a step forward to building a standard candle. The plateau emission properties are appealing for the search for new correlations because the plateau features are more regular (e.g. they have no variability as is shown in the prompt emission) than the prompt ones. Indeed, once the prompt is available, the plateau emission, if present, is visible thanks to the Swift satellite XRT instrument. 

The plateau emission is involved in many relevant correlations between prompt and AG parameters in the GRB emission. We here mention the correlation between the X-ray end time of the plateau emission in the rest-frame ($\log T^{*}_{\rm a,X}$) and its correspondent luminosity in the same band ($\log L_{\rm a,X}$), the so-called Dainotti relation \citep{Dainotti2008,Dainotti2013,Dainotti2015,Dainotti2017a}.
This correlation means that the more luminous the plateau emission, the shorter its duration. Since the slope of this correlation is -1, this means that the energy reservoir of the plateau is constant.
There are also the prompt-AG relations between $L_{\rm a,X}$ and the 1-second X-ray peak luminosity of the GRB ($L_{\rm peak}$), see \citet{Dainotti2011b,Dainotti2015}. 
The former has been recently discovered also in the optical wavelengths \citep{DainottiLivermore2020}.
Extending the 2-dimensional relation into a 3-dimensional relation adding the $L_{\rm peak}$ to $L_{\rm a,X}$ and $T^{*}_{\rm a,X}$, a 3-dimensional relation can be found, called the fundamental plane relation \citep{Dainotti2016,Dainotti2017b,Dainotti2017plateau,Dainotti2020a,Cao2022a,Cao2022b}.

Many GRB correlations regarding the prompt emission have been discovered during the last 20 years: the Amati relation \citep{Amati2002,Ghisellini2006,Campana2007,Amati2008MG,Martone2017}, the Yonetoku relation \citep{Yonetoku2004,Ito2019}, the Ghirlanda relation \citep{LiangZhang2005,Ghirlanda2004,Ghirlanda2010}, the correlation between the intrinsic brightness and the decay rate of Swift GRBs detected with the UVOT instrument \citep{Oates2012}, the Tsutsui relation \citep{Tsutsui2013a,Tsutsui2013b}. However, a discussion on the prompt relations concerning the GRB prompt variables and their selection biases have been discussed in \citep{Shahmoradi2015}. Among these, the Amati, the Ghirlanda, and the Tsutsui relations hold also for the case of GRB-SNe, while for the AG relations we here mention that the 2-dimensional Dainotti relations in X-rays and in optical \citep{Dainotti2008,Dainotti2013,Dainotti2015,DainottiLivermore2020} are fulfilled also for the GRB-SNe cases.

Regarding the physical interpretation of the relations which involve the plateau emission, the so-called
Dainotti 2-dimensional and 3-dimensional correlations, these have been interpreted as resulting from a newly born spinning down magnetar \citep{Duncan2001,DallOsso2011,Rowlinson2013,Rowlinson2014,Rea2015,Stratta:2018xza} with an intense magnetic field ($B\sim10^{14}\div10^{15}\,{\rm G}$) or as the fallback accretion \citep{Kumar2008a,Kumar2008b,Cannizzo2009,Cannizzo2011}. Within the first model, a magnetar powers the AG plateau phase and collapses into a BH at the end of this phase, transitioning into the power-law decline typically seen in late AG LCs. In the second scenario, the Dainotti 2-dimensional relation suggests that the mass reservoir of the accretion disk is constant \citep{Cannizzo2011}.
The prompt-afterglow plateau correlation has been interpreted within the standard fireball model accounting for a variation of the micro-physical parameters, see \citet{VanEerten2014a,VanEerten2014b}.
We can also refer to additional models which represent alternatives to these current ones, for example, the prior emission model \citep{Yamazaki2008} and the supercritical pile GRB model \citep{Sultana2013}.

The aforementioned relations concerning the GRB-SNe events are limited to the cross-relation of only GRB properties among themselves, without taking into account the SN observables. Indeed, the connection between GRB and SN properties has been highlighted in the other three relevant correlations, which are listed below.
First, a correlation between the spectrum peak energy of the GRB, $E_{\rm p}$, and the peak bolometric magnitude of the SN was found by \citet{LiXinLi2006}: if an average luminosity SN has an accompanying GRB, the latter is expected to have soft spectra with a peak in the X-ray/UV wavelengths. 
The second is between the stretch, $s$, and luminosity factor, $k$, of SNe associated with GRBs \citep{Cano2014a}. This relation is similar to the peak luminosity-decline rate relationship (or Phillips relation, \citealt{Phillips1993}) used to establish SNe Type Ia as standard candles. For a schematic picture of these quantities, see the lower panel of Figure \ref{fig:GRBSNcartoon}.

The relation between the peak bolometric magnitude and $E_{\rm p}$ was also confirmed in \citet{Lu2018}. From this relation, \citet{Lu2018} derive the one between $E_{\rm p}$ and the SN nickel mass ($M_{\rm Ni}$) naturally, given the proportionality of the peak bolometric magnitude to $M_{\rm Ni}$ \citep{Arnett1982}, but with a probability of non-correlated data $\sim23\%$.

So far, only the above-mentioned few correlations have been investigated between the properties of GRBs and SNe, but with no restrictive metrics about the probability of chance occurrence. In addition, the quest for correlations in the GRB-SNe class has involved only the GRB prompt parameters to date.
Thus, we propose an extensive search in the literature among 91 pairs of variables between GRBs and SNe parameters to seek possible new correlations, significantly leveraging the GRB plateau properties. 
We here stress that this work is focused on the search for new correlations between GRB and SNe parameters and not between the GRB-GRB and SN-SN relations.

In this way, the AG emission plays a central role in searching for GRB-SNe correlations. 
In general, this work aims to investigate newly found GRB-SNe correlations, to advance the goal of standardizing GRB-SNe giving clues on their emission mechanism, and advance towards the goal of their use GRBs as cosmological probes.
More specifically, the more we know about the existence and reliability of the correlations, the more we can reveal and interpret their connection with the GRB emission mechanism. Once the phenomenon is known, then the variable which does not carry the information on the distance luminosity can be employed to use correlations as standard candles.

The paper is structured as follows: in Section \ref{sec:notations} we present the observables pertinent both to GRBs and SNe to date; in Section \ref{sec:sample}, we describe the data samples, the further classification of the investigated events, and the data selection before the analysis; in Section \ref{sec:EPresults} we show the results after the correction for selection biases and redshift evolution through the Efron \& Petrosian method (hereafter called the EP method, \citealt{EfronPetrosian1992}); in Section \ref{sec:fitting} we introduce the metrics for the existence of the correlations; in Section \ref{sec:results}, we provide the results of the analysis and we empirically discuss their possible physical interpretation. Finally, we conclude with a summary of findings in Section \ref{sec:conclusion}. In Appendix \ref{sec:scatterplot} the complete scatter matrices and all the possible cross-related pairs of variables are reported.
We here report also the correlations with GRB-GRB and SN-SN parameters for completeness, although we do not study nor discuss them.
In Appendix \ref{sec:ODR} we show the weighted fitting procedure with errors on both the variables. Lastly, in the Appendix \ref{sec:EP}, we provide a more detailed introduction of the reliable statistical EP method, which allows for overcoming biases and selection effects in the newly found correlations.

\section{Notations}\label{sec:notations}
For clarity, the nomenclature adopted in this paper is summarized in this Section. Together with the redshift of the GRB-SNe, we have gathered from the literature the values for 9 GRB parameters and 10 SN parameters. 
\begin{itemize}
    \item $z$: the redshift of the GRB-SN;
    \item $T^{*}_{90}$: the rest-frame time during which a GRB emits $90\%$ of its energy, from $5\%$ to $95\%$, expressed in ${\rm s}$;
    \item $E_{\gamma,\rm iso}$, $L_{\gamma,\rm iso}$: the isotropic energy and luminosity of the GRB expressed in ${\rm erg}$ and ${\rm erg/s}$, respectively;
    \item $E^{*}_{\rm p}$: the peak spectral energy of the GRB, reported in ${\rm keV}$, in the rest-frame: $E^{*}_{\rm p}=E_{\rm p}(1+z)$;
    \item $L_{\rm a,opt}$: the luminosity at the end of the GRB plateau emission, observed in the optical wavelengths and measured in ${\rm erg/s}$;
    \item $T^{*}_{\rm a,opt}$: the rest-frame time at the end of the GRB plateau emission, observed in the optical wavelengths and measured in ${\rm s}$;
    \item $L_{\rm a,X}$: the luminosity at the end of the GRB plateau emission, observed in the X-rays and measured in ${\rm erg/s}$;
    \item $T^{*}_{\rm a,X}$: the rest-frame time at the end of the GRB plateau emission, observed in the X-rays and measured in ${\rm s}$;  
    \item $\theta_{\rm jet}$: the jet opening angle of the GRB estimated with the jet-break time, expressed in degrees ($^\circ$);
    \item $T^{*}_{\rm jet}$: the rest-frame jet break time of the GRB, expressed in ${\rm days}$;
    \item $L_{\rm p,bol}$: the peak bolometric luminosity of the SN, measured in ${\rm erg/s}$;
    \item $\Delta m_{\rm 15,bol}$: the magnitude decline of the SN 15 days after its peak in the bolometric LC, in ${\rm mag}$, defined in the rest-frame \citep{Cano2017a};
    \item $t^{*}_{\rm p}$: the rest-frame time for the SN peak in the bolometric LC, after the GRB trigger, expressed in ${\rm days}$;
    \item $E_{\rm K}$: the kinetic energy of the SN, expressed in ${\rm erg}$;
    \item $M_{\rm ej}$: the ejecta mass of the SN, expressed in solar masses, ${\rm M_{\odot}}$;
    \item $M_{\rm Ni}$: the mass of nickel-56 ($^{56}{\rm Ni}$) produced in the SN explosion, expressed in $M_{\rm \odot}$;
    \item $v_{\rm ph}$: the photospheric velocity of the SN, measured in ${\rm km/s}$;
    \item $k_{\rm avg}$: the luminosity factor of the SN, averaged between the different bands, and compared to the template 1998bw SN lightcurve;
    \item $s_{\rm avg}$: the stretch factor of the SN, averaged between the different bands, and compared to the template 1998bw SN lightcurve.
    \end{itemize}

\noindent The superscript ($^{*}$) denotes that the quantity is estimated in the rest-frame only when the variables are not already by definition in the rest-frame, such as the GRB luminosities and isotropic energy. For example, concerning the time variables, we need to consider the cosmological evolution and divide by $(1+z)$, while we multiply by $(1+z)$ for peak energy. The superscript ($'$) denotes the variable de-evolved when redshift evolution is removed. We mention that the redshift $z$ is the only parameter common to the GRB and SN parameter sets.

Regarding the classes of GRBs, we here summarize the events pertinent to our analysis, and we quote their respective acronyms, which appear in Table \ref{subchartable0}:

\begin{itemize}
\item ULGRBs: the ultra-long GRBs, \citep{Nakauchi2013,Virgili2013,Levan2014,Piro2014,Schady2017,Perna2018,Moriya2020UL,Tsvetkova2021} with an unusual long duration, $T^{*}_{90}\geq 10^{3}\,s$ \citep{Gendre2019};
\item XRFs: the X-ray Flashes, for which the spectra is uncommonly soft and fluences in the X-ray band ($2$--$30$\,keV) are bigger than the ones in the $\gamma$-ray ($30$--$400$\,keV) band \citep{Heise2003,Sakamoto2004,Gendre2007}; 
\item llGRBs: low-luminosity events, where $L_{\gamma,\rm iso}<10^{48}\,{\rm erg/s}$ \citep{Liang2007,Virgili2009,Zhang2018NAT,Zhang2018}; 
\item INTs: the intermediate luminosity GRBs, where $10^{48}\,{\rm erg/s}<L_{\gamma,\rm iso}<10^{49.5}\,{\rm erg/s}$ \citep{Schulze2014,Cano2017a}. 
\end{itemize}
We here remind the reader that the rest of the GRBs with luminosities $L_{\gamma, \rm iso}>10^{49.5}\,{\rm erg/s}$ are considered typical GRBs and indicated with the label GRBs. We briefly define the classes of SNe that are present in the current sample of GRB-SNe associations: 

\begin{itemize}
    \item SNe Ib: show no silicon nor hydrogen line, but include helium line in their spectra;
    \item SNe Ic: the silicon, hydrogen, and helium lines are absent; together with the Ib, these mostly originate from the death of massive stars;
    \item SNe Ic-BL (broad-lined Ic): Ic with broader lines than the usual Ic spectra \citep{Chen2017,Modjaz2009};
    \item SLSNe (superluminous SNe): events that are around 100 times brighter than a Core-Collapse Supernova (CC SN) \citep{Nicholl2021,Tanaka2012}.
   \end{itemize}

In Figure \ref{fig:GRBSNcartoon}, we show a schematic GRB picture showing the most relevant variables for GRBs (see the upper panel of Figure \ref{fig:GRBSNcartoon}) and a plot of two SN LCs (the template SN 1998bw and the transient SN 2003lw) associated with GRBs where the parameters of the LC have been highlighted (see the lower panel of Figure \ref{fig:GRBSNcartoon}).

In our example, we consider the SN 2003lw for the definition of the parameters, while the average stretch and luminosity factors, $s_{\rm avg}$ and $k_{\rm avg}$, are shown concerning the template LC of SN 1998bw.

\section{Methodology}\label{sec:analysis}
This Section provides a complete description of the methods adopted to investigate the GRB-SNe correlations, describing the sample selection cut, the fitting procedures, and the selection bias corrections applied.

\subsection{The data sample selection} \label{sec:sample}
\noindent To conduct our analyses, we searched the literature for GRB-SN associations and gathered reported properties for both components.

Of the hundreds of GRBs annually observed, only $\sim2$-$7\%$ are spectroscopically or photometrically linked with SNe \citep{Rossi2021}. To date, $\sim40$ GRB-SNe associations have been identified by LC bumps, while other independent 28 GRB-SNe associations (differently from the 40 SNe) have been spectroscopically confirmed. In many cases, the connection has been made by cross-checking the times and locations of GRBs and SNe in their respective catalogs \citep{Wang1998,Bosnjak2006}, expanding to more than 100 possible associations. 

Taking as our primary reference the GRB-SN events discussed in \citet{Cano2017a}, our search uncovered 106 possible GRB-SN events spanning the 30-year time frame from 1991 to February 2021.
It is important to stress that all the SN LCs associated with GRBs present in this sample consider the event SN 1998bw as the template.
When we do not find out values in \citet{Cano2017a} we refer to other values tabulated in the literature referenced in Table \ref{chartable04}. Concerning the $v_{\rm ph}$ parameter, when we do not find the uncertainty for the reported value, we substitute it with "-" in the Table \ref{chartable04}, although a standard deviation $\sigma_{v_{\rm ph}}=8000\,km/s$ is reported in \citet{Cano2017a}. This was estimated considering the distribution of the $v_{\rm ph}$ values.
We also stress that we consider in the analysis only bolometric variables and not the ones in the given bands (e.g. the V-band absolute magnitude of the SN, $M_V$).

We report many SNe associated with GRBs in tables even if we do not have the associated estimated SN properties. We have reported them in any case since the associations that have only GRB properties are used to investigate the evolutionary effects through the statistical EP method. Thus, these tables are complete so that any reader can safely reproduce and compare the results.  
From the list of 106 probable associations, we remove 35 events due to the lack of data, the uncertain association, or the association with SNe II (and even with SN Ia). These latter associations with other spectroscopical classes than the Ib/c have arisen solely by a spatial and temporal coincidence of two different events and are based on old data (in the years $\sim$1991), given that GRBs are associated with SNe Ib/c and not with SNe Ia nor SNe II.

Because of these uncertainties or unreliable observable parameters, the following 35 events were removed: 
GRB 920321/SN 1992Q, 
GRB 951107C/SN 1995bc,
GRB 970514/SN 1997cy, 
GRB 971221/SN 1997ey, 
GRB 980525/SN 1998ce, 
GRB 980910/SN 1999E, 
GRB 990902/SN 1999dp, 
GRB 011121/SN 2001ke, 
GRB 920613/SN 1992ae, 
GRB 920628/SN 1992at,
GRB 920708/SN 1992al, 
GRB 920925/SN 1992bg,
GRB 930524/SN 1993R,
GRB 950917/SN 1995ac,
GRB 961029/SN 1996bx,
GRB 970907/SN 1997dg,
GRB 971218/SN 1998B, 
GRB 980503/SN 1998ck,
GRB 990527/SN 1999ct,
GRB 990719/SN 1999dg,
GRB 990810,
GRB 991015/SN 1999ef,
GRB 991123/SN 1999gj,
GRB 000319,
GRB 000415/SN 2000ca,
GRB 010921,
GRB 020410,
GRB 080503,
GRB 090426,
iPTF15dld,
SN 2020bvc,
GRB 210610B,
GRB 211015A, 
GRB 211023A/AT 2021acco, 
and GRB 211211A (a recent transient associated with a Kilonova, \citealt{Rastinejad2022KN}).

Removing these GRBs leaves us with a total sample of 71 GRBs, tabulated in Table \ref{chartable04}.
This sample includes 58 associated GRB-SNe events according to the grading scheme of \cite{Hjorth2012}.
This scheme can be summarized in 5 classes A,B,C,D, and E: Class A) have reliable spectroscopic evidence of the SNe; B) have a well-defined LC bump as well as some spectroscopic evidence resembling a GRB-SN; C) shows a bump which is evident and consistent with other GRB-SNe located at the spectroscopic redshift of the GRB; D) here a bump is present but the SN properties are not fully compatible with other GRB-SNe, or the bump is not well sampled or there is no spectroscopic redshift of the GRB; E) a bump, is either not significant or inconsistent with other GRB-SNe.

In the sample we have $9$ A-graded, $4$ AB-graded, $14$ B-graded, $1$ BC-graded, $11$ C-graded, $1$ CD-graded, $8$ D-graded, $2$ DE-graded, $8$ E-graded events. The remaining $13$ events included in the 71 GRBs sample are ungraded.

Previous studies have focused on A and B classes only \citep{Lu2018}. 

In the current analysis, we retain all the possible GRB-SNe associations with sufficient data coverage. We also consider the plateau emission in the AG for 15 GRBs in the X-ray of the GRBs associated with SNe gathered in our sample. The luminosity at the end of the plateau emission is obtained through the LCs fitting with the Willingale model \citep{Willingale2007}, while the other parameters at play have been collected from the literature. In our sample, there are $15$ and $20$ GRBs with X-ray and optical plateau emission, respectively.

We found in our sample that the majority belong to the Ic class ($17$ events) and SNe Ic-BL ($7$ events), but there are several SNe Ib ($3$ events), and SNe with no clear distinction between Ib and Ic classes ($5$ events). We also have one single case of SLSN.
After gathering all the properties for the reliable GRB-SNe associations, a further sample selection cut is performed by removing the two shock breakout events.
The mechanisms for the production of LGRBs and shock breakout events are different: the former is likely generated from a collapsar, while the latter is probably linked to failed jets, namely events where the ultra-relativistic jets do not pierce into the stellar envelope and release their energy inside the progenitor, causing a relativistic shock breakout emission. 
The role of relativistic jets in the emission of GRB-SNe has been widely discussed in \citet{Aloy1999,Zhang2002jet,Umeda2005,Tominaga2007,Nomoto2008,Nagataki2009jet,Granot2014,Kumar2015,Eisenberg2022}.

Two events, given their possible nature of shock breakouts, have been reported in our GRB-SNe sample but for the safety of separation from the other classes have been excluded from the analysis: GRB 060218/SN 2006aj \citep{Campana2006,Ferrero2006,Pian2006,Soderberg2006,Sollerman2006, Dainotti2007,Waxman2007,Campana2008,Zhang2022-SN2006aj} and GRB 080109/SN 2008D \citep{Li080109,Mazzali2008,Modjaz2009}. The former has an energy many hundred times smaller than other cosmological GRBs, while the latter has an SN with relatively high energy if compared with the other CC SNe. 
For the other GRB-SNe events characterized by a relatively low luminosity, a consensus on their progenitor mechanism does not exist despite many of their features being typical of the shock breakout events. Indeed, a shock breakout scenario can be found also in GRBs that are not low luminous, like the case of GRB 120422A/SN 2012bz \citep{Schulze2014}, so this is not necessarily a typical behavior of the llGRBs. We here recall that the other events classified as llGRBs are GRB 980425/SN 1998bw \citep{Galama1998,Pian2004}, GRB 100316D/SN 2010bh \citep{Starling2010}, and GRB 171205A/SN 2017iuk \citep{Barthelmy2011,Delia2018,Suzuki2019}. The subsequent analysis will be then performed on the sample of 69 GRB-SNe, removing the two shock breakouts GRB 060218/SN 2006aj and GRB 080109/SN 2008D from the aforementioned sample of 71 GRB-SNe associations.

In our sample, the GRB variables are affected by a mean ratio of errors over measurements $\Delta x_{GRB}/x_{GRB}\sim 21\%$, while the SN ones show larger relative errors leading to a mean $\Delta x_{SN}/x_{SN}\sim39\%$. The extreme cases are the log-luminosity at the end of the optical GRB plateau with the smallest ratio in our sample ($\Delta \log_{10}{L_{a,opt}}/\log_{10}{L_{a,opt}}\sim0.1\%$) and the SN kinetic energy with the largest relative uncertainties ($\Delta E_K/E_K\sim94\%$). The details of these estimations are reported in Table \ref{tab:errors}: in the first column, we report the variable, in the second we report the mean of the ratios between the symmetrized 1 $\sigma$ errors and the values, while in the third column we report the number of data points involved in the estimation (namely, the number of events that have both the central value and the error for the given variable).

\begin{table}[h]
    \centering
    \begin{tabular}{c|c|c}
        Variable & $<\Delta x/x>$ & N. of GRBs\\
        \hline 
        $T^{*}_{90}$ & 15\% & 54 \\
        $E_{\gamma,\rm iso}$ & 51\% & 51 \\
        $E^{*}_{\rm p}$ & 52\% & 55 \\
        $\log_{10}L_{\rm a,opt}$ & 0.1\% & 20 \\
        $\log_{10}T^{*}_{\rm a,opt}$ & 1\% & 20 \\
        $\log_{10}L_{\rm a,X}$ & 0.2\% & 15 \\
        $\log_{10}T^{*}_{\rm a,X}$ &  2\% & 15 \\
        $\theta_{\rm jet}$ & 17\% & 16 \\
        $T^{*}_{\rm jet}$ & 50\% & 16 \\
        $\log_{10}L_{\rm p,bol}$ & 1\% & 22 \\
        $E_{\rm K}$ & 94\% & 34 \\
        $M_{\rm ej}$ & 58\% & 34 \\
        $M_{\rm Ni}$ & 47\% & 36 \\
        $v_{\rm ph}$ & 35\% & 22 \\
        $k_{\rm avg}$ & 22\% & 35 \\
        $s_{\rm avg}$ & 15\% & 33 \\
    \end{tabular}
    \caption{The table contains the variable name, the mean of relative errors for the variables, indicated with $<\Delta x/x>$, and the number of GRBs present for each variable.}
    \label{tab:errors}
\end{table}

\subsection{The application of the EP method to the variables}\label{sec:EPresults}

\begin{figure}[h]
    \hspace{35ex}
    \includegraphics[scale=0.70]{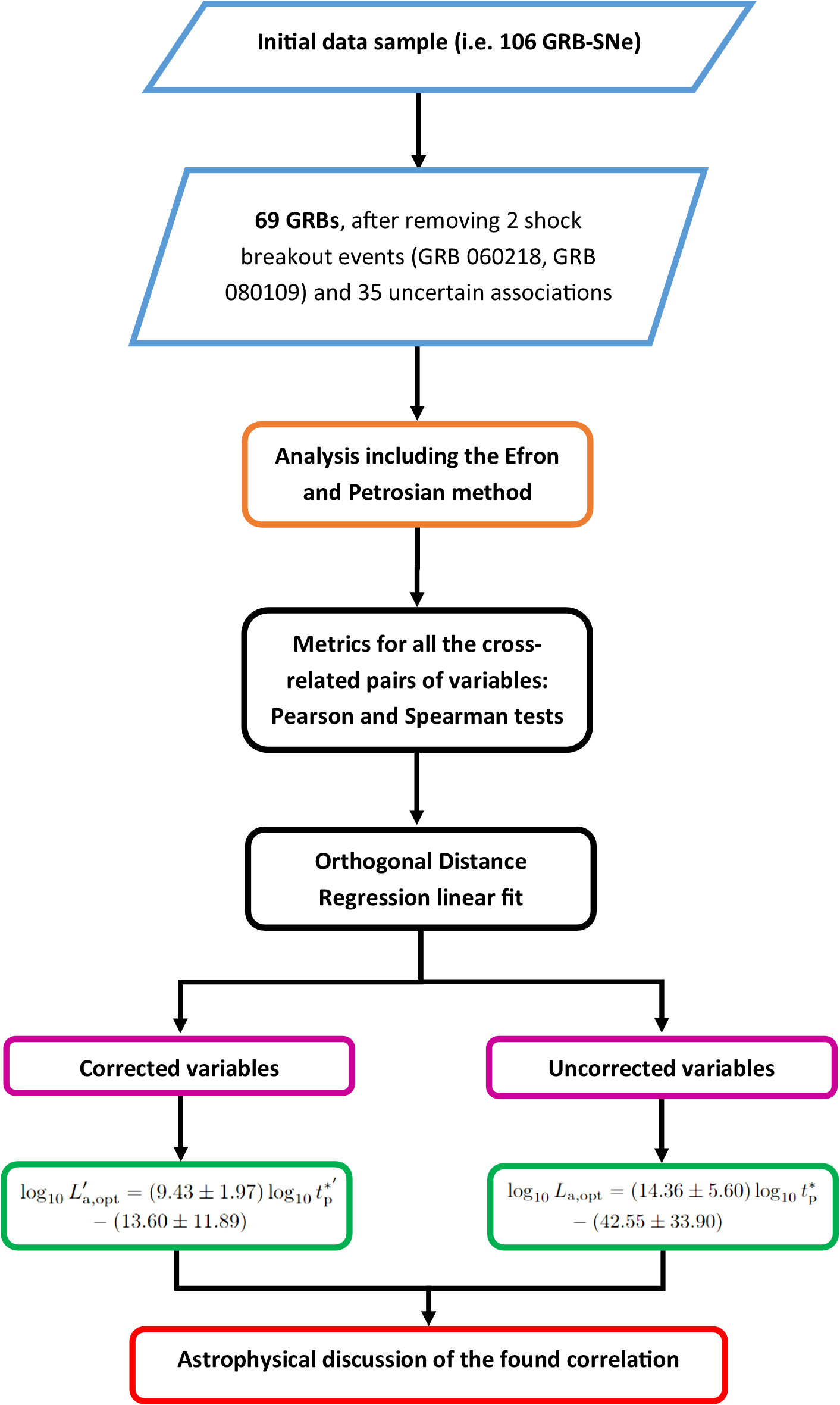}
    \caption{The flow chart here presented summarizes the steps of the current research for probable correlations in the sample of GRB-SNe associations.}
    \label{fig:flowchart}
\end{figure}

The EP method \citep{EfronPetrosian1992} is a non-parametric method that involves an adapted version of Kendall's $\tau$ statistics \citep{kendall1948rank}. The method can be used to address the problem of the statistical dependence of the astrophysical variables on the redshift: in this way, the presence of hidden biases or selection effects for the investigated quantities can be revealed.
Indeed, the EP could even reveal intrinsic correlations which would otherwise be hidden and masked out by the presence of redshift evolution and biases. On the other hand, the method can reveal if there are correlations that are simply induced by selection biases and redshift evolution, and they are not intrinsic to the GRB physics. 

More specifically, the EP method defines evolutionary functions, $g(z)=(1+z)^k$ where $k$ is the evolutionary coefficient so that the new de-evolved variables indicated with $'$ will be divided by this function.
Despite the continuous improvement of instruments, telescopes, and facilities with the subsequent enhanced precision and sensitivity in measurements, the biases still affect the data, given that each satellite has its instrumental selection effect. One of the principal biases is the so-called Malmquist bias \citep{Malmquist}: according to this, at larger distances (or redshifts), it is easier to observe the more luminous objects rather than the fainter ones. 
The presence of biases in astrophysics affects also cosmological analysis \citep{DainottiDeSimone2021,DainottiDeSimone2022}, so it is important to tackle them.\\
The EP method's description is present in the Appendix \ref{sec:EP}. We have applied it to several selected variables which are expected to undergo redshift evolution, both for GRB and SNe parameters.
A complete correction of all the variables in the GRB-SNe sample can be found in Table \ref{tab:EP_coefficients}. 
We here stress that we used the GRB-SNe sample to evaluate the evolutionary effects to allow the limiting variables, such as fluxes, times, or fluences sample to be as much as representative of the current sample. We also note that when the sample is small we use the full LGRBs class and the results of the evolutionary parameters are compatible within 1 $\sigma$.

\begin{table}[h]
    \centering
    \begin{tabular}{|c|c|c|c|}
    \hline
    Quantity & EP coefficient & Symmetrized error & Computed in\\
    \hline
    $T^{*}_{90}$ & $k_{T^{*}_{90}} = -0.78^{+0.89}_{-0.95}$ & $\delta k_{T^{*}_{90}}=0.92$ & This paper\\
    $E_{\gamma,\rm iso}$ & $k_{E_{\gamma,\rm iso}} = 2.30\pm0.50$ & $\delta k_{E_{\gamma,\rm iso}}=0.50$ & \cite{Lloyd2019}\\
    $L_{\gamma,\rm iso}$ & $k_{L_{\gamma,\rm iso}} = 3.50\pm0.50$ & $\delta k_{L_{\gamma,\rm iso}}=0.50$ & \cite{Lloyd2019}\\
    $E^{*}_{\rm p}$ & $k_{E^{*}_{\rm p}}=0.75\pm0.25$ & $\delta k_{E^{*}_{\rm p}}=0.25$ & \cite{Xu2021} \\
    $L_{\rm a,opt}$ & $k_{L_{\rm a,opt}} = 3.96\pm0.58$ & $\delta k_{L_{\rm a,opt}}=0.58$ & \cite{DainottiLevine2021} \\
    $T^{*}_{\rm a,opt}$ & $k_{T^{*}_{\rm a,opt}} = -2.11\pm0.49$ & $\delta k_{T^{*}_{\rm a,opt}}=0.49$ & \cite{DainottiLevine2021}\\
    $L_{a,X}$ & $k_{L_{a,X}} = 2.42\pm0.58$ & $\delta l_{L_{a,X}}=0.58$ & \cite{DainottiLevine2021}\\
    $T^{*}_{a,X}$ & $k_{T^{*}_{a,X}} = -0.85\pm0.30$ & $\delta k_{T^{*}_{a,X}}=0.30$ & \cite{DainottiLevine2021}\\  
    $\theta_{\rm jet}$ & $k_{\theta_{\rm jet}}=0.75\pm0.25$ & $\delta k_{\theta_{\rm jet}}=0.25$ & \cite{Lloyd2020}\\
    $T^{*}_{\rm jet}$ & - & - & (*)\\
    $L_{\rm p,bol}$ & $k_{L_{\rm p,bol}}=0.77^{+0.80}_{-0.25}$ & $\delta k_{L_{\rm p,bol}}=0.52$ & This paper \\
    $\Delta m_{\rm 15,bol}$ & - & - & (*)\\
    $t^{*}_{\rm p}$ & $k_{t^{*}_{\rm p}}=-0.43^{+0.19}_{-0.24}$ & $\delta k_{t^{*}_{\rm p}}=0.22$ & This paper\\
    $E_K$ & $k_{E_K} = -0.10^{+0.55}_{-0.90}$ & $\delta k_{E_K}=0.72$ & This paper\\
    $M_{\rm ej}$ & $k_{M_{\rm ej}} = -0.45^{+0.81}_{-0.52}$ & $\delta k_{M_{\rm ej}}=0.66$ & This paper\\
    $M_{\rm Ni}$ & $k_{M_{\rm Ni}}=0.07^{+0.77}_{-1.04}$ & $\delta k_{M_{\rm Ni}}=0.90$ & This paper \\
    $v_{\rm ph}$ & - & - & (*)\\
    $k_{\rm avg}$ & - & - & (*)\\
    $s_{\rm avg}$ & $k_{s_{\rm avg}}=0.16^{+0.21}_{-0.22}$ & $\delta k_{s_{\rm avg}}=0.22$ & This paper\\
    \hline
    \end{tabular}
    \caption{Summarizing of the EP coefficients for the GRB-SNe parameters present in this paper. (*) means that the EP correction cannot be applied due to the paucity or the sparsity of the data.}
    \label{tab:EP_coefficients}
\end{table}

We here report the corrections related to variables found for $t^{*}_{\rm p}$, which is one of the variables present in our correlations, while for $v_{\rm ph}$ there is no evolution. For the plateau properties in the X-rays and optical bands, in particular $L_{\rm a,opt}$, the method does not allow precise estimates of the coefficients given the small sample related to the GRB-SN connection, so we adopt the values from a recent study in which the sample size is much larger, and we assume that the coefficients of the evolution do not change between the different samples \citep{DainottiLevine2021}. Thus, we take the evolutionary coefficients as $k_{L_{\rm a,opt}} = 3.96^{+0.58}_{-0.58}$.
Concerning the $T^{*}_{90}$, the determined $k_{T^{*}_{90}} = -0.84^{+0.73}_{-0.67}$ is compatible in 1 $\sigma$ with the one of \citet{Lloyd2019}, thus validating our approach in the current analysis. For the variables $M_{\rm Ni}$ and $M_{\rm ej}$ we find $k_{M_{\rm Ni}}=-0.07^{+0.77}_{-1.04}$ and $k_{M_{\rm ej}}=-0.45^{+0.81}_{-0.52}$, respectively. Differently from the other variables, in the $M_{\rm Ni}$ and $M_{\rm ej}$ cases it is expected that no evolutionary effect is present: indeed, the coefficients are compatible with zero in 1 $\sigma$. Thus, we simply report them in the Table \ref{tab:EP_coefficients} but in the subsequent analysis, we will assume $k_{M_{\rm Ni}}=0$ and $k_{M_{\rm ej}}=0$. Finally, for what it concerns the $s_{\rm avg}$ and $k_{\rm avg}$ parameters, an evolution is not expected and we will assume $k_{s_{\rm avg}}=0$, $k_{k_{\rm avg}}=0$; nevertheless, we performed a consistency check on the $s_{\rm avg}$ parameter since it is not unlikely that the stretch parameter may undergo a redshift evolution effect, as pointed out in \citet{Nicolas2021} for the SNe Ia. Thus, we estimated the following value of the correction parameter for the stretch: $k_{s_{\rm avg}}=0.16^{+0.21}_{-0.22}$. Being this value compatible with zero in less than 1 $\sigma$, our assumption of $k_{s_{\rm avg}}=0$ is legit and will be taken into account for the rest of the analysis.

The flowchart shown in Figure \ref{fig:flowchart} explains the steps of this research and how the EP method enters into play.

\subsection{The metrics of our analysis}\label{sec:fitting} 
This Section introduces the metrics and the statistical tests used for our analysis: the Pearson and the Spearman, as also shown in the third step of the flowchart in Figure \ref{fig:flowchart}.
We will denote Pearson's correlation coefficient and p-value as $r$ and $P_P$, respectively, while for Spearman's rank coefficient and p-value we will write $\rho$ and $P_\rho$, respectively. The p-values $P_P$ and $P_\rho$ express the probability that the correlation may be drawn by chance, see for results Section \ref{sec:results}.
The Pearson correlation coefficient, $r$, \citep{Pearson} is a measure of how well a linear correlation can describe the data.


Spearman's $\rho$ \citep{Spearman} measures how well a monotonic function can describe the relationship between two lists of data, basing the result on the rank differences. 



The coefficients $r$ and $\rho$ both represent the correlation between the two distributions, and they both range from $-1$ to $+1$ (from the anticorrelation to the correlation case, respectively), but while the Pearson indicates the presence of a linear relation, the Spearman takes into account the ties and the ordering of the distribution, highlighting a monotonic function for the data, if present.

Both the Pearson and the Spearman have been performed with a null hypothesis $\mathcal{H}_0$ that the vectors are independent, and as the complementary hypothesis, $\mathcal{H}_a$ that they are not. 
A small p-value suggests that it is improbable that $\mathcal{H}_0$ is true; namely, it is unlikely that the two vectors are independent. In other words, if they are not independent, the smaller the p-value, the higher the probability that the correlation is not drawn by chance. 
We assess the presence of the correlations with the metrics indicated by the Pearson and/or Spearman correlation and their relative probability that the correlation is driven by chance. Specifically, the correlations must fulfill at least one of the two following conditions: (I) $|r|\geq0.50$ and $P_P\leq0.05$ for the Pearson's test, and/or (II) $|\rho| \geq 0.50$ and $P_\rho\leq0.05$ for the Spearman's test.

We here remind the reader that this work is focused on the search for new correlations between GRB and SNe parameters and not between the GRB-GRB and SN-SN relations. Thus, we limit ourselves to reporting the GRB-GRB and the SN-SN in Table \ref{chartable02} for the cases without EP-correction and in Table \ref{chartable03} for the cases that include the EP-correction without further discussing their physical meaning, since the majority of them have been discussed in the literature already. Despite many of the GRB-SN relations satisfying the criteria (I) and/or (II), we do not investigate them due to the paucity of data (less than 5 data points) or the absence of the EP-correction coefficient for one of the variables. The GRB-SNe relations which fulfill the metrics, but have not been discussed, are $\Delta m_{\rm 15,bol}-\log_{10}T^{*}_{\rm jet}$ (both with and without EP-correction) since the correlation is drawn by only 3 data points, $\log_{10}v_{\rm ph}-\log_{10}\theta'_{\rm jet}$, and $\log_{10}v_{\rm ph}-\log_{10}L'_{\gamma,\rm iso}$ since the $v_{\rm ph}$ variable has no EP-correction coefficient. Nevertheless, we cannot exclude a priori the existence of these correlations that may be, indeed, confirmed with the future observations of GRB-SNe.

In all the analyses we tested the criteria (I) and (II) and we did not find any candidate correlation that fulfills criterion (II). However, since the Tables \ref{chartable02}, \ref{chartable03} show all the relations, we present anyway the analysis performed with this criterion giving in the table both the Pearson and Spearman correlation coefficients.
Using the sample of 69 GRB-SNe, we begin our analysis by comparing each property of the GRB-SN sample against every other. To this end, we cross-correlate all the variables and we compute the correlation coefficients $r$, $P_r$, $\rho$, and $P_{\rho}$ through the \textit{scipy.stats} module\footnote{For Python package SciPy, see \url{https://docs.scipy.org/doc/scipy/reference/generated}}.  

After the metrics have been fulfilled we need to determine the slope and the normalization through a fitting procedure. For the correlations which respect at least one condition between (I) and (II) we perform a weighted linear regression (namely, with the model $y=ax+c$, being $a$ the slope of the correlation and $c$ the intercept), using the Orthogonal Distance Regression method (hereafter ODR, \citealt{Boggs1989}). The ODR method deals with the problem of best fitting for data when the uncertainties are present not only on the dependent variable but also on the independent variable. For a quick introduction to the ODR method, see Appendix \ref{sec:ODR}. The ODR regression is performed with the \textit{scipy.odr} package \footnote{For the SciPy ODR package, see \url{https://docs.scipy.org/doc/scipy/reference/odr.html}}.
The error on each of the variables involved in the ODR fitting is estimated through the formula of error propagation, involving the uncertainties related to the EP evolutionary coefficients together with the 1 $\sigma$ uncertainties of the variables, when present.

It is important to discuss the impact of the errors in the analysis of the candidate correlation. All the GRB plateau observables in our sample are affected by errors $<2\%$ (see Table \ref{tab:errors}), thus enforcing the reliability of the results in the research of the correlation between the SN and the GRB AG properties.
It must be stressed that the uncertainties of the variables do not affect the metrics of our analysis and, thus, do not affect the identification of possible new correlations. This is due to the fact that to perform the Pearson and Spearman tests, we consider the central values of the parameters and not their uncertainties, focusing the subsequent investigation on the possible correlations that satisfy criterion (I).

The ODR fitting is applied to the following cases for each correlation: the sample with all the GRBs that obey the found correlations and the subsample of the correlations containing only the A, AB, and B-graded events, the so-called AB subsample.
It is important to focus on correlations fulfilled by a sufficient number of at least 5 data points.
We summarize in Table \ref{subchartable0} the values of the parameters which show a clear correlation. 
In Appendix \ref{sec:scatterplot}, we report the results of the cross-checking for all the possible pairs of variables present in our GRB-SNe sample.
The Pearson and Spearman coefficients for the tested pairs of the EP-corrected parameters are reported in the Appendix \ref{sec:scatterplot}, together with the tables mentioned above containing the cases without the EP corrections.

All the steps of the current analysis are summarized in the flow chart of Figure \ref{fig:flowchart}. Although it is important to stress that the relevant correlations are the ones that have been corrected for selection effects and redshift evolution, in this work we report also the correlations found in the cases with the observed variables without the EP-correction and the hybrid cases where the EP correction is present only for one of the variables in each pair. In this way, we provide a useful reference for future comparison between the corrected and uncorrected correlations.

\begin{longrotatetable}
\begin{deluxetable*}{lllllllllllllllll}
\tablecaption{GRB-SNe parameters present in our correlations\label{subchartable0}}
\tablewidth{600pt}
\tabletypesize{\scriptsize}
\tablehead{
\colhead{GRB ID} & \colhead{SN ID} & 
\colhead{GRB Type} & \colhead{SN Type}  & \colhead{$\log_{10}L_{\rm a,opt}$} & \colhead{$t^{*}_{\rm p}$} \\
\colhead{} & \colhead{} &
\colhead{} & \colhead{} & 
\colhead{($erg/s$)} & \colhead{(days)}
}

\startdata
910423 & 1991aa & \nodata & Ib & \nodata & $\cdots$ & ~ \\ 
960221 & 1996N & \nodata & Ib & \nodata & $\cdots$ & ~ \\ 
960925 & 1996at & \nodata & Ic-Ib/c & \nodata & $\cdots$ & ~ \\ 
961218 & 1997B & \nodata & Ic-Ib/c & \nodata & $\cdots$ & ~ \\ 
970228 & \nodata & GRB & \nodata & \nodata & $\cdots$ & ~ \\ 
970508 & \nodata & \nodata & Ib/c & \nodata & $16.79\pm 3.30$ & ~ \\ 
971013 & 1997dq & \nodata & Ib & \nodata & $\cdots$ & ~ \\ 
971115 & 1997ef & \nodata & Ic & \nodata & $\cdots$ & ~ \\ 
971120 & 1997ei & \nodata & Ic & \nodata & $\cdots$ & ~ \\ 
980326 & \nodata & GRB & \nodata & \nodata & $\cdots$ & ~ \\ 
980425 & 1998bw & llGRB & Ic & \nodata & 15.16 & ~ \\ 
980703 & \nodata & \nodata & \nodata & \nodata & $\cdots$ & ~ \\ 
990712 & \nodata & GRB & \nodata & \nodata & $\cdots$ & ~ \\ 
991002 & 1999eb & \nodata & \nodata & \nodata & $\cdots$ & ~ \\ 
991021 & 1999ex & \nodata & Ic & \nodata & $\cdots$ & ~ \\ 
991208 & \nodata & GRB & \nodata & \nodata & $\cdots$ & ~ \\ 
000114 & 2000C & \nodata & Ic & \nodata & $\cdots$ & ~ \\ 
000418 & \nodata & \nodata & \nodata & \nodata & $\cdots$ & ~ \\ 
000911 & \nodata & GRB & \nodata & \nodata & $\cdots$ & ~ \\ 
020405 & \nodata & GRB & \nodata & \nodata & $\cdots$ & ~ \\ 
020903 & \nodata & llGRB & \nodata & \nodata & $\cdots$ & ~ \\ 
021211 & 2002lt & GRB & Ic & \nodata & $\cdots$ & ~ \\ 
030329 & 2003dh & GRB & Ic-BL & $44.11\pm0.09$ & 12.75 & ~ \\ 
030723 & \nodata & XRF & \nodata & \nodata & $\cdots$ & ~ \\ 
031203 & 2003lw & INT & Ic & \nodata & 17.33 & ~ \\ 
040924 & \nodata & GRB & \nodata & $45.26\pm0.13$ & $\cdots$ & ~ \\ 
041006 & \nodata & GRB & \nodata & $44.76\pm0.07$ & $\cdots$ & ~ \\ 
050416A & \nodata & GRB & \nodata & $43.70\pm0.08$ & $\cdots$ & ~ \\ 
050525A & 2005nc & GRB & Ic & $45.51\pm0.04$ & $\cdots$ & ~ \\ 
050824 & \nodata & GRB & \nodata & $44.87\pm0.06$ & $\cdots$ & ~ \\ 
060218 & 2006aj & llGRB & Ic-BL & \nodata & $\cdots$ & ~ \\ 
060729 & \nodata & GRB & \nodata & $44.88\pm0.03$ & $\cdots$ & ~ \\ 
060904B & \nodata & GRB & \nodata & $45.25\pm0.05$ & $\cdots$ & ~ \\ 
070419A & \nodata & INT & \nodata & $45.05\pm0.14$ & $\cdots$ & ~ \\ 
071025 & \nodata & \nodata & \nodata & \nodata & 11.9 & ~ \\ 
071112C & \nodata & \nodata & \nodata & \nodata & $\cdots$ & ~ \\ 
080109 & 2008D & XRF & Ib & \nodata & 12 & ~ \\ 
080319B & \nodata & GRB & \nodata & \nodata & $\cdots$ & ~ \\ 
081007A & 2008hw & GRB & Ic & $43.98\pm0.10$ & 15 & ~ \\ 
090618 & \nodata & GRB & \nodata & $45.17\pm0.01$ & 8.76 & ~ \\ 
091127 & 2009nz & GRB & Ic-BL & $44.83\pm0.04$ & $\cdots$ & ~ \\ 
100316D & 2010bh & llGRB & Ic & $42.06\pm0.11$ & 11.8 & ~ \\ 
100418A & \nodata & GRB & \nodata & $44.19\pm0.12$ & $\cdots$ & ~ \\ 
101219B & 2010ma & GRB & Ic & $44.97\pm0.02$ & $\cdots$ & ~ \\ 
101225A & \nodata & ULGRB & \nodata & $42.43\pm0.06$ & 22.7 & ~ \\ 
111209A & 2011kl & ULGRB & SLSN & $45.10\pm0.08$ & 14.45 & ~ \\ 
111211A & \nodata & \nodata & \nodata & \nodata & 13.6 & ~ \\
111228A & \nodata & GRB & \nodata & $45.52\pm0.01$ & $\cdots$ & ~ \\ 
120422A & 2012bz & GRB & Ic & \nodata & $\cdots$ &  ~ \\ 
120714B & 2012eb & INT & Ib/c & \nodata & 13 & ~ \\ 
120729A & \nodata & GRB & \nodata & \nodata & 12.94 & ~ \\ 
130215A & 2013ez & GRB & Ic & \nodata & 11.9 & ~ \\ 
130427A & 2013cq & GRB & Ic & \nodata & $\cdots$ & ~ \\ 
130702A & 2013dx & INT & Ic & $43.59\pm0.02$ & $\cdots$ & ~ \\ 
130831A & 2013fu & GRB & Ib/c & $45.52\pm0.03$ & $\cdots$ & ~ \\ 
140206A & \nodata & \nodata & \nodata & \nodata & $\cdots$ & ~ \\ 
140606B & iPTF14bfu & GRB & Ic-BL & \nodata & $\cdots$ & ~ \\ 
150818A & \nodata & INT & \nodata & \nodata & $\cdots$ & ~ \\ 
161219B & 2016jca & INT & Ic & \nodata & $\cdots$ & ~ \\ 
161228B & iPTF17cw & \nodata & Ic-BL & \nodata & $\cdots$ & ~ \\ 
171010A & 2017htp & \nodata & Ic-BL & \nodata & $\cdots$ & ~ \\ 
171205A & 2017iuk & \nodata & Ic-BL & \nodata & $\cdots$ & ~ \\ 
180720B & \nodata & \nodata & \nodata & \nodata & $\cdots$ & ~ \\ 
180728A & 2018fip & XRF & Ic & \nodata & $\cdots$ & ~ \\ 
190114C & 2019jrj & \nodata & \nodata & \nodata & $\cdots$ & ~ \\ 
190829A & 2019oyw & \nodata & Ic-BL & \nodata & $\cdots$ & ~ \\ 
200826A & \nodata & GRB & \nodata & \nodata & $\cdots$ & ~ \\ 
210210A & \nodata & GRB & \nodata & \nodata & $\cdots$ & ~ \\ 
\enddata
\end{deluxetable*}
\end{longrotatetable}

\section{An hint of the relation between $L_{\rm \lowercase{a,opt}}$ and $\lowercase{t^{*}_{\rm p}}$?} \label{sec:results}

We here present the results of the only possible correlation we found in our analysis: $\log_{10}L'_{\rm a,opt}$ vs. $\log_{10}t^{*'}_{\rm p}$.
We point out that this correlation fulfills only the criterion that reports $|r|\geq0.50$ and $P_P\leq0.05$, namely the condition (I). For this correlation, the EP correction coefficient exists for both variables, thus allowing the subsequent analysis of the $\log_{10}L_{\rm a,opt}-\log_{10}t^{*}_{\rm p}$ relation. Furthermore, being this relation drawn by more than 5 data points, it can be further tested.

Our sample includes 9 GRB-SN events for which both $L_{\rm a,opt}$ and $t^{*}_{\rm p}$ values were reported in the literature. The 9 GRB-SN events present in this correlation are: GRB 030329/SN 2003dh (Ic/HN), GRB 081007A/SN 2008hw (Ic), GRB 091127/SN 2009nz (Ic-BL), GRB 100316D/SN 2010bh (Ic), GRB 101219B/SN 2010ma (Ic), GRB 111209A/SN 2011kl (SLSN), GRB 111228A, GRB 130702A/SN 2013dx (Ic), and GRB 130831A/SN 2013fu (Ib/c). The fitting equation for the EP-corrected full sample is the following:

 \begin{equation}
     \log_{10}L'_{\rm a,opt} = (9.43\pm1.97)\log_{10}t^{*'}_{\rm p} - (13.60\pm11.89),
 \end{equation}

while for the EP-corrected AB subsample, the fitting EP-corrected equation is

 \begin{equation}
     \log_{10}L'_{\rm a,opt} = (10.80\pm1.98)\log_{10}t^{*'}_{\rm p} - (21.74\pm11.93).
 \end{equation}

\noindent The full sample case shows the Pearson's correlation coefficient $r = 0.71$ and a p-value of $P_P = 0.03$ in the EP-corrected case with the full sample fitting. Regarding instead the AB sample, Pearson's correlation coefficient is $r=0.75$ and its p-value is $P_P=0.03$. The fitting parameters for this correlation are reported in Table \ref{tab:fitting_logLaopt_logtprest}. 

 \begin{table}[h]
     \centering

     \begin{tabular}{|c|c|c|c|}
     \hline
     \multicolumn{4}{|c|}{\textbf{Fitting without the EP method}}\\
     \hline
        \bf{Sample} & \bf{ODR method} & $\mathbf{r}$ & $\mathbf{P_P}$ \\
        Full & $(14.36\pm5.60)x+(-42.55\pm33.90)$ & 0.70 & 0.04 \\
        AB only & $(21.50\pm8.82)x+(-85.36\pm53.14)$ & 0.70 & 0.05 \\

    \hline
    
    \multicolumn{4}{|c|}{\textbf{Fitting with the EP method}}\\
     \hline
        \bf{Sample} & \bf{ODR method} & $\mathbf{r}$ & $\mathbf{P_P}$  \\
        Full & $(9.43\pm1.97)x+(-13.60\pm11.89)$ & 0.71 & 0.03 \\
        AB only & $(10.80\pm1.98)x+(-21.74\pm11.93)$ & 0.75 & 0.03 \\

    \hline
    \end{tabular}

    \caption{The fitting parameters of the correlation $\log_{10}L_{\rm a,opt}-\log_{10}t^{*}_{\rm p}$ in the cases without and with the EP correction. The columns, from left to right, contain the sample, the ODR method fitting, Pearson's correlation coefficient $r$, and Pearson's p-value $P_P$. The fitting relations are expressed as: $y=(a\pm \sigma a)x+(c\pm \sigma c)$, where $x=\log_{10}t^{*}_{\rm p}$, $y=\log_{10}L_{\rm a,opt}$, $a\pm \sigma a$ is the slope with its error, and $c \pm \sigma c$ is the intercept with its error.}
    \label{tab:fitting_logLaopt_logtprest}
\end{table}

\begin{figure*}[h]
\graphicspath{{06-03/}}

\gridline{\fig{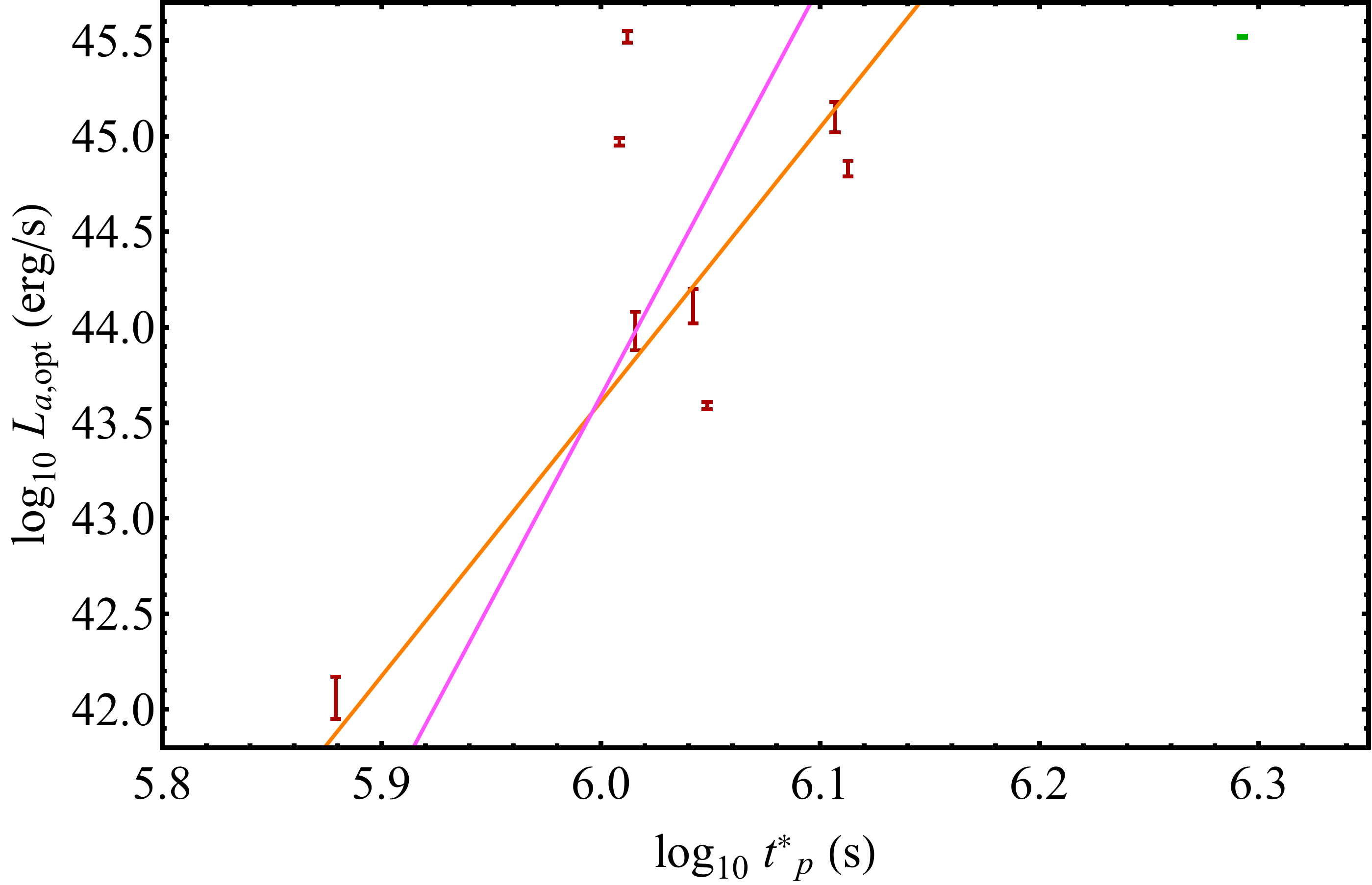}{0.50\textwidth}{(a) The $\log_{10}L_{\rm a,opt}$-$\log_{10}t^{*}_{p}$ correlation without the EP correction.}
\fig{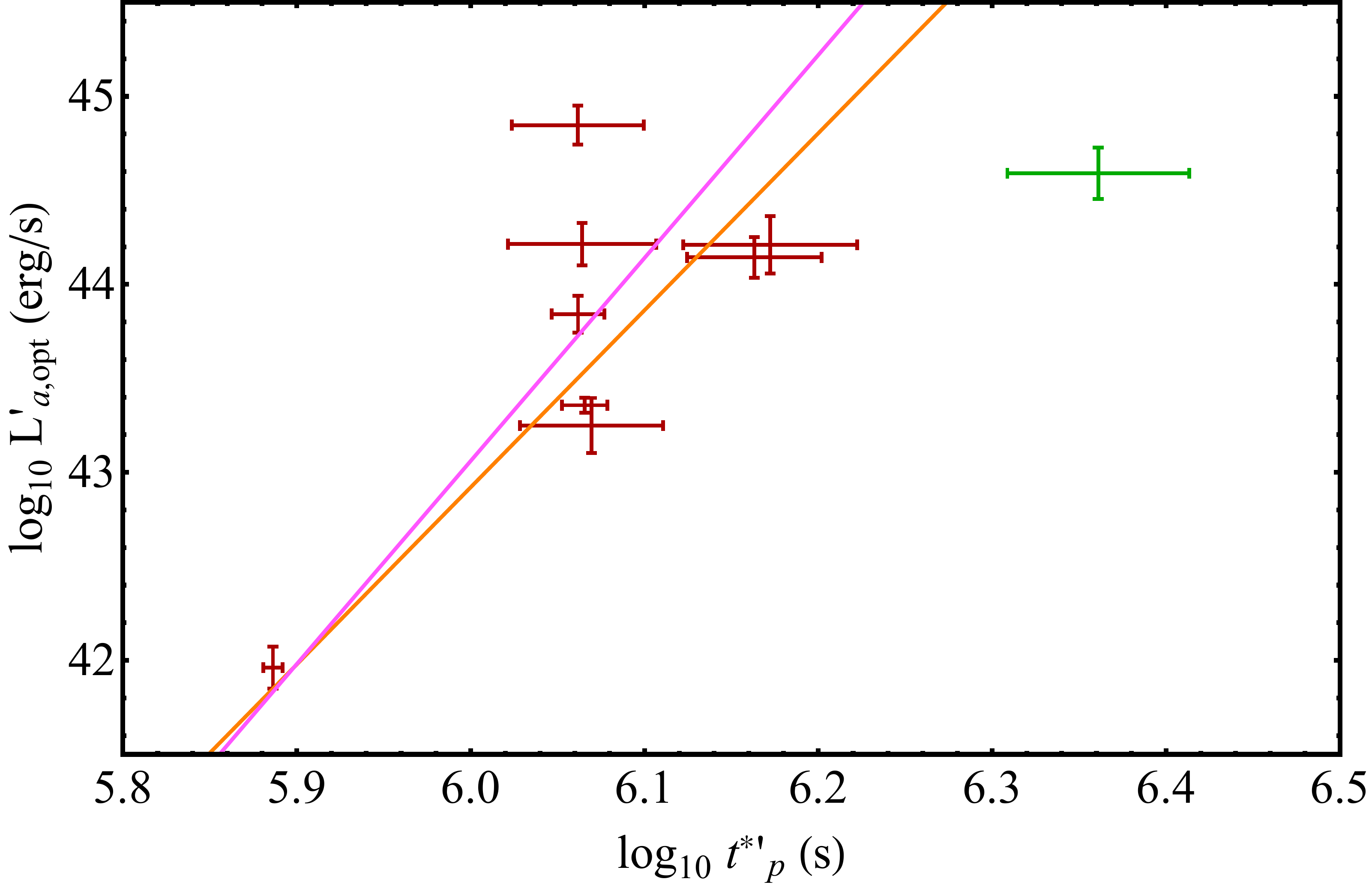}{0.50\textwidth}{(b) The $\log_{10}L'_{\rm a,opt}$-$\log_{10}t^{*'}_{p}$ correlation after the EP correction.}}

\caption{\textbf{(a)} The correlation between $\log_{10}L_{\rm a,opt}$ and $\log_{10}t^{*}_{p}$ before the EP correction. \textbf{(b)} The correlation between $\log_{10}L'_{\rm a,opt}$ and $\log_{10}t^{*'}_{p}$ after correcting both the variables with the EP method. The color-coded is the following: red refers to the AB subsample and green to the remaining E-graded event (GRB 111228A). For the fitting lines, orange indicates the full sample fitting and magenta the AB subsample fitting. The lines correspond to the fitting with ODR method parameters.}
\label{fig:logLaopt_logtprest}
\end{figure*}

We here note that this correlation contains all the points from the AB sample together with an E-graded event (GRB 111228A), which draws the correlation together with the other present data points.
This correlation is plotted in Figure \ref{fig:logLaopt_logtprest}. 
This correlation exists by evaluating the flatness of the slope in 5.4 $\sigma$ when we consider the entire sample and 4.7 $\sigma$ when we consider the AB sample.

\subsection{The physical discussion of the $\log_{10}L_{\rm a,opt}-\log_{10}t^{*}_{\rm p}$ and the other relevant correlations}
The peak time of a SN can be connected to the diffusion time through the Arnett model (\citealt{Arnett1982}, see \citealt{Cano2011} for the implementation of the cobalt contribution):

\begin{equation}
\begin{aligned}
L(t) &= M_{\rm Ni} e^{-x^2} \biggr((\epsilon_{\rm Ni}-\epsilon_{\rm Co})\int^{x}_{0}2ke^{-2ky+k^2}dk + \int^{x}_{0} 2ke^{-2ky+2ks+k^2}dk \biggr)
\end{aligned}
\label{eq:Arnett}
\end{equation}

where $x\equiv t/\tau_{m}$, $y\equiv\tau_{m}/(2\tau_{\rm Ni})$, $s=(\tau_{m}(\tau_{\rm Co}-\tau_{\rm Ni})/(2\tau_{\rm Co}\tau_{\rm Ni}))$, 
and $\tau_{m}=(k/\beta c)^{1/2}(M^{3}_{\rm ej}/E_{\rm K})^{1/4}$ is the effective diffusion time, given the opacity $k=0.07\rm\,cm^{2}\,g^{-1}$. 
It is important to stress that, in the Arnett formulation, the diffusion time $\tau_{m}$ is proportional to the peak time $t^{*}_{\rm p}$ with a multiplying factor that depends on the temperatures and the opacity \citep{Khatami2019}. 
Thanks to this property, given the found correlation between $L_{\rm a,opt}$ and $t^{*}_{\rm p}$, we can expect the brightest GRB optical plateaus to be connected with the longer diffusion times of the associated SNe.

Other probable correlations have been highlighted between the SN $v_{\rm ph}$ and the GRB $\theta_{\rm jet}$ and between $v_{\rm ph}$ and the GRB $L_{\gamma, \rm iso}$ but, in these cases, since the photospheric velocity $v_{\rm ph}$ had no correction for selection biases through the EP method (see Section \ref{sec:EPresults}), it was not possible to achieve a reliable comparison between the uncorrected and corrected versions, being only $\theta_{\rm jet}$ and $L_{\gamma, \rm iso}$ corrected in the correlation. So, we retained only the probable correlation $\log_{10}L'_{\rm a,opt}$-$\log_{10}t^{*'}_{\rm p}$ for further testing.

Finally, we would like to discuss the relevance of the correlation between $E_{\rm p}$ and $M_{\rm Ni}$ \citep{Lu2018}: in the case without the EP-correction, the Pearson correlation coefficient and p-value are $r=0.33$ and $P_P=0.06$, respectively, while for the Spearman we have $\rho=0.31$ and $P_\rho=0.08$, respectively. In the EP-corrected analysis, this relation shows $r=0.34$ and $P_P=0.05$ for the Pearson test, while $\rho=0.34$ and $P_\rho=0.05$ for the Spearman. Comparing our results with the ones from \citep{Lu2018}, these values show that the correlation coefficient (considering the EP-corrected case) drops from $r=0.77$ (in \citealt{Lu2018}) to $0.34$ (our results). This indicates a percentage decrease of $56\%$ in $r$, while the p-values in our cases are smaller ($P_P=0.05$ and $P_\rho=0.05$), thus having a percentage decrease of $78\%$ in our results.
These new results have been achieved with a sample size of 33 GRBs vs. the 20 GRBs used in \citet{Lu2018}, thus increasing the sample size by $40\%$. 
The fact that the p-value of this correlation drops substantially encourages the analysis of this correlation with the forthcoming GRB-SNe events in the next few years. However, we remark that the correlation coefficients from Pearson and Spearman tests in the EP-corrected case are $r=0.34$ and $\rho=0.34$, respectively: this is the reason why this correlation has not been presented as the main result since it does not fulfill our metrics.

\section{Summary and Conclusions} \label{sec:conclusion}
This research aims to study the GRB-SN connection with the largest possible sample of physical variables presented in the literature.
To this end, we compiled an extensive collection of GRB-SN properties and searched for correlations, see Appendix \ref{sec:scatterplot} for the scatter matrix plots of all the variables.
More specifically, we have analyzed 9 and 10 variables among the SNe and GRB properties, respectively. We have tested 91 bi-dimensional correlations among GRB and SN properties for parameters both in the cases without and with selection biases corrections. For completeness, we have reported a total of 171 correlations which also include GRB-GRB and SN-SN correlations investigated with their Pearson, Spearman, and their probability to be drawn by chance, both in the cases without and with EP correction (see Table \ref{allrelationsGRBSNe} and \ref{allrelationsGRBSNeEP}, respectively).

Indeed, the novelty of our approach compared to other previous works in the literature is that we apply a reliable statistical method, the EP method, to overcome selection biases due to instrumental thresholds and redshift evolution, which otherwise would remain hidden within the correlations. In addition, we have also applied two metrics for selecting the correlations before and after the EP method. To summarize, only a correlation revealed with the metrics in (I), namely $|r|\geq0.50$ and $P_P\leq0.05$ for the Pearson, has been found. 

We unveil a probable correlation between the optical luminosity at the end of the plateau emission and the rest-frame peak time of the SN. This correlation holds for a sample of 9 GRB-SNe and suggests that the SNe with a larger rest-frame peak time accompany the brightest GRB optical plateau emissions. Furthermore, the SNe with a later peak rising is associated with longer diffusion times, thus connecting the GRB optical plateau luminosity with the scaling of the SN LC. Thus, brighter GRBs at the end of the optical plateau emission are characterized by a larger diffusion time in their associated SNe. The slope of this correlation is non-zero at 5.4 $\sigma$ for the full sample and at 4.8 $\sigma$ for the AB sample. 
As a general remark, this correlation yields, however for small sample size, thus a larger sample is needed to secure our conclusions further. With such a future enlarged sample in our hands, we will be able to cast light on the GRB-SNe diversity and on the models that describe the progenitor of GRB-SNe.

Currently, no further significant correlations have been highlighted with these metrics.
A possible reason for which other correlations that may be expected have not been found is the wide-spanning of the GRB prompt parameters, in many orders of magnitude with respect to the SNe parameters. Here we refer especially to the GRB prompt isotropic energy vs. the peak bolometric luminosities of the associated SNe. While the GRB isotropic energy can be observed between $10^{47} erg$ and $10^{54} erg$ for GRBs associated with SNe, the peak bolometric luminosities of the associated SNe  remain in the order of $10^{52}-10^{53}erg\,s^{-1}$, showing that is unlikely for a correlation to be present between the two parameters (Della Valle et al., 2022 in prep.). The same argument applies in the comparison between the isotropic energy of the GRB and the SN peak time, the latter being in the order of $10^{6} s$. In view of these considerations, it's clear that the AG parameters in contrast to the prompt emission-only parameters, deserve a central role in the process of looking for correlations.

The metrics adopted here are stricter than previous metrics used in the literature. We point out hints about the existence of a new correlation in the GRB-SNe connections. 
The important key result is that the found correlation is intrinsic and not merely due to the result of selection effects or redshift evolution since we have tested it through the EP method.

We also point out that the correction for selection biases and redshift evolution, performed for the first time in the literature for this GRB-SNe sample, allows estimating the intrinsic behavior of physical parameters. 

Indeed, our method of cross-correlating all quantities in principle could have revealed new correlations after applying the EP method, which otherwise would have remained hidden.
In summary, this study paves the way to a) a broad investigation among all the existing parameters in the literature for both the SNe and GRBs; b) the use of reliable statistical methods to pinpoint possible hidden correlations; c) showing the challenge in finding new correlations given the paucity of the data we currently have at our disposal.

\noindent 
\section*{Acknowledgments}
This work was partly supported by the U.S. Department of Energy, Office of Science, Office of Workforce Development for Teachers and Scientists (WDTS) under the Science Undergraduate Laboratory Internships (SULI) program. This work used data supplied by the UK Swift Science Data Centre at the University of Leicester. We are grateful to Nicki O'Shea for the data collection and analysis performed during the SULI program. 
In the end, we would like to thank Dr. Enrique Cuellar (†) for the kindness and availability he proved to have during his career: without him, there would not have been such a fundamental collaboration with the SULI program.

\newpage

\begin{appendices}

\section{Scatter matrices and tables}\label{sec:scatterplot}
\begin{figure}[h]
\centering
\includegraphics[width=1.0\textwidth]{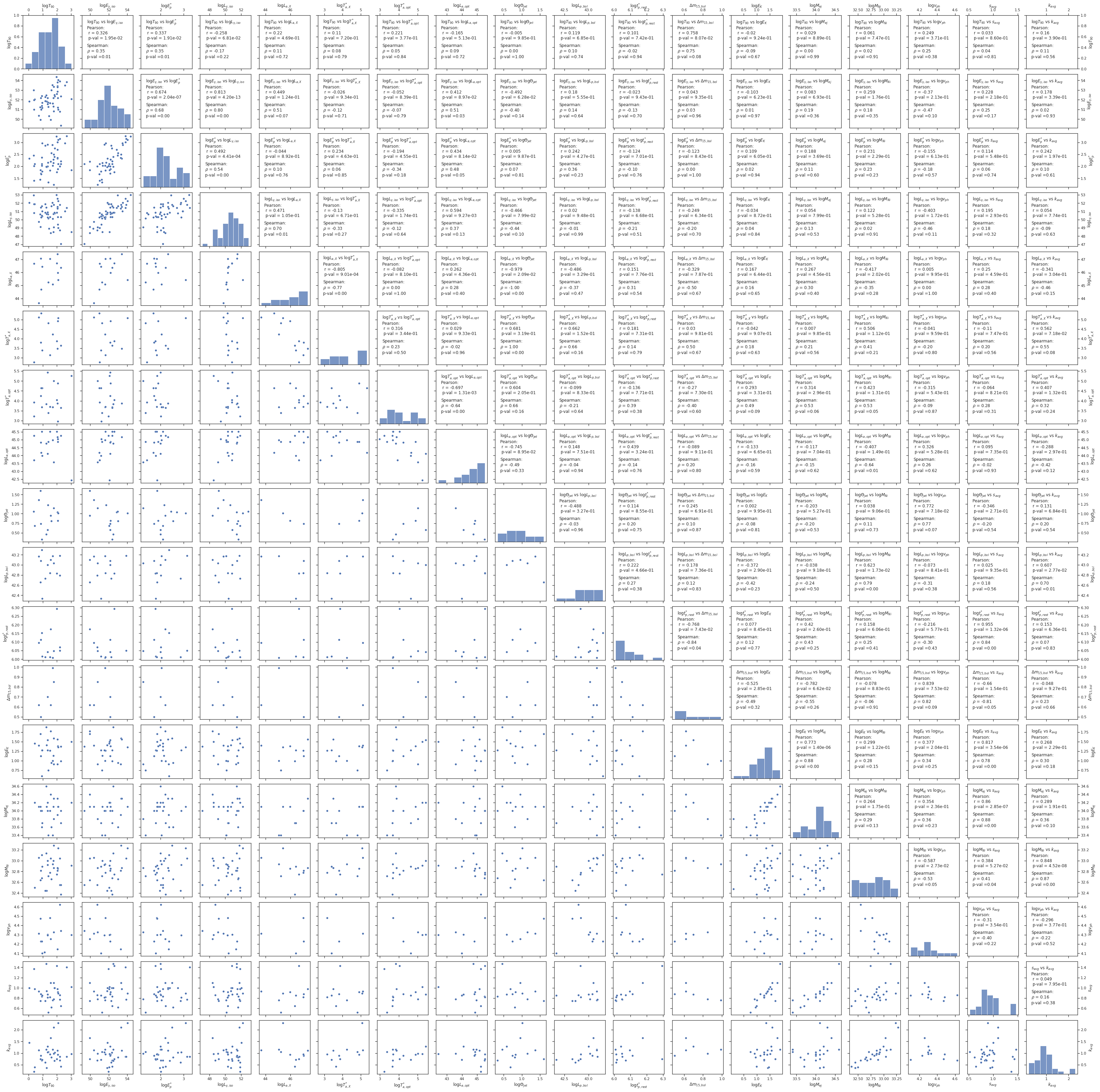}\\     
\caption{The scatter matrix for the variables not corrected with EP, removing only the two shock breakout events (GRB 060218/SN 2006aj and GRB 080109/SN 2008D). For brevity, we here write $\log$ in place of $\log_{10}$.}
    \label{fig:matrixnoEP}
\end{figure}

\begin{figure}[h]
\centering
\includegraphics[width=1.0\textwidth]{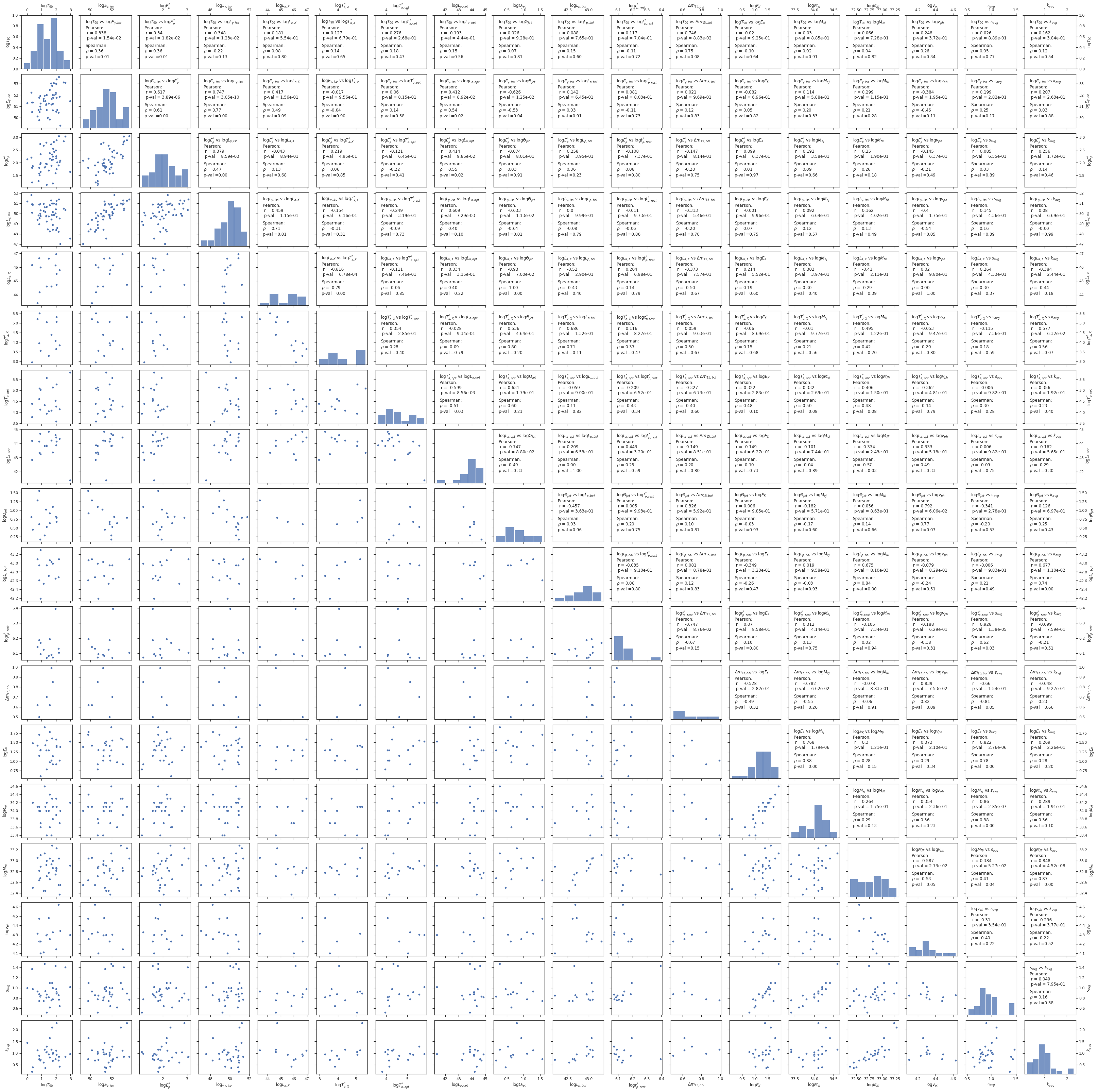}\\
    \caption{The scatter matrix with the EP-corrected variables, removing only the two shock breakout events (GRB 060218/SN 2006aj and GRB 080109/SN 2008D). For brevity, we here write $\log$ in place of $\log_{10}$.}
    \label{fig:matrixEP}
\end{figure}

In this Appendix, we report the complete versions of the scatter matrices, namely Figure \ref{fig:matrixnoEP} for the cases without EP and Figure \ref{fig:matrixEP} for the cases including EP. We here also report the full tables with all the pairs of cross-related parameters. For the cases without the EP corrections, Table \ref{allrelationsGRBSNe} summarizes the Pearson and Spearman's coefficients for each pair of variables between GRB and SN properties. Concerning the cases where the EP corrections are included, Table \ref{allrelationsGRBSNeEP} contains the Pearson and Spearman's coefficients between GRB and SN properties.

\newpage

\footnotesize

\begin{longrotatetable}
\small\addtolength{\tabcolsep}{-2pt}
\begin{deluxetable}{cccccccccccccc}
\tablecaption{The cross-relation of GRB and SN observables in the case without the EP correction.\label{chartable02}}
\tabletypesize{\scriptsize}

\tablehead{
\colhead{\textbf{Relation}} & \colhead{$\mathbf{r}$ } & 
\colhead{$\mathbf{P_P}$} & \colhead{$\mathbf{\rho}$} & 
\colhead{$\mathbf{P_\rho}$}&
\colhead{Notes} & \colhead{N. of points} &
\colhead{\textbf{Relation}} & \colhead{$\mathbf{r}$ } & 
\colhead{$\mathbf{P_P}$} & \colhead{$\mathbf{\rho}$} & 
\colhead{$\mathbf{P_\rho}$}&
\colhead{Notes} & \colhead{N. of points}}
\startdata
$\log_{10}T^{*}_{90}$ vs. $\log_{10}E_{\gamma,\rm iso}$ & 0.3 & 0.03 & 0.34 & 0.01 & 5 & 54 &
$\log_{10}T^{*}_{90}$ vs. $\log_{10}E^{*}_{\rm p}$ & 0.25 & 0.08 & 0.27 & 0.06 & 5 & 52\\
$\log_{10}E_{\gamma,\rm iso}$ vs. $\log_{10}E^{*}_{\rm p}$ & 0.7 & $<0.01$ & 0.72 & $<0.01$ & 2 & 50 &
$\log_{10}T^{*}_{90}$ vs. $\log_{10}L_{\gamma,\rm iso}$ & -0.36 & 0.01 & -0.27 & 0.04 & 5 & 55\\
$\log_{10}E_{\gamma,\rm iso}$ vs. $\log_{10}L_{\gamma,\rm iso}$ & 0.82 & $<0.01$ & 0.78 & $<0.01$ & 5 & 54 &
$\log_{10}E^{*}_{\rm p}$ vs. $\log_{10}L_{\gamma,\rm iso}$ & 0.55 & $<0.01$ & 0.57 & $<0.01$ & 4 & 51 \\
$\log_{10}T^{*}_{90}$ vs. $\log_{10}L_{\rm p,bol}$ & 0.32 & 0.21 & 0.19 & 0.47 & - & 17 &
$\log_{10}E_{\gamma,\rm iso}$ vs. $\log_{10}L_{\rm p,bol}$ & 0.35 & 0.2 & 0.29 & 0.3 & - & 15 \\
$\log_{10}E^{*}_{\rm p}$ vs. $\log_{10}L_{\rm p,bol}$ & 0.41 & 0.12 & 0.39 & 0.13 & 7 & 16 &
$\log_{10}L_{\gamma,\rm iso}$ vs. $\log_{10}L_{\rm p,bol}$ & 0.13 & 0.62 & 0.07 & 0.8 & - & 16\\
$\log_{10}T^{*}_{90}$ vs. $\log_{10}t^{*}_{\rm p}$ & -0.18 & 0.51 & -0.08 & 0.76 & - & 16 &
$\log_{10}E_{\gamma,\rm iso}$ vs. $\log_{10}t^{*}_{\rm p}$ & -0.06 & 0.83 & -0.15 & 0.61 & - & 14\\ 
$\log_{10}E^{*}_{\rm p}$ vs. $\log_{10}t^{*}_{\rm p}$ & 0.11 & 0.71 & 0.06 & 0.82 & - & 15 &
$\log_{10}L_{\gamma,\rm iso}$ vs. $\log_{10}t^{*}_{\rm p}$ & 0.17 & 0.54 & -0.1 & 0.72 & - & 15\\
$\log_{10}L_{\rm p,bol}$ vs. $\log_{10}t^{*}_{\rm p}$ & 0.3 & 0.26 & 0.37 & 0.15 & - & 16 &
$\log_{10}T^{*}_{90}$ vs. $\Delta m_{\rm 15,bol}$ & 0.48 & 0.19 & 0.53 & 0.14 & - & 9\\
$\log_{10}E_{\gamma,\rm iso}$ vs. $\Delta m_{\rm 15,bol}$ & -0.03 & 0.94 & -0.07 & 0.86 & - & 8 &
$\log_{10}E^{*}_{\rm p}$ vs. $\Delta m_{\rm 15,bol}$ & -0.16 & 0.71 & -0.24 & 0.57 & - & 8 \\
$\log_{10}L_{\gamma,\rm iso}$ vs. $\Delta m_{\rm 15,bol}$ & -0.42 & 0.26 & -0.38 & 0.31 & - & 9 &
$\log_{10}L_{\rm p,bol}$ vs. $\Delta m_{\rm 15,bol}$ & -0.15 & 0.7 & -0.18 & 0.65 & - & 9\\
$\log_{10}t^{*}_{\rm p}$ vs. $\Delta m_{\rm 15,bol}$ & -0.63 & 0.07 & -0.65 & 0.06 & - & 9 &
$\log_{10}T^{*}_{90}$ vs. $\log_{10}E_{\rm K}$ & 0.08 & 0.68 & -0.04 & 0.83 & - & 30 \\
$\log_{10}E_{\gamma,\rm iso}$ vs. $\log_{10}E_{\rm K}$ & -0.06 & 0.78 & 0.03 & 0.87 & - & 28 &
$\log_{10}E^{*}_{\rm p}$ vs. $\log_{10}E_{\rm K}$ & 0.09 & 0.63 & 0.07 & 0.71 & - & 29 \\ 
$\log_{10}L_{\gamma,\rm iso}$ vs. $\log_{10}E_{\rm K}$ & -0.03 & 0.87 & -0.02 & 0.93 & - & 29 &
$\log_{10}L_{\rm p,bol}$ vs. $\log_{10}E_{\rm K}$ & -0.1 & 0.75 & -0.08 & 0.79 & - & 13\\
$\log_{10}t^{*}_{\rm p}$ vs. $\log_{10}E_{\rm K}$ & 0.17 & 0.6 & 0.24 & 0.46 & - & 12 & 
$\Delta m_{\rm 15,bol}$ vs. $\log_{10}E_{\rm K}$ & -0.4 & 0.28 & -0.49 & 0.18 & - & 9 \\
$\log_{10}T^{*}_{90}$ vs. $\log_{10}M_{\rm ej}$ & -0.13 & 0.51 & -0.12 & 0.53 & - & 30 &
$\log_{10}E_{\gamma,\rm iso}$ vs. $\log_{10}M_{\rm ej}$ & -0.07 & 0.73 & 0.0 & 0.99 & - & 28\\ 
$\log_{10}E^{*}_{\rm p}$ vs. $\log_{10}M_{\rm ej}$ & 0.07 & 0.71 & 0.03 & 0.87 & - & 29 &
$\log_{10}L_{\gamma,\rm iso}$ vs. $\log_{10}M_{\rm ej}$ & 0.05 & 0.79 & 0.08 & 0.67 & - & 29\\
$\log_{10}L_{\rm p,bol}$ vs. $\log_{10}M_{\rm ej}$ & -0.02 & 0.94 & -0.15 & 0.62 & - & 13 &
$\log_{10}t^{*}_{\rm p}$ vs. $\log_{10}M_{\rm ej}$ & 0.47 & 0.12 & 0.47 & 0.12 & - & 12 \\
$\Delta m_{\rm 15,bol}$ vs. $\log_{10}M_{\rm ej}$ & -0.73 & 0.02 & -0.7 & 0.04 & - & 9 &
$\log_{10}E_{\rm K}$ vs. $\log_{10}M_{\rm ej}$ & 0.73 & $<0.01$ & 0.81 & $<0.01$ & - & 32\\
$\log_{10}T^{*}_{90}$ vs. $\log_{10}M_{\rm Ni}$ & 0.13 & 0.48 & 0.06 & 0.72 & - & 34 &
$\log_{10}E_{\gamma,\rm iso}$ vs. $\log_{10}M_{\rm Ni}$ & 0.26 & 0.16 & 0.23 & 0.21 & - & 32\\
$\log_{10}E^{*}_{\rm p}$ vs. $\log_{10}M_{\rm Ni}$ & 0.33 & 0.06 & 0.31 & 0.08 & 6 & 33 &
$\log_{10}L_{\gamma,\rm iso}$ vs. $\log_{10}M_{\rm Ni}$ & 0.18 & 0.32 & 0.05 & 0.78 & - & 33 \\
$\log_{10}L_{\rm p,bol}$ vs. $\log_{10}M_{\rm Ni}$ & 0.69 & $<0.01$ & 0.82 & $<0.01$ & - & 17 &
$\log_{10}t^{*}_{\rm p}$ vs. $\log_{10}M_{\rm Ni}$ & 0.43 & 0.1 & 0.43 & 0.1 & - & 16 \\
$\Delta m_{\rm 15,bol}$ vs. $\log_{10}M_{\rm Ni}$ & -0.25 & 0.51 & -0.36 & 0.34 & - & 9 &
$\log_{10}E_{\rm K}$ vs. $\log_{10}M_{\rm Ni}$ & 0.34 & 0.06 & 0.34 & 0.05 & - & 32 \\
$\log_{10}M_{\rm ej}$ vs. $\log_{10}M_{\rm Ni}$ & 0.26 & 0.15 & 0.27 & 0.13 & -  & 32 &
$\log_{10}T^{*}_{90}$ vs. $\log_{10}v_{\rm ph}$ & 0.34 & 0.17 & 0.47 & 0.05 & - & 18\\
$\log_{10}E_{\gamma,\rm iso}$ vs. $\log_{10}v_{\rm ph}$ & -0.17 & 0.54 & -0.24 & 0.38 & - & 15 &
$\log_{10}E^{*}_{\rm p}$ vs. $\log_{10}v_{\rm ph}$ & -0.24 & 0.37 & -0.19 & 0.49  & - & 16\\
$\log_{10}L_{\gamma,\rm iso}$ vs. $\log_{10}v_{\rm ph}$ & -0.4 & 0.13 & -0.44 & 0.09 & $\times$ & 16 &
$\log_{10}L_{\rm p,bol}$ vs. $\log_{10}v_{\rm ph}$ & -0.13 & 0.66 & -0.3 & 0.31 & - & 13 \\
$\log_{10}t^{*}_{\rm p}$ vs. $\log_{10}v_{\rm ph}$ & -0.48 & 0.12 & -0.5 & 0.1 & -  & 12 &
$\Delta m_{\rm 15,bol}$ vs. $\log_{10}v_{\rm ph}$ & 0.67 & 0.07 & 0.77 & 0.03 & - & 8 \\ 
$\log_{10}E_{\rm K}$ vs. $\log_{10}v_{\rm ph}$ & 0.25 & 0.35 & 0.23 & 0.4 & - & 16 &
$\log_{10}M_{\rm ej}$ vs. $\log_{10}v_{\rm ph}$ & 0.04 & 0.87 & 0.06 & 0.83 & - & 16 \\ 
$\log_{10}M_{\rm Ni}$ vs. $\log_{10}v_{\rm ph}$ & -0.59 & 0.01 & -0.51 & 0.04 & - & 17 &
$\log_{10}T^{*}_{90}$ vs. $k_{\rm avg}$ & 0.24 & 0.16 & 0.16 & 0.36 & - & 35 \\ 
$\log_{10}E_{\gamma,\rm iso}$ vs. $k_{\rm avg}$ & 0.24 & 0.17 & 0.12 & 0.5 & - & 34 &
$\log_{10}E^{*}_{\rm p}$ vs. $k_{\rm avg}$ & 0.33 & 0.05 & 0.21 & 0.22 & - & 34 \\ 
$\log_{10}L_{\gamma,\rm iso}$ vs. $k_{\rm avg}$ & 0.11 & 0.51 & -0.02 & 0.9 & - & 35 &
$\log_{10}L_{\rm p,bol}$ vs. $k_{\rm avg}$ & 0.74 & $<0.01$ & 0.77 & $<0.01$ & - & 16 \\ 
$\log_{10}t^{*}_{\rm p}$ vs. $k_{\rm avg}$ & 0.34 & 0.22 & 0.34 & 0.21 & - & 15 &
$\Delta m_{\rm 15,bol}$ vs. $k_{\rm avg}$ & -0.19 & 0.62 & -0.18 & 0.65 & - & 9 \\ 
$\log_{10}E_{\rm K}$ vs. $k_{\rm avg}$ & 0.33 & 0.1 & 0.34 & 0.09 & - & 26 &
$\log_{10}M_{\rm ej}$ vs. $k_{\rm avg}$ & 0.25 & 0.22 & 0.28 & 0.17 & - & 26 \\ 
$\log_{10}M_{\rm Ni}$ vs. $k_{\rm avg}$ & 0.86 & $<0.01$ & 0.89 & $<0.01$ & - & 30 &
$\log_{10}v_{\rm ph}$ vs. $k_{\rm avg}$ & -0.36 & 0.21 & -0.31 & 0.28 & - & 14 \\ 
$\log_{10}T^{*}_{90}$ vs. $s_{\rm avg}$ & -0.06 & 0.73 & 0.0 & 0.98 & - & 35 &
$\log_{10}E_{\gamma,\rm iso}$ vs. $s_{\rm avg}$ & 0.16 & 0.38 & 0.21 & 0.24 & - & 34 \\ 
$\log_{10}E^{*}_{\rm p}$ vs. $s_{\rm avg}$ & 0.15 & 0.39 & 0.13 & 0.47 & - & 34 &
$\log_{10}L_{\gamma,\rm iso}$ vs. $s_{\rm avg}$ & 0.23 & 0.17 & 0.13 & 0.44 & - & 35 \\ 
$\log_{10}L_{\rm p,bol}$ vs. $s_{\rm avg}$ & 0.21 & 0.43 & 0.37 & 0.16 & - & 16 &
$\log_{10}t^{*}_{\rm p}$ vs. $s_{\rm avg}$ & 0.96 & $<0.01$ & 0.9 & $<0.01$ & - & 15 \\ 
$\Delta m_{\rm 15,bol}$ vs. $s_{\rm avg}$ & -0.51 & 0.17 & -0.63 & 0.07 & - & 9 &
$\log_{10}E_{\rm K}$ vs. $s_{\rm avg}$ & 0.81 & $<0.01$ & 0.8 & $<0.01$ & - & 26 \\ 
$\log_{10}M_{\rm ej}$ vs. $s_{\rm avg}$ & 0.83 & $<0.01$ & 0.83 & $<0.01$ & - & 26 &
$\log_{10}M_{\rm Ni}$ vs. $s_{\rm avg}$ & 0.47 & 0.01 & 0.5 & $<0.01$ & - & 30 \\ 
$\log_{10}v_{\rm ph}$ vs. $s_{\rm avg}$ & -0.49 & 0.08 & -0.48 & 0.09 & - & 14 &
$k_{\rm avg}$ vs. $s_{\rm avg}$ & 0.12 & 0.48 & 0.24 & 0.17 & 3 & 35 \\ 
$\log_{10}T^{*}_{90}$ vs. $\log_{10}T^{*}_{\rm a,opt}$ & 0.4 & 0.08 & 0.27 & 0.25 & - & 20 &
$\log_{10}E_{\gamma,\rm iso}$ vs. $\log_{10}T^{*}_{\rm a,opt}$ & -0.04 & 0.88 & -0.05 & 0.84 & - & 19 \\ 
$\log_{10}E^{*}_{\rm p}$ vs. $\log_{10}T^{*}_{\rm a,opt}$ & -0.31 & 0.2 & -0.34 & 0.15 & - & 19 &
$\log_{10}L_{\gamma,\rm iso}$ vs. $\log_{10}T^{*}_{\rm a,opt}$ & -0.61 & $<0.01$ & -0.27 & 0.24 & 1 & 20 \\ 
$\log_{10}L_{\rm p,bol}$ vs. $\log_{10}T^{*}_{\rm a,opt}$ & -0.28 & 0.47 & -0.38 & 0.31 & - & 9 &
$\log_{10}t^{*}_{\rm p}$ vs. $\log_{10}T^{*}_{\rm a,opt}$ & -0.53 & 0.14 & -0.12 & 0.77 & - & 9 \\ 
$\Delta m_{\rm 15,bol}$ vs. $\log_{10}T^{*}_{\rm a,opt}$ & 0.07 & 0.9 & -0.14 & 0.79 & - & 6 &
$\log_{10}E_{\rm K}$ vs. $\log_{10}T^{*}_{\rm a,opt}$ & 0.13 & 0.64 & 0.32 & 0.25 & - & 15 \\ 
$\log_{10}M_{\rm ej}$ vs. $\log_{10}T^{*}_{\rm a,opt}$ & 0.11 & 0.7 & 0.33 & 0.24 & - & 15 &
$\log_{10}M_{\rm Ni}$ vs. $\log_{10}T^{*}_{\rm a,opt}$ & -0.12 & 0.65 & 0.25 & 0.35 & - & 16 \\ 
$\log_{10}v_{\rm ph}$ vs. $\log_{10}T^{*}_{\rm a,opt}$ & 0.29 & 0.49 & 0.12 & 0.78 & - & 8 &
$k_{\rm avg}$ vs. $\log_{10}T^{*}_{\rm a,opt}$ & 0.06 & 0.83 & 0.17 & 0.53 & - & 17 \\ 
$s_{\rm avg}$ vs. $\log_{10}T^{*}_{\rm a,opt}$ & -0.28 & 0.28 & 0.07 & 0.8 & - & 17 &
$\log_{10}T^{*}_{90}$ vs. $\log_{10}L_{\rm a,opt}$ & -0.28 & 0.24 & 0.0 & 0.99 & - & 20 \\ 
$\log_{10}E_{\gamma,\rm iso}$ vs. $\log_{10}L_{\rm a,opt}$ & 0.41 & 0.08 & 0.52 & 0.02 & 1 & 19 &
$\log_{10}E^{*}_{\rm p}$ vs. $\log_{10}L_{\rm a,opt}$ & 0.5 & 0.03 & 0.54 & 0.02 & 1 & 19 \\ 
$\log_{10}L_{\gamma,\rm iso}$ vs. $\log_{10}L_{\rm a,opt}$ & 0.74 & $<0.01$ & 0.44 & 0.05 & 1 & 20 &
$\log_{10}L_{\rm p,bol}$ vs. $\log_{10}L_{\rm a,opt}$ & 0.33 & 0.39 & 0.3 & 0.43 & - & 9 \\ 
$\log_{10}t^{*}_{\rm p}$ vs. $\log_{10}L_{\rm a,opt}$ & 0.7 & 0.04 & 0.33 & 0.38 & $*$ & 9 &
$\Delta m_{\rm 15,bol}$ vs. $\log_{10}L_{\rm a,opt}$ & -0.29 & 0.58 & -0.14 & 0.79 & - & 6 \\ 
$\log_{10}E_{\rm K}$ vs. $\log_{10}L_{\rm a,opt}$ & 0.01 & 0.96 & 0.0 & 0.99 & - & 15 &
$\log_{10}M_{\rm ej}$ vs. $\log_{10}L_{\rm a,opt}$ & 0.03 & 0.91 & 0.01 & 0.97 & - & 15 \\ 
$\log_{10}M_{\rm Ni}$ vs. $\log_{10}L_{\rm a,opt}$ & 0.15 & 0.57 & -0.27 & 0.31 & - & 16 &
$\log_{10}v_{\rm ph}$ vs. $\log_{10}L_{\rm a,opt}$ & -0.19 & 0.65 & -0.07 & 0.87 & - & 8 \\ 
$k_{\rm avg}$ vs. $\log_{10}L_{\rm a,opt}$ & 0.06 & 0.81 & -0.2 & 0.45 & - & 17 &
$s_{\rm avg}$ vs. $\log_{10}L_{\rm a,opt}$ & 0.31 & 0.23 & 0.18 & 0.5 & - & 17 \\ 
$\log_{10}T^{*}_{\rm a,opt}$ vs. $\log_{10}L_{\rm a,opt}$ & -0.8 & $<0.01$ & -0.68 & $<0.01$ & 8 & 20 &
$\log_{10}T^{*}_{90}$ vs. $\log_{10}T^{*}_{\rm a,X}$ & 0.02 & 0.96 & 0.01 & 0.97 & - & 14 \\ 
$\log_{10}E_{\gamma,\rm iso}$ vs. $\log_{10}T^{*}_{\rm a,X}$ & -0.05 & 0.85 & -0.09 & 0.77 & - & 14 &
$\log_{10}E^{*}_{\rm p}$ vs. $\log_{10}T^{*}_{\rm a,X}$ & 0.19 & 0.54 & 0.11 & 0.72 & - & 13 \\ 
$\log_{10}L_{\gamma,\rm iso}$ vs. $\log_{10}T^{*}_{\rm a,X}$ & -0.11 & 0.72 & -0.35 & 0.21 & - & 14 &
$\log_{10}L_{\rm p,bol}$ vs. $\log_{10}T^{*}_{\rm a,X}$ & 0.53 & 0.22 & 0.68 & 0.09 & - & 7 \\ 
$\log_{10}t^{*}_{\rm p}$ vs. $\log_{10}T^{*}_{\rm a,X}$ & 0.18 & 0.7 & 0.04 & 0.94 & - & 7 &
$\Delta m_{\rm 15,bol}$ vs. $\log_{10}T^{*}_{\rm a,X}$ & -0.05 & 0.95 & 0.4 & 0.6 & - & 4 \\ 
$\log_{10}E_{\rm K}$ vs. $\log_{10}T^{*}_{\rm a,X}$ & -0.1 & 0.78 & 0.14 & 0.69 & - & 11 &
$\log_{10}M_{\rm ej}$ vs. $\log_{10}T^{*}_{\rm a,X}$ & 0.0 & 0.99 & 0.18 & 0.6 & - & 11 \\ 
$\log_{10}M_{\rm Ni}$ vs. $\log_{10}T^{*}_{\rm a,X}$ & 0.37 & 0.24 & 0.39 & 0.21 & - & 12 &
$\log_{10}v_{\rm ph}$ vs. $\log_{10}T^{*}_{\rm a,X}$ & -0.04 & 0.95 & -0.1 & 0.87 & - & 5 \\ 
$k_{\rm avg}$ vs. $\log_{10}T^{*}_{\rm a,X}$ & 0.46 & 0.13 & 0.55 & 0.06 & - & 12 &
$s_{\rm avg}$ vs. $\log_{10}T^{*}_{\rm a,X}$ & -0.13 & 0.69 & 0.15 & 0.63 & - & 12 \\ 
$\log_{10}T^{*}_{\rm a,opt}$ vs. $\log_{10}T^{*}_{\rm a,X}$ & 0.31 & 0.32 & 0.24 & 0.44 & - & 12 &
$\log_{10}L_{\rm a,opt}$ vs. $\log_{10}T^{*}_{\rm a,X}$ & 0.03 & 0.94 & -0.07 & 0.83 & - & 12 \\ 
$\log_{10}T^{*}_{90}$ vs. $\log_{10}L^{*}_{a,X}$ & 0.48 & 0.08 & 0.37 & 0.19 & - & 14 &
$\log_{10}E_{\gamma,\rm iso}$ vs. $\log_{10}L^{*}_{a,X}$ & 0.56 & 0.04 & 0.57 & 0.03 & 1 & 14 \\ 
$\log_{10}E^{*}_{\rm p}$ vs. $\log_{10}L^{*}_{a,X}$ & 0.2 & 0.51 & 0.24 & 0.43 & - & 13 &
$\log_{10}L_{\gamma,\rm iso}$ vs. $\log_{10}L^{*}_{a,X}$ & 0.28 & 0.33 & 0.42 & 0.14 & - & 14 \\ 
$\log_{10}L_{\rm p,bol}$ vs. $\log_{10}L^{*}_{a,X}$ & -0.02 & 0.97 & 0.14 & 0.76 & - & 7 &
$\log_{10}t^{*}_{\rm p}$ vs. $\log_{10}L^{*}_{a,X}$ & 0.16 & 0.73 & 0.43 & 0.34 & - & 7 \\ 
$\Delta m_{\rm 15,bol}$ vs. $\log_{10}L^{*}_{a,X}$ & -0.1 & 0.9 & 0.0 & 1.0 & - & 4 &
$\log_{10}E_{\rm K}$ vs. $\log_{10}L^{*}_{a,X}$ & 0.41 & 0.22 & 0.37 & 0.26 & - & 11 \\ 
$\log_{10}M_{\rm ej}$ vs. $\log_{10}L^{*}_{a,X}$ & 0.24 & 0.48 & 0.21 & 0.54 & - & 11 &
$\log_{10}M_{\rm Ni}$ vs. $\log_{10}L^{*}_{a,X}$ & -0.04 & 0.9 & -0.04 & 0.9 & - & 12 \\ 
$\log_{10}v_{\rm ph}$ vs. $\log_{10}L^{*}_{a,X}$ & 0.08 & 0.9 & 0.3 & 0.62 & - & 5 &
$k_{\rm avg}$ vs. $\log_{10}L^{*}_{a,X}$ & -0.05 & 0.87 & -0.18 & 0.57 & - & 12 \\ 
$s_{\rm avg}$ vs. $\log_{10}L^{*}_{a,X}$ & 0.34 & 0.28 & 0.42 & 0.17 & - & 12 &
$\log_{10}T^{*}_{\rm a,opt}$ vs. $\log_{10}L^{*}_{a,X}$ & -0.04 & 0.9 & 0.04 & 0.9 & - & 12 \\ 
$\log_{10}L_{\rm a,opt}$ vs. $\log_{10}L^{*}_{a,X}$ & 0.3 & 0.35 & 0.34 & 0.28 & - & 12 &
$\log_{10}T^{*}_{\rm a,X}$ vs. $\log_{10}L^{*}_{a,X}$ & -0.74 & $<0.01$ & -0.69 & 0.01 & 1 & 14 \\ 
$\log_{10}T^{*}_{90}$ vs. $\log_{10} \theta_{\rm jet}$ & 0.05 & 0.86 & 0.02 & 0.95 & - & 17 &
$\log_{10}E_{\gamma,\rm iso}$ vs. $\log_{10} \theta_{\rm jet}$ & -0.4 & 0.11 & -0.36 & 0.15 & - & 17 \\ 
$\log_{10}E^{*}_{\rm p}$ vs. $\log_{10}\theta_{\rm jet}$ & -0.01 & 0.98 & 0.01 & 0.96 & - & 16 &
$\log_{10}L_{\gamma,\rm iso}$ vs. $\log_{10}\theta_{\rm jet}$ & -0.41 & 0.1 & -0.44 & 0.08 & - & 17 \\ 
$\log_{10}L_{\rm p,bol}$ vs. $\log_{10}\theta_{\rm jet}$ & -0.39 & 0.34 & -0.19 & 0.65 & - & 8 &
$\log_{10}t^{*}_{\rm p}$ vs. $\log_{10}\theta_{\rm jet}$ & 0.13 & 0.78 & -0.04 & 0.94 & - & 7 \\ 
$\Delta m_{\rm 15,bol}$ vs. $\log_{10}\theta_{\rm jet}$ & 0.26 & 0.58 & 0.4 & 0.38 & - & 7 &
$\log_{10}E_{\rm K}$ vs. $\log_{10}\theta_{\rm jet}$ & 0.02 & 0.95 & -0.05 & 0.86 & - & 14 \\ 
$\log_{10}M_{\rm ej}$ vs. $\log_{10}\theta_{\rm jet}$ & -0.19 & 0.51 & -0.19 & 0.51 & - & 14 &
$\log_{10}M_{\rm Ni}$ vs. $\log_{10}\theta_{\rm jet}$ & 0.07 & 0.8 & 0.16 & 0.59 & - & 14 \\ 
$\log_{10}v_{\rm ph}$ vs. $\log_{10}\theta_{\rm jet}$ & 0.74 & 0.04 & 0.69 & 0.06 & $\times$ & 8 &
$k_{\rm avg}$ vs. $\log_{10}\theta_{\rm jet}$ & 0.14 & 0.64 & 0.2 & 0.48 & - & 14 \\ 
$s_{\rm avg}$ vs. $\log_{10}\theta_{\rm jet}$ & -0.31 & 0.29 & -0.11 & 0.71 & - & 14 &
$\log_{10}T^{*}_{\rm a,opt}$ vs. $\log_{10}\theta_{\rm jet}$ & 0.47 & 0.29 & 0.54 & 0.22 & - & 7 \\ 
$\log_{10}L_{\rm a,opt}$ vs. $\log_{10}\theta_{\rm jet}$ & -0.55 & 0.2 & -0.32 & 0.48 & - & 7 &
$\log_{10}T^{*}_{\rm a,X}$ vs. $\log_{10}\theta_{\rm jet}$ & 0.58 & 0.3 & 0.9 & 0.04 & - & 5 \\ 
$\log_{10}L^{*}_{a,X}$ vs. $\log_{10}\theta_{\rm jet}$ & -0.74 & 0.15 & -0.7 & 0.19 & - & 5 &
$\log_{10}T^{*}_{90}$ vs. $\log_{10}T^{*}_{\rm jet}$ & 0.03 & 0.91 & -0.03 & 0.91 & - & 20 \\ 
$\log_{10}E_{\gamma,\rm iso}$ vs. $\log_{10}T^{*}_{\rm jet}$ & -0.14 & 0.56 & -0.18 & 0.45 & - & 20 &
$\log_{10}E^{*}_{\rm p}$ vs. $\log_{10}T^{*}_{\rm jet}$ & -0.52 & 0.02 & -0.54 & 0.02 & - & 19 \\ 
$\log_{10}L_{\gamma,\rm iso}$ vs. $\log_{10}T^{*}_{\rm jet}$ & -0.19 & 0.42 & -0.37 & 0.11 & - & 20 &
$\log_{10}L_{\rm p,bol}$ vs. $\log_{10}T^{*}_{\rm jet}$ & -0.32 & 0.6 & -0.3 & 0.62 & - & 5 \\ 
$\log_{10}t^{*}_{\rm p}$ vs. $\log_{10}T^{*}_{\rm jet}$ & -0.21 & 0.74 & -0.7 & 0.19 & - & 5 &
$\Delta m_{\rm 15,bol}$ vs. $\log_{10}T^{*}_{\rm jet}$ & 0.81 & 0.39 & 1.0 & $<0.01$ & $\times$ & 3 \\ 
$\log_{10}E_{\rm K}$ vs. $\log_{10}T^{*}_{\rm jet}$ & 0.08 & 0.82 & 0.27 & 0.42 & - & 11 &
$\log_{10}M_{\rm ej}$ vs. $\log_{10}T^{*}_{\rm jet}$ & -0.21 & 0.53 & -0.1 & 0.77 & - & 11 \\ 
$\log_{10}M_{\rm Ni}$ vs. $\log_{10}T^{*}_{\rm jet}$ & -0.24 & 0.44 & -0.11 & 0.74 & - & 12 &
$\log_{10}v_{\rm ph}$ vs. $\log_{10}T^{*}_{\rm jet}$ & -0.44 & 0.46 & -0.3 & 0.62 & - & 5 \\ 
$k_{\rm avg}$ vs. $\log_{10}T^{*}_{\rm jet}$ & -0.3 & 0.34 & -0.27 & 0.39 & - & 12 &
$s_{\rm avg}$ vs. $\log_{10}T^{*}_{\rm jet}$ & 0.13 & 0.68 & 0.21 & 0.51 & - & 12 \\ 
$\log_{10}T^{*}_{\rm a,opt}$ vs. $\log_{10}T^{*}_{\rm jet}$ & 0.66 & 0.1 & 0.71 & 0.07 & - & 7 &
$\log_{10}L_{\rm a,opt}$ vs. $\log_{10}T^{*}_{\rm jet}$ & -0.36 & 0.43 & -0.43 & 0.34 & - & 7 \\ 
$\log_{10}T^{*}_{\rm a,X}$ vs. $\log_{10}T^{*}_{\rm jet}$ & -0.38 & 0.53 & -0.3 & 0.62 & - & 5 &
$\log_{10}L^{*}_{a,X}$ vs. $\log_{10}T^{*}_{\rm jet}$ & 0.39 & 0.52 & 0.1 & 0.87 & - & 5 \\ 
$\log_{10}\theta_{\rm jet}$ vs. $\log_{10}T^{*}_{\rm jet}$ & 0.42 & 0.26 & 0.47 & 0.21 & - & 9 &
    \enddata
    \label{allrelationsGRBSNe}

\end{deluxetable}
\tablecomments{\textbf{The legend of the notes is the following: 1) the relation is from \citet{Dainotti2008}; 2) the relation is from \citet{Amati2002}; 3) the relation is from \citet{Cano2013}; 4) the relation is from  \citet{Yonetoku2004}; 5) the relation is discussed in \citet{Shahmoradi2015} ; 6) the relation is from \citet{Lu2018} ; 7) the relation is from \citet{Li2006}; 8) the relation is from \citet{DainottiLivermore2020}; $*$) the relation has been found in the current work; $\times$) this correlation is not investigated due to the paucity of data (less than 5 data points) or the absence of the EP-correction coefficient for one of the two variables. Since in the current Table and in Table \ref{chartable03} the p-value$=0.0$ comes from the precision we choose (which is on the second significant digit) all those p-values smaller than this threshold are indicated with  are indeed smaller $<0.01$.}}
\end{longrotatetable}

\begin{longrotatetable}
\begin{deluxetable}{cccccccccccccccc}
\tablecaption{The cross-relation of GRB and SN observables in the case with the EP correction.\label{chartable03}}
\tabletypesize{\scriptsize}

\tablehead{
\colhead{\textbf{Relation}} & \colhead{$\mathbf{r}$ } & 
\colhead{$\mathbf{P_P}$} & \colhead{$\mathbf{\rho}$} & 
\colhead{$\mathbf{P_\rho}$}&
\colhead{\textbf{Relation}} & \colhead{$\mathbf{r}$ } & 
\colhead{$\mathbf{P_P}$} & \colhead{$\mathbf{\rho}$} & 
\colhead{$\mathbf{P_\rho}$}&
\colhead{\textbf{Relation}} & \colhead{$\mathbf{r}$ } & 
\colhead{$\mathbf{P_P}$} & \colhead{$\mathbf{\rho}$} & 
\colhead{$\mathbf{P_\rho}$}}

\startdata
$\log_{10}T^{*'}_{90}$ vs. $\log_{10}E'_{\gamma,\rm iso}$ & 0.41 & $<0.01$ & 0.42 & $<0.01$ & 
$\log_{10}T^{*'}_{90}$ vs. $\log_{10}E^{*'}_{\rm p}$ & 0.33 & 0.02 & 0.35 & 0.01 & 
$\log_{10}E'_{\gamma,\rm iso}$ vs. $\log_{10}E^{*'}_{\rm p}$ & 0.65 & $<0.01$ & 0.65 & $<0.01$ \\
$\log_{10}T^{*'}_{90}$ vs. $\log_{10}L'_{\gamma,\rm iso}$ & -0.33 & 0.01 & -0.23 & 0.09 & 
$\log_{10}E'_{\gamma,\rm iso}$ vs. $\log_{10}L'_{\gamma,\rm iso}$ & 0.76 & $<0.01$ & 0.74 & $<0.01$ & 
$\log_{10}E^{*'}_{\rm p}$ vs. $\log_{10}L'_{\gamma,\rm iso}$ & 0.44 & $<0.01$ & 0.49 & $<0.01$ \\
$\log_{10}T^{*'}_{90}$ vs. $\log_{10}L'_{\rm p,bol}$ & 0.26 & 0.31 & 0.21 & 0.42 & 
$\log_{10}E'_{\gamma,\rm iso}$ vs. $\log_{10}L'_{\rm p,bol}$ & 0.24 & 0.38 & 0.22 & 0.42 & 
$\log_{10}E^{*'}_{\rm p}$ vs. $\log_{10}L'_{\rm p,bol}$ & 0.36 & 0.17 & 0.38 & 0.15 \\
$\log_{10}L'_{\gamma,\rm iso}$ vs. $\log_{10}L'_{\rm p,bol}$ & 0.0 & 0.99 & -0.05 & 0.86 & 
$\log_{10}T^{*'}_{90}$ vs. $\log_{10}t^{*'}_{\rm p}$ & -0.12 & 0.66 & -0.09 & 0.75 & 
$\log_{10}E'_{\gamma,\rm iso}$ vs. $\log_{10}t^{*'}_{\rm p}$ & 0.18 & 0.54 & 0.07 & 0.82 \\
$\log_{10}E^{*'}_{\rm p}$ vs. $\log_{10}t^{*'}_{\rm p}$ & 0.17 & 0.54 & 0.32 & 0.25 & 
$\log_{10}L'_{\gamma,\rm iso}$ vs. $\log_{10}t^{*'}_{\rm p}$ & 0.35 & 0.2 & 0.1 & 0.73 & 
$\log_{10}L'_{\rm p,bol}$ vs. $\log_{10}t^{*'}_{\rm p}$ & 0.12 & 0.66 & 0.21 & 0.44 \\
$\log_{10}T^{*'}_{90}$ vs. $\Delta m_{\rm 15,bol}$ & 0.48 & 0.2 & 0.64 & 0.07 & 
$\log_{10}E'_{\gamma,\rm iso}$ vs. $\Delta m_{\rm 15,bol}$ & -0.04 & 0.92 & -0.04 & 0.93 & 
$\log_{10}E^{*'}_{\rm p}$ vs. $\Delta m_{\rm 15,bol}$ & -0.16 & 0.7 & -0.36 & 0.39 \\
$\log_{10}L'_{\gamma,\rm iso}$ vs. $\Delta m_{\rm 15,bol}$ & -0.46 & 0.21 & -0.43 & 0.25 & 
$\log_{10}L'_{\rm p,bol}$ vs. $\Delta m_{\rm 15,bol}$ & -0.18 & 0.65 & -0.24 & 0.53 & 
$\log_{10}t^{*'}_{\rm p}$ vs. $\Delta m_{\rm 15,bol}$ & -0.58 & 0.1 & -0.59 & 0.1 \\
$\log_{10}T^{*'}_{90}$ vs. $\log_{10}E'_{\rm K}$ & 0.09 & 0.65 & -0.08 & 0.69 & 
$\log_{10}E'_{\gamma,\rm iso}$ vs. $\log_{10}E'_{\rm K}$ & -0.03 & 0.87 & 0.08 & 0.69 & 
$\log_{10}E^{*'}_{\rm p}$ vs. $\log_{10}E'_{\rm K}$ & 0.09 & 0.64 & 0.08 & 0.69 \\
$\log_{10}L'_{\gamma,\rm iso}$ vs. $\log_{10}E'_{\rm K}$ & -0.01 & 0.97 & 0.0 & 0.98 & 
$\log_{10}L'_{\rm p,bol}$ vs. $\log_{10}E'_{\rm K}$ & -0.13 & 0.67 & -0.04 & 0.9 & 
$\log_{10}t^{*'}_{\rm p}$ vs. $\log_{10}E'_{\rm K}$ & 0.22 & 0.49 & 0.32 & 0.31 \\
$\Delta m_{\rm 15,bol}$ vs. $\log_{10}E'_{\rm K}$ & -0.41 & 0.28 & -0.49 & 0.18 & 
$\log_{10}T^{*'}_{90}$ vs. $\log_{10}M_{\rm ej}$ & -0.13 & 0.5 & -0.15 & 0.43 & 
$\log_{10}E'_{\gamma,\rm iso}$ vs. $\log_{10}M_{\rm ej}$ & -0.05 & 0.8 & 0.01 & 0.96 \\
$\log_{10}E^{*'}_{\rm p}$ vs. $\log_{10}M_{\rm ej}$ & 0.08 & 0.68 & 0.03 & 0.88 & 
$\log_{10}L'_{\gamma,\rm iso}$ vs. $\log_{10}M_{\rm ej}$ & 0.09 & 0.66 & 0.08 & 0.68 & 
$\log_{10}L'_{\rm p,bol}$ vs. $\log_{10}M_{\rm ej}$ & 0.03 & 0.92 & -0.07 & 0.81 \\
$\log_{10}t^{*'}_{\rm p}$ vs. $\log_{10}M_{\rm ej}$ & 0.32 & 0.3 & 0.22 & 0.5 & 
$\Delta m_{\rm 15,bol}$ vs. $\log_{10}M_{\rm ej}$ & -0.73 & 0.02 & -0.7 & 0.04 & 
$\log_{10}E'_{\rm K}$ vs. $\log_{10}M_{\rm ej}$ & 0.73 & $<0.01$ & 0.81 & $<0.01$ \\
$\log_{10}T^{*'}_{90}$ vs. $\log_{10}M_{\rm Ni}$ & 0.13 & 0.48 & 0.07 & 0.7 & 
$\log_{10}E'_{\gamma,\rm iso}$ vs. $\log_{10}M_{\rm Ni}$ & 0.29 & 0.11 & 0.25 & 0.17 & 
$\log_{10}E^{*'}_{\rm p}$ vs. $\log_{10}M_{\rm Ni}$ & 0.34 & 0.05 & 0.34 & 0.05 \\
$\log_{10}L'_{\gamma,\rm iso}$ vs. $\log_{10}M_{\rm Ni}$ & 0.2 & 0.27 & 0.14 & 0.44 & 
$\log_{10}L'_{\rm p,bol}$ vs. $\log_{10}M_{\rm Ni}$ & 0.68 & $<0.01$ & 0.83 & $<0.01$ & 
$\log_{10}t^{*'}_{\rm p}$ vs. $\log_{10}M_{\rm Ni}$ & 0.34 & 0.2 & 0.31 & 0.25 \\
$\Delta m_{\rm 15,bol}$ vs. $\log_{10}M_{\rm Ni}$ & -0.25 & 0.51 & -0.36 & 0.34 & 
$\log_{10}E'_{\rm K}$ vs. $\log_{10}M_{\rm Ni}$ & 0.34 & 0.05 & 0.34 & 0.06 & 
$\log_{10}M_{\rm ej}$ vs. $\log_{10}M_{\rm Ni}$ & 0.26 & 0.15 & 0.27 & 0.13 \\
$\log_{10}T^{*'}_{90}$ vs. $\log_{10}v_{\rm ph}$ & 0.34 & 0.17 & 0.42 & 0.08 & 
$\log_{10}E'_{\gamma,\rm iso}$ vs. $\log_{10}v_{\rm ph}$ & -0.18 & 0.51 & -0.23 & 0.4 & 
$\log_{10}E^{*'}_{\rm p}$ vs. $\log_{10}v_{\rm ph}$ & -0.23 & 0.39 & -0.23 & 0.4 \\
$\log_{10}L'_{\gamma,\rm iso}$ vs. $\log_{10}v_{\rm ph}$ & -0.41 & 0.12 & -0.49 & 0.05 & 
$\log_{10}L'_{\rm p,bol}$ vs. $\log_{10}v_{\rm ph}$ & -0.13 & 0.68 & -0.29 & 0.34 & 
$\log_{10}t^{*'}_{\rm p}$ vs. $\log_{10}v_{\rm ph}$ & -0.45 & 0.14 & -0.36 & 0.25 \\
$\Delta m_{\rm 15,bol}$ vs. $\log_{10}v_{\rm ph}$ & 0.67 & 0.07 & 0.77 & 0.03 & 
$\log_{10}E'_{\rm K}$ vs. $\log_{10}v_{\rm ph}$ & 0.25 & 0.36 & 0.19 & 0.48 & 
$\log_{10}M_{\rm ej}$ vs. $\log_{10}v_{\rm ph}$ & 0.04 & 0.87 & 0.06 & 0.83 \\
$\log_{10}M_{\rm Ni}$ vs. $\log_{10}v_{\rm ph}$ & -0.59 & 0.01 & -0.51 & 0.04 & 
$\log_{10}T^{*'}_{90}$ vs. $k_{\rm avg}$ & 0.25 & 0.15 & 0.16 & 0.35 & 
$\log_{10}E'_{\gamma,\rm iso}$ vs. $k_{\rm avg}$ & 0.27 & 0.13 & 0.13 & 0.46 \\
$\log_{10}E^{*'}_{\rm p}$ vs. $k_{\rm avg}$ & 0.34 & 0.05 & 0.26 & 0.14 & 
$\log_{10}L'_{\gamma,\rm iso}$ vs. $k_{\rm avg}$ & 0.12 & 0.48 & 0.04 & 0.8 & 
$\log_{10}L'_{\rm p,bol}$ vs. $k_{\rm avg}$ & 0.74 & $<0.01$ & 0.76 & $<0.01$ \\
$\log_{10}t^{*'}_{\rm p}$ vs. $k_{\rm avg}$ & 0.26 & 0.35 & 0.19 & 0.51 & 
$\Delta m_{\rm 15,bol}$ vs. $k_{\rm avg}$ & -0.19 & 0.62 & -0.18 & 0.65 & 
$\log_{10}E'_{\rm K}$ vs. $k_{\rm avg}$ & 0.33 & 0.1 & 0.33 & 0.1 \\
$\log_{10}M_{\rm ej}$ vs. $k_{\rm avg}$ & 0.25 & 0.22 & 0.28 & 0.17 & 
$\log_{10}M_{\rm Ni}$ vs. $k_{\rm avg}$ & 0.86 & $<0.01$ & 0.89 & $<0.01$ & 
$\log_{10}v_{\rm ph}$ vs. $k_{\rm avg}$ & -0.36 & 0.21 & -0.31 & 0.28 \\
$\log_{10}T^{*'}_{90}$ vs. $s_{\rm avg}$ & -0.05 & 0.77 & -0.04 & 0.82 & 
$\log_{10}E'_{\gamma,\rm iso}$ vs. $s_{\rm avg}$ & 0.05 & 0.76 & 0.14 & 0.44 & 
$\log_{10}E^{*'}_{\rm p}$ vs. $s_{\rm avg}$ & 0.08 & 0.64 & 0.04 & 0.81 \\
$\log_{10}L'_{\gamma,\rm iso}$ vs. $s_{\rm avg}$ & 0.13 & 0.45 & 0.06 & 0.74 & 
$\log_{10}L'_{\rm p,bol}$ vs. $s_{\rm avg}$ & 0.18 & 0.5 & 0.25 & 0.34 & 
$\log_{10}t^{*'}_{\rm p}$ vs. $s_{\rm avg}$ & 0.88 & $<0.01$ & 0.68 & 0.01 \\
$\Delta m_{\rm 15,bol}$ vs. $s_{\rm avg}$ & -0.51 & 0.16 & -0.56 & 0.12 & 
$\log_{10}E'_{\rm K}$ vs. $s_{\rm avg}$ & 0.82 & $<0.01$ & 0.8 & $<0.01$ & 
$\log_{10}M_{\rm ej}$ vs. $s_{\rm avg}$ & 0.85 & $<0.01$ & 0.84 & $<0.01$ \\
$\log_{10}M_{\rm Ni}$ vs. $s_{\rm avg}$ & 0.48 & 0.01 & 0.5 & 0.01 & 
$\log_{10}v_{\rm ph}$ vs. $s_{\rm avg}$ & -0.49 & 0.08 & -0.49 & 0.08 & 
$k_{\rm avg}$ vs. $s_{\rm avg}$ & 0.13 & 0.46 & 0.21 & 0.22 \\
$\log_{10}T^{*'}_{90}$ vs. $\log_{10}T^{*'}_{\rm a,opt}$ & 0.39 & 0.09 & 0.32 & 0.17 & 
$\log_{10}E'_{\gamma,\rm iso}$ vs. $\log_{10}T^{*'}_{\rm a,opt}$ & 0.07 & 0.79 & 0.15 & 0.53 & 
$\log_{10}E^{*'}_{\rm p}$ vs. $\log_{10}T^{*'}_{\rm a,opt}$ & -0.2 & 0.41 & -0.23 & 0.35 \\
$\log_{10}L'_{\gamma,\rm iso}$ vs. $\log_{10}T^{*'}_{\rm a,opt}$ & -0.52 & 0.02 & -0.27 & 0.26 & 
$\log_{10}L'_{\rm p,bol}$ vs. $\log_{10}T^{*'}_{\rm a,opt}$ & -0.14 & 0.71 & 0.07 & 0.86 & 
$\log_{10}t^{*'}_{\rm p}$ vs. $\log_{10}T^{*'}_{\rm a,opt}$ & -0.6 & 0.09 & -0.57 & 0.11 \\
$\Delta m_{\rm 15,bol}$ vs. $\log_{10}T^{*'}_{\rm a,opt}$ & 0.05 & 0.93 & -0.03 & 0.96 & 
$\log_{10}E'_{\rm K}$ vs. $\log_{10}T^{*'}_{\rm a,opt}$ & 0.16 & 0.56 & 0.34 & 0.22 & 
$\log_{10}M_{\rm ej}$ vs. $\log_{10}T^{*'}_{\rm a,opt}$ & 0.14 & 0.63 & 0.29 & 0.3 \\
$\log_{10}M_{\rm Ni}$ vs. $\log_{10}T^{*'}_{\rm a,opt}$ & -0.1 & 0.73 & 0.24 & 0.36 & 
$\log_{10}v_{\rm ph}$ vs. $\log_{10}T^{*'}_{\rm a,opt}$ & 0.22 & 0.6 & 0.17 & 0.69 & 
$k_{\rm avg}$ vs. $\log_{10}T^{*'}_{\rm a,opt}$ & 0.05 & 0.84 & 0.12 & 0.65 \\
$s_{\rm avg}$ vs. $\log_{10}T^{*'}_{\rm a,opt}$ & -0.19 & 0.47 & 0.1 & 0.71 & 
$\log_{10}T^{*'}_{90}$ vs. $\log_{10}L'_{\rm a,opt}$ & -0.24 & 0.31 & 0.05 & 0.84 & 
$\log_{10}E'_{\gamma,\rm iso}$ vs. $\log_{10}L'_{\rm a,opt}$ & 0.4 & 0.09 & 0.54 & 0.02 \\
$\log_{10}E^{*'}_{\rm p}$ vs. $\log_{10}L'_{\rm a,opt}$ & 0.45 & 0.06 & 0.59 & 0.01 & 
$\log_{10}L'_{\gamma,\rm iso}$ vs. $\log_{10}L'_{\rm a,opt}$ & 0.7 & $<0.01$ & 0.43 & 0.06 & 
$\log_{10}L'_{\rm p,bol}$ vs. $\log_{10}L'_{\rm a,opt}$ & 0.23 & 0.54 & 0.07 & 0.86 \\
$\log_{10}t^{*'}_{\rm p}$ vs. $\log_{10}L'_{\rm a,opt}$ & 0.71 & 0.03 & 0.27 & 0.49 & 
$\Delta m_{\rm 15,bol}$ vs. $\log_{10}L'_{\rm a,opt}$ & -0.32 & 0.53 & 0.03 & 0.96 & 
$\log_{10}E'_{\rm K}$ vs. $\log_{10}L'_{\rm a,opt}$ & -0.02 & 0.93 & -0.02 & 0.94 \\
$\log_{10}M_{\rm ej}$ vs. $\log_{10}L'_{\rm a,opt}$ & 0.01 & 0.96 & 0.05 & 0.85 & 
$\log_{10}M_{\rm Ni}$ vs. $\log_{10}L'_{\rm a,opt}$ & 0.11 & 0.69 & -0.27 & 0.32 & 
$\log_{10}v_{\rm ph}$ vs. $\log_{10}L'_{\rm a,opt}$ & -0.08 & 0.86 & 0.17 & 0.69 \\
$k_{\rm avg}$ vs. $\log_{10}L'_{\rm a,opt}$ & 0.08 & 0.75 & -0.08 & 0.76 & 
$s_{\rm avg}$ vs. $\log_{10}L'_{\rm a,opt}$ & 0.2 & 0.44 & 0.07 & 0.79 & 
$\log_{10}T^{*'}_{\rm a,opt}$ vs. $\log_{10}L'_{\rm a,opt}$ & -0.7 & $<0.01$ & -0.57 & 0.01 \\
$\log_{10}T^{*'}_{90}$ vs. $\log_{10}T^{*'}_{\rm a,X}$ & 0.04 & 0.9 & 0.13 & 0.66 & 
$\log_{10}E'_{\gamma,\rm iso}$ vs. $\log_{10}T^{*'}_{\rm a,X}$ & -0.04 & 0.88 & -0.02 & 0.93 & 
$\log_{10}E^{*'}_{\rm p}$ vs. $\log_{10}T^{*'}_{\rm a,X}$ & 0.18 & 0.56 & 0.11 & 0.72 \\
$\log_{10}L'_{\gamma,\rm iso}$ vs. $\log_{10}T^{*'}_{\rm a,X}$ & -0.13 & 0.67 & -0.31 & 0.27 & 
$\log_{10}L'_{\rm p,bol}$ vs. $\log_{10}T^{*'}_{\rm a,X}$ & 0.58 & 0.17 & 0.71 & 0.07 & 
$\log_{10}t^{*'}_{\rm p}$ vs. $\log_{10}T^{*'}_{\rm a,X}$ & 0.12 & 0.8 & 0.29 & 0.53 \\
$\Delta m_{\rm 15,bol}$ vs. $\log_{10}T^{*'}_{\rm a,X}$ & -0.01 & 0.99 & 0.4 & 0.6 & 
$\log_{10}E'_{\rm K}$ vs. $\log_{10}T^{*'}_{\rm a,X}$ & -0.1 & 0.76 & 0.12 & 0.73 & 
$\log_{10}M_{\rm ej}$ vs. $\log_{10}T^{*'}_{\rm a,X}$ & -0.01 & 0.97 & 0.18 & 0.6 \\
$\log_{10}M_{\rm Ni}$ vs. $\log_{10}T^{*'}_{\rm a,X}$ & 0.37 & 0.24 & 0.4 & 0.2 & 
$\log_{10}v_{\rm ph}$ vs. $\log_{10}T^{*'}_{\rm a,X}$ & -0.04 & 0.94 & -0.1 & 0.87 & 
$k_{\rm avg}$ vs. $\log_{10}T^{*'}_{\rm a,X}$ & 0.48 & 0.12 & 0.57 & 0.05 \\
$s_{\rm avg}$ vs. $\log_{10}T^{*'}_{\rm a,X}$ & -0.14 & 0.66 & 0.09 & 0.78 & 
$\log_{10}T^{*'}_{\rm a,opt}$ vs. $\log_{10}T^{*'}_{\rm a,X}$ & 0.35 & 0.26 & 0.3 & 0.34 & 
$\log_{10}L'_{\rm a,opt}$ vs. $\log_{10}T^{*'}_{\rm a,X}$ & -0.03 & 0.93 & -0.13 & 0.68 \\
$\log_{10}T^{*'}_{90}$ vs. $\log_{10}L^{*'}_{a,X}$ & 0.45 & 0.1 & 0.26 & 0.37 & 
$\log_{10}E'_{\gamma,\rm iso}$ vs. $\log_{10}L^{*'}_{a,X}$ & 0.54 & 0.05 & 0.55 & 0.04 & 
$\log_{10}E^{*'}_{\rm p}$ vs. $\log_{10}L^{*'}_{a,X}$ & 0.19 & 0.54 & 0.27 & 0.37 \\
$\log_{10}L'_{\gamma,\rm iso}$ vs. $\log_{10}L^{*'}_{a,X}$ & 0.25 & 0.4 & 0.37 & 0.2 & 
$\log_{10}L'_{\rm p,bol}$ vs. $\log_{10}L^{*'}_{a,X}$ & -0.08 & 0.86 & 0.11 & 0.82 & 
$\log_{10}t^{*'}_{\rm p}$ vs. $\log_{10}L^{*'}_{a,X}$ & 0.24 & 0.6 & 0.36 & 0.43 \\
$\Delta m_{\rm 15,bol}$ vs. $\log_{10}L^{*'}_{a,X}$ & -0.13 & 0.87 & 0.0 & 1.0 & 
$\log_{10}E'_{\rm K}$ vs. $\log_{10}L^{*'}_{a,X}$ & 0.44 & 0.18 & 0.39 & 0.23 & 
$\log_{10}M_{\rm ej}$ vs. $\log_{10}L^{*'}_{a,X}$ & 0.27 & 0.42 & 0.21 & 0.54 \\
$\log_{10}M_{\rm Ni}$ vs. $\log_{10}L^{*'}_{a,X}$ & -0.04 & 0.9 & 0.01 & 0.98 & 
$\log_{10}v_{\rm ph}$ vs. $\log_{10}L^{*'}_{a,X}$ & 0.09 & 0.89 & 0.3 & 0.62 & 
$k_{\rm avg}$ vs. $\log_{10}L^{*'}_{a,X}$ & -0.09 & 0.78 & -0.16 & 0.62 \\
$s_{\rm avg}$ vs. $\log_{10}L^{*'}_{a,X}$ & 0.34 & 0.28 & 0.48 & 0.12 & 
$\log_{10}T^{*'}_{\rm a,opt}$ vs. $\log_{10}L^{*'}_{a,X}$ & -0.05 & 0.87 & 0.01 & 0.98 & 
$\log_{10}L'_{\rm a,opt}$ vs. $\log_{10}L^{*'}_{a,X}$ & 0.34 & 0.28 & 0.36 & 0.25 \\
$\log_{10}T^{*'}_{\rm a,X}$ vs. $\log_{10}L^{*'}_{a,X}$ & -0.75 & $<0.01$ & -0.71 & $<0.01$ & 
$\log_{10}T^{*'}_{90}$ vs. $\log_{10}\theta'_{\rm jet}$ & 0.02 & 0.95 & 0.03 & 0.9 & 
$\log_{10}E'_{\gamma,\rm iso}$ vs. $\log_{10}\theta'_{\rm jet}$ & -0.54 & 0.02 & -0.54 & 0.03 \\
$\log_{10}E^{*'}_{\rm p}$ vs. $\log_{10}\theta'_{\rm jet}$ & -0.09 & 0.73 & -0.09 & 0.75 & 
$\log_{10}L'_{\gamma,\rm iso}$ vs. $\log_{10}\theta'_{\rm jet}$ & -0.58 & 0.02 & -0.65 & $<0.01$ & 
$\log_{10}L'_{\rm p,bol}$ vs. $\log_{10}\theta'_{\rm jet}$ & -0.48 & 0.23 & -0.17 & 0.69 \\
$\log_{10}t^{*'}_{\rm p}$ vs. $\log_{10}\theta'_{\rm jet}$ & -0.1 & 0.83 & -0.04 & 0.94 & 
$\Delta m_{\rm 15,bol}$ vs. $\log_{10}\theta'_{\rm jet}$ & 0.31 & 0.5 & 0.31 & 0.5 & 
$\log_{10}E'_{\rm K}$ vs. $\log_{10}\theta'_{\rm jet}$ & 0.01 & 0.98 & -0.02 & 0.93 \\
$\log_{10}M_{\rm ej}$ vs. $\log_{10}\theta'_{\rm jet}$ & -0.15 & 0.6 & -0.13 & 0.67 & 
$\log_{10}M_{\rm Ni}$ vs. $\log_{10}\theta'_{\rm jet}$ & 0.07 & 0.81 & 0.16 & 0.58 & 
$\log_{10}v_{\rm ph}$ vs. $\log_{10}\theta'_{\rm jet}$ & 0.72 & 0.04 & 0.61 & 0.11 \\
$k_{\rm avg}$ vs. $\log_{10}\theta'_{\rm jet}$ & 0.11 & 0.71 & 0.24 & 0.42 & 
$s_{\rm avg}$ vs. $\log_{10}\theta'_{\rm jet}$ & -0.2 & 0.49 & -0.09 & 0.77 & 
$\log_{10}T^{*'}_{\rm a,opt}$ vs. $\log_{10}\theta'_{\rm jet}$ & 0.56 & 0.2 & 0.54 & 0.22 \\
$\log_{10}L'_{\rm a,opt}$ vs. $\log_{10}\theta'_{\rm jet}$ & -0.65 & 0.11 & -0.32 & 0.48 & 
$\log_{10}T^{*'}_{\rm a,X}$ vs. $\log_{10}\theta'_{\rm jet}$ & 0.49 & 0.4 & 0.8 & 0.1 & 
$\log_{10}L^{*'}_{a,X}$ vs. $\log_{10}\theta'_{\rm jet}$ & -0.73 & 0.16 & -0.7 & 0.19 \\
$\log_{10}T^{*'}_{90}$ vs. $\log_{10}T^{*}_{\rm jet}$ & 0.01 & 0.98 & -0.06 & 0.79 & 
$\log_{10}E'_{\gamma,\rm iso}$ vs. $\log_{10}T^{*}_{\rm jet}$ & -0.09 & 0.72 & -0.16 & 0.49 & 
$\log_{10}E^{*'}_{\rm p}$ vs. $\log_{10}T^{*}_{\rm jet}$ & -0.53 & 0.02 & -0.54 & 0.02 \\
$\log_{10}L'_{\gamma,\rm iso}$ vs. $\log_{10}T^{*}_{\rm jet}$ & -0.09 & 0.71 & -0.3 & 0.19 & 
$\log_{10}L'_{\rm p,bol}$ vs. $\log_{10}T^{*}_{\rm jet}$ & -0.4 & 0.51 & -0.1 & 0.87 & 
$\log_{10}t^{*'}_{\rm p}$ vs. $\log_{10}T^{*}_{\rm jet}$ & 0.14 & 0.82 & -0.2 & 0.75 \\
$\Delta m_{\rm 15,bol}$ vs. $\log_{10}T^{*}_{\rm jet}$ & 0.81 & 0.39 & 1.0 & $<0.01$ & 
$\log_{10}E'_{\rm K}$ vs. $\log_{10}T^{*}_{\rm jet}$ & 0.07 & 0.84 & 0.27 & 0.42 & 
$\log_{10}M_{\rm ej}$ vs. $\log_{10}T^{*}_{\rm jet}$ & -0.21 & 0.53 & -0.1 & 0.77 \\
$\log_{10}M_{\rm Ni}$ vs. $\log_{10}T^{*}_{\rm jet}$ & -0.24 & 0.44 & -0.11 & 0.74 & 
$\log_{10}v_{\rm ph}$ vs. $\log_{10}T^{*}_{\rm jet}$ & -0.44 & 0.46 & -0.3 & 0.62 & 
$k_{\rm avg}$ vs. $\log_{10}T^{*}_{\rm jet}$ & -0.3 & 0.34 & -0.27 & 0.39 \\
$s_{\rm avg}$ vs. $\log_{10}T^{*}_{\rm jet}$ & 0.19 & 0.56 & 0.28 & 0.38 & 
$\log_{10}T^{*'}_{\rm a,opt}$ vs. $\log_{10}T^{*}_{\rm jet}$ & 0.72 & 0.07 & 0.79 & 0.04 & 
$\log_{10}L'_{\rm a,opt}$ vs. $\log_{10}T^{*}_{\rm jet}$ & -0.26 & 0.57 & -0.29 & 0.53 \\
$\log_{10}T^{*'}_{\rm a,X}$ vs. $\log_{10}T^{*}_{\rm jet}$ & -0.4 & 0.51 & -0.3 & 0.62 & 
$\log_{10}L^{*'}_{a,X}$ vs. $\log_{10}T^{*}_{\rm jet}$ & 0.44 & 0.46 & 0.1 & 0.87 & 
$\log_{10}\theta'_{\rm jet}$ vs. $\log_{10}T^{*}_{\rm jet}$ & 0.47 & 0.2 & 0.47 & 0.21 \\
    \enddata
    \label{allrelationsGRBSNeEP}

\tablecomments{Since the checked correlations of this table are the same as Table \ref{allrelationsGRBSNe}, the notes' column and the number of data points will be omitted. See the previous Table \ref{allrelationsGRBSNe} for the references.}
\end{deluxetable}
\end{longrotatetable}

\newpage
\section{The ODR method}\label{sec:ODR}
The ODR method \citep{Boggs1989} deals with the problem of best fitting for data when the uncertainties are present not only on the dependent variable ($x$) but also on the independent variables ($y$) that are present in the correlations. 

Let us call $(x_i,y_i)$, with $i=1,...,N$, the pairs of independent and dependent observed variables, respectively. If we consider $(X_i,Y_i)$ as the true values of the variables and $(\delta_i,\epsilon_i)$ the random errors associated with $x_i$ and $y_i$, respectively, we can say that $x_i=X_i-\delta_i$ and $y_i=Y_i-\epsilon_i$. Furthermore, it can be assumed that $Y_i$ is a smooth function of $X_i$ and a set of parameters $\pmb{\beta}$, namely $Y_i=f(X_i,\pmb{\beta})$: this condition can be recast as $y_i=f(x_i+\delta_i,\pmb{\beta})-\epsilon_i$. An ODR problem is based on the minimization of the orthogonal distance, $R_i$, from the point $(x_i,y_i)$ to the curve expressed by $f(\hat{x},\hat{\pmb{\beta}})$ (where the $\hat{}$ notation denotes a generic choice for the values). If the hypothesis of Gaussian distributions for the errors $\delta_i$ and $\epsilon_i$ is assumed, then the ODR problem can be written as:

\begin{equation}
    R^{2}_i=\min_{\hat{\pmb{\beta}},\hat{\delta}}\sum_{i=1}^{N}w^{2}_{i}[(f(x_i+\hat{\delta_i},\hat{\beta})-y_i)^2+d^{2}_{i}\hat{\delta_i}^2],
    \label{eq:ODR}
\end{equation}

where the constraint given by $y_i=f(x_i+\hat{\delta_i},\hat{\pmb{\beta}})-\hat{\epsilon_i}$ substituted the $\hat{\epsilon_i}$, $w_i\geq0$ and $d_i>0$ are the weights for the problem in the general form. In our case, we consider the following expressions for the weights: $w_i=1/\sigma_{\epsilon_i}$ and $d_i=\sigma_{\epsilon_i}/\sigma_{d_i}$ \citep{Boggs1989}. We here remark that the ODR method corresponds to a fitting with the weights on only one variable when the other variable has no associated uncertainties and, thus, no defined weights. 

\begin{figure}[hbt!]
    \graphicspath{{06-03/}}
    \centering
    \gridline{\fig{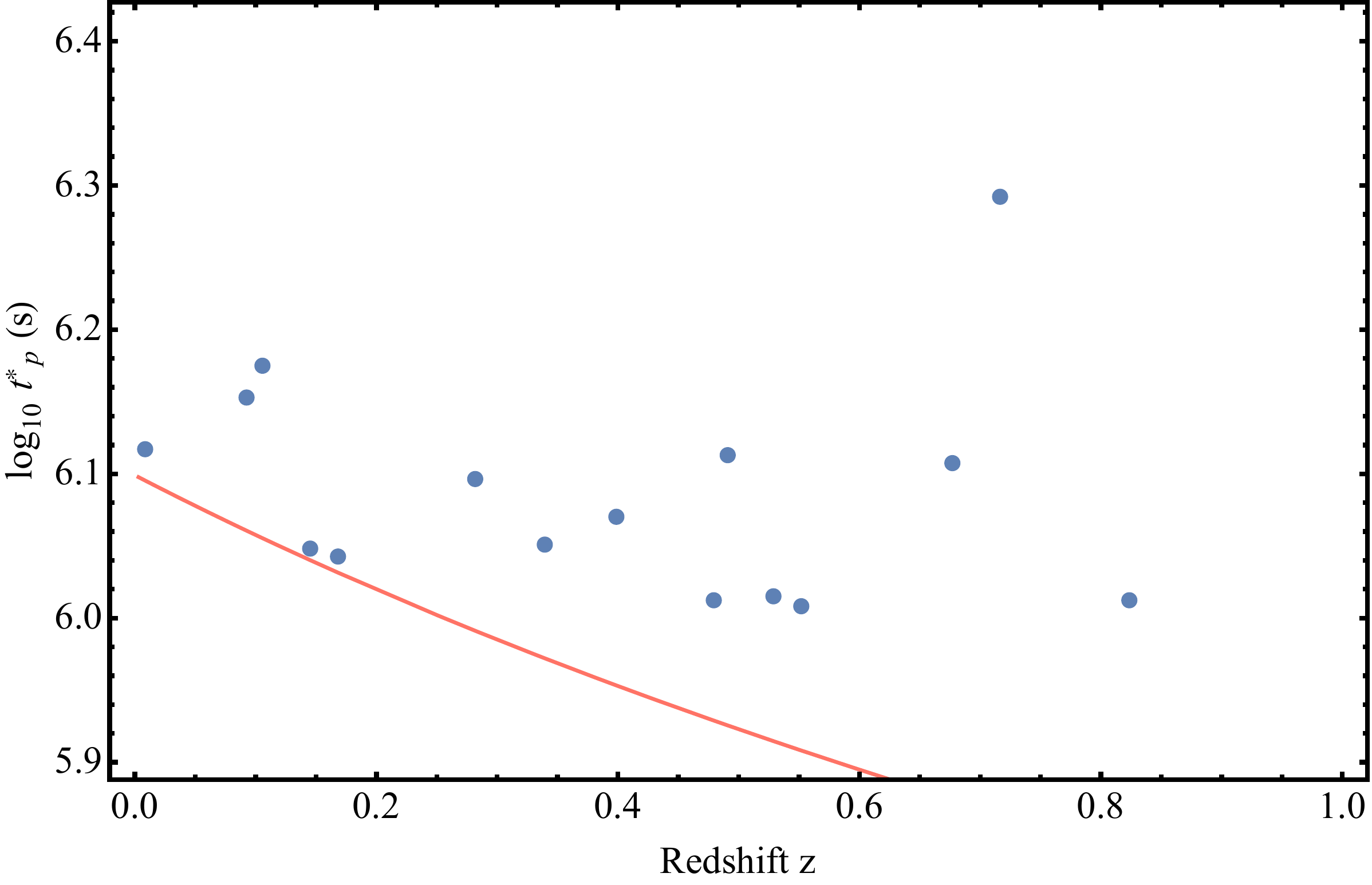}{0.40\textwidth}{(a) Limiting curve of $\log_{10}t^{*}_{\rm p}$.}
    \fig{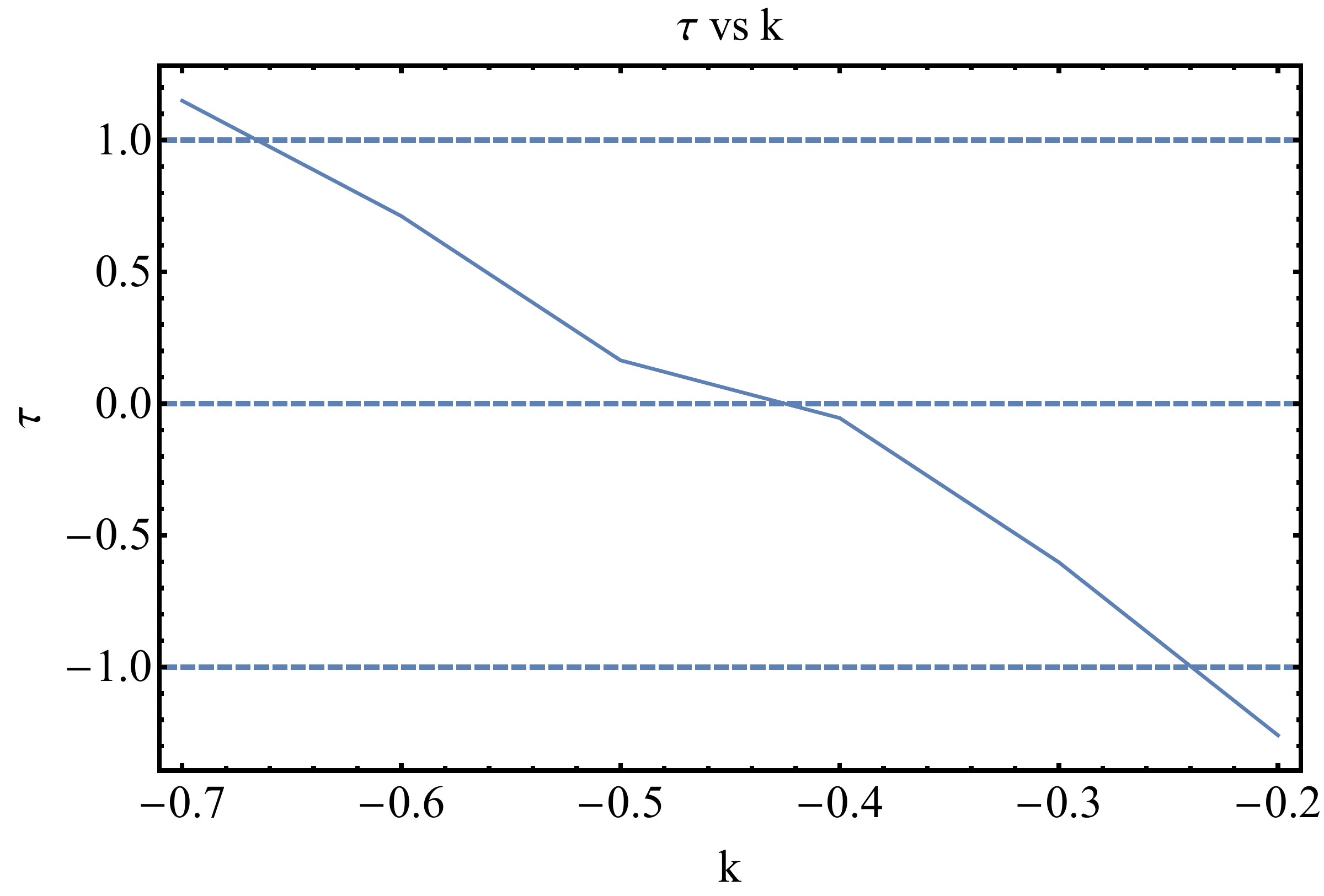}{0.40\textwidth}{(b) $\tau$-$k$ plot for $t^{*}_{\rm p}$.}}
    \caption{\textbf{(a)} The limiting curve applied to the variable $t^{*'}_{p}$ expressed as the $\log_{10}$ of its values in $s$. \textbf{(b)} The plot $\tau$-$k$ for this variable: it can be seen that $k_{t^{*'}_{p}}$ value is obtained at the level $\tau=0$.}
    \label{fig:EP}
\end{figure}

\section{The EP method}\label{sec:EP}
\noindent 
GRB correlations are affected by biases resulting from sample size and selection effects due to instrumental thresholds \citep{Dainotti2013, Dainotti2015,Dainotti2017a,Dainotti2020a,DainottiLevine2021,DainottiNielson2022}. To determine whether the GRB-SNe correlations plotted in the scatter matrix are intrinsic or affected by these biases, we utilize the EP method \citep{EfronPetrosian1992} to create new corrected and unbiased data, the so-called \textit{local variables},  denoted with the symbol $'$. Previous studies have successfully applied this method to remove biases due to the incomplete data and to remove the cosmological evolution of the variables \citep{Lloyd2000,Kocevski2006,Petrosian2009,Dainotti2013,Dainotti2015,Dainotti2017a,Dainotti2018,Dainotti2020a}. The EP method employs a modification of the Kendall $\tau$ rank test \citep{kendall1948rank} to determine the dependence of a selected variable, $x$, on the GRB-SN's redshift, $z$, namely the \textit{evolution of variable} (see \citet{Dainotti2013} and references therein, i.e. \citet{Singal2011}). The test statistic is defined as

\begin{equation}
    \tau = \frac{\Sigma_{i}\left(\mathcal{R}_{i}-\mathcal{E}_{i}\right)}{\sqrt{\Sigma_{i}\mathcal{V}_{i}}},
\end{equation}

where $\mathcal{R}_{i}$ is the rank of the selected variable, $x_{i}$, of the $i$-th data point, $\left(z_{i},x_{i}\right)$, within its \textit{associated set}. For truncated data, the associated set relative to a given data point consists of the points having a lower redshift that would have been observed at the redshift of the given GRB characterized by $z_i$ and $L_i$ and with a luminosity (or energy) higher than the minimum one determined by that given GRB. Thus, for the $i$-th data point, for luminosity-like and energy-like properties, the associated set for determining the evolution of variable $x$ is:
\begin{equation}
    J_{i} \equiv \left\{j \colon z_{j} < z_{\rm max} \left(x_{i}\right) \right\} \vee \left\{j \colon x_{j} > x_{i} \right\},
\end{equation}
where $z_{max}\left(x_{i}\right)$ is the maximum redshift at which the $i$-th object could be placed and still be included in the survey for the given variable. For time-like properties, the associated set is:
\begin{equation}
    J_{i} \equiv \{j \colon z_{j} > z_{\rm min}\left(x_{i}\right) \} \vee \{j \colon x_{j} > x_{i} \},
\end{equation}
where $z_{min}\left(x_{i}\right)$ is the minimum redshift at which the $i$-th object could be placed and still be included in the survey. 
If $x$ is not dependent on $z$, the rank $\mathcal{R}_{i}$ has then a continuous distribution in the $(0,1)$ interval, an the expectation value $\mathcal{E}_{i} = \frac{1}{2}\left(i+1\right)$ and variance $\mathcal{V}_{i} = \frac{1}{12}\left(i^{2}+1\right)$. The mean and variance are estimated separately for each associated set and summed to find a single value for $\tau$, and independence is rejected at the $n\sigma$ level if $|\tau|>n$.

For a selected variable $x$, the $\tau_{x}$ test statistic represents the degree of correlation for the truncated sample and is used to find the best-fit parametrization describing the evolution of $x$ and, in particular, the slope of the evolutionary function, $k_{x}$. The found value of $k_{x}$ is then used in a correlation function $g\left(z\right)$ to de-evolve the selected variable and thereby determine the local variable, $x' = x/g\left(z\right)$. Here, we choose the simple correlation function, $f\left(z\right) = \left(1+z\right)^{k}$, since previous investigations tested more complex functions and obtained results compatible with those found using the simple function \citep{Dainotti2017a}. The local variable is then defined as $x' = x/\left(1+z\right)^{k_{x}}$.

We plot a limiting curve alongside the data based on observational limitations for the variables of interest. 
For the time, we consider the limiting time as $T^{* \rm lim}=T_{\rm lim}/(1+z)$ where $T_{\rm lim}$ is the observed limiting time, see the left panel of Figure \ref{fig:EP}.  

For luminosity and energy properties, we determine the minimum observable value for a given redshift using a limiting flux, $F_{\rm lim}$, or fluence, $f_{\rm lim}$, to calculate a limiting luminosity, $L_{\rm lim} = 4\pi D_{L}^{2}\left(z\right)F_{\rm lim}$, or limiting energy, $E_{\rm lim} = 4\pi D^{2}_{L}\left(z\right)f_{\rm lim}$, respectively, where $D_{L}^{2}\left(z\right)$ is the luminosity distance of a given GRB-SN. The limiting sensitivity of Swift satellite XRT, $F_{\rm lim} = 10^{-14} \rm \,erg\,cm^{-2}\,s^{-1}$, is not high enough to describe the truncation of our sample. Seeking a compromise between retaining $\gtrsim 90\%$ of the plotted GRB-SNe and a reasonable estimation of the truncation, we keep on restricting the limiting value until the curve has good agreement with the data. For other reported properties, we begin with the minimum value of the data set, and likewise further restrict the value until the limiting curve is representative of the data. Finally, the Kendall $\tau$ is computed for several values of the slope of the evolutionary functions, $k$, until we find values of $k$ for which $\tau=0$ with $1\mhyphen\sigma$ uncertainty given by $|\tau_x|\leq1$ (see the right panel of Figure \ref{fig:EP}).

\end{appendices}

\normalsize
\bibliography{GRBSN_Bibliography.bib}{}

\begin{thebibliography}{}
\expandafter\ifx\csname natexlab\endcsname\relax\def\natexlab#1{#1}\fi
\providecommand{\url}[1]{\href{#1}{#1}}
\providecommand{\dodoi}[1]{doi:~\href{http://doi.org/#1}{\nolinkurl{#1}}}
\providecommand{\doeprint}[1]{\href{http://ascl.net/#1}{\nolinkurl{http://ascl.net/#1}}}
\providecommand{\doarXiv}[1]{\href{https://arxiv.org/abs/#1}{\nolinkurl{https://arxiv.org/abs/#1}}}

\bibitem[{Abbott {et~al.}(2017)Abbott, Abbott, Abbott, Acernese, Ackley, Adams,
  Adams, Addesso, Adhikari, Adya, Affeldt, Afrough, Agarwal, Agathos, Agatsuma,
  Aggarwal, Aguiar, Aiello, Ain, Ajith, Allen, Allen, Allocca, Altin, Amato,
  Ananyeva, Anderson, Anderson, Angelova, Antier, Appert, Arai, Araya, Areeda,
  Arnaud, Arun, Ascenzi, Ashton, Ast, Aston, Astone, Atallah, Aufmuth, Aulbert,
  AultONeal, Austin, Avila-Alvarez, Babak, Bacon, Bader, Bae, Baker,
  Baldaccini, Ballardin, Ballmer, Banagiri, Barayoga, Barclay, Barish, Barker,
  Barkett, Barone, Barr, Barsotti, Barsuglia, Barta, Barthelmy, Bartlett,
  Bartos, Bassiri, Basti, Batch, Bawaj, Bayley, Bazzan, B{\'{e}}csy, Beer,
  Bejger, Belahcene, Bell, Berger, Bergmann, Bero, Berry, Bersanetti,
  Bertolini, Betzwieser, Bhagwat, Bhandare, Bilenko, Billingsley, Billman,
  Birch, Birney, Birnholtz, Biscans, Biscoveanu, Bisht, Bitossi, Biwer,
  Bizouard, Blackburn, Blackman, Blair, Blair, Blair, Bloemen, Bock, Bode,
  Boer, Bogaert, Bohe, Bondu, Bonilla, Bonnand, Boom, Bork, Boschi, Bose,
  Bossie, Bouffanais, Bozzi, Bradaschia, Brady, Branchesi, Brau, Briant,
  Brillet, Brinkmann, Brisson, Brockill, Broida, Brooks, Brown, Brown, Brunett,
  Buchanan, Buikema, Bulik, Bulten, Buonanno, Buskulic, Buy, Byer, Cabero,
  Cadonati, Cagnoli, Cahillane, Bustillo, Callister, Calloni, Camp, Canepa,
  Canizares, Cannon, Cao, Cao, Capano, Capocasa, Carbognani, Caride, Carney,
  Diaz, Casentini, Caudill, Cavagli{\`{a}}, Cavalier, Cavalieri, Cella, Cepeda,
  Cerd{\'{a}}-Dur{\'{a}}n, Cerretani, Cesarini, Chamberlin, Chan, Chao,
  Charlton, Chase, Chassande-Mottin, Chatterjee, Chatziioannou, Cheeseboro,
  Chen, Chen, Chen, Cheng, Chia, Chincarini, Chiummo, Chmiel, Cho, Cho, Chow,
  Christensen, Chu, Chua, Chua, Chung, Chung, Ciani, Ciolfi, Cirelli, Cirone,
  Clara, Clark, Clearwater, Cleva, Cocchieri, Coccia, Cohadon, Cohen, Colla,
  Collette, Cominsky, Jr., Conti, Cooper, Corban, Corbitt,
  Cordero-Carri{\'{o}}n, Corley, Cornish, Corsi, Cortese, Costa, Coughlin,
  Coughlin, Coulon, Countryman, Couvares, Covas, Cowan, Coward, Cowart, Coyne,
  Coyne, Creighton, Creighton, Cripe, Crowder, Cullen, Cumming, Cunningham,
  Cuoco, Canton, D{\'{a}}lya, Danilishin, D'Antonio, Danzmann, Dasgupta, Costa,
  Dattilo, Dave, Davier, Davis, Daw, Day, De, DeBra, Degallaix, Laurentis,
  Del{\'{e}}glise, Pozzo, Demos, Denker, Dent, Pietri, Dergachev, Rosa, DeRosa,
  Rossi, DeSalvo, de~Varona, Devenson, Dhurandhar, D{\'{\i}}az, Fiore,
  Giovanni, Girolamo, Lieto, Pace, Palma, Renzo, Doctor, Dolique, Donovan,
  Dooley, Doravari, Dorrington, Douglas, {\'{A}}lvarez, Downes, Drago,
  Dreissigacker, Driggers, Du, Ducrot, Dupej, Dwyer, Edo, Edwards, Effler,
  Ehrens, Eichholz, Eikenberry, Eisenstein, Essick, Estevez, Etienne, Etzel,
  Evans, Evans, Factourovich, Fafone, Fair, Fairhurst, Fan, Farinon, Farr,
  Farr, Fauchon-Jones, Favata, Fays, Fee, Fehrmann, Feicht, Fejer,
  Fernandez-Galiana, Ferrante, Ferreira, Ferrini, Fidecaro, Finstad, Fiori,
  Fiorucci, Fishbach, Fisher, Fitz-Axen, Flaminio, Fletcher, Fong, Font,
  Forsyth, Forsyth, Fournier, Frasca, Frasconi, Frei, Freise, Frey, Frey,
  Fries, Fritschel, Frolov, Fulda, Fyffe, Gabbard, Gadre, Gaebel, Gair,
  Gammaitoni, Ganija, Gaonkar, Garcia-Quiros, Garufi, Gateley, Gaudio, Gaur,
  Gayathri, Gehrels, Gemme, Genin, Gennai, George, George, Gergely, Germain,
  Ghonge, Ghosh, Ghosh, Ghosh, Giaime, Giardina, Giazotto, Gill, Glover, Goetz,
  Goetz, Gomes, Goncharov, Gonz{\'{a}}lez, Castro, Gopakumar, Gorodetsky,
  Gossan, Gosselin, Gouaty, Grado, Graef, Granata, Grant, Gras, Gray, Greco,
  Green, Gretarsson, Griswold, Groot, Grote, Grunewald, Gruning, Guidi, Guo,
  Gupta, Gupta, Gushwa, Gustafson, Gustafson, Halim, Hall, Hall, Hamilton,
  Hammond, Haney, Hanke, Hanks, Hanna, Hannam, Hannuksela, Hanson, Hardwick,
  Harms, Harry, Harry, Hart, Haster, Haughian, Healy, Heidmann, Heintze,
  Heitmann, Hello, Hemming, Hendry, Heng, Hennig, Heptonstall, Heurs, Hild,
  Hinderer, Hoak, Hofman, Holt, Holz, Hopkins, Horst, Hough, Houston, Howell,
  Hreibi, Hu, Huerta, Huet, Hughey, Husa, Huttner, Huynh-Dinh, Indik, Inta,
  Intini, Isa, Isac, Isi, Iyer, Izumi, Jacqmin, Jani, Jaranowski, Jawahar,
  Jim{\'{e}}nez-Forteza, Johnson, Jones, Jones, Jonker, Ju, Junker, Kalaghatgi,
  Kalogera, Kamai, Kandhasamy, Kang, Kanner, Kapadia, Karki, Karvinen,
  Kasprzack, Katolik, Katsavounidis, Katzman, Kaufer, Kawabe,
  K{\'{e}}f{\'{e}}lian, Keitel, Kemball, Kennedy, Kent, Key, Khalili, Khan,
  Khan, Khan, Khazanov, Kijbunchoo, Kim, Kim, Kim, Kim, Kim, Kim, Kimbrell,
  King, King, Kinley-Hanlon, Kirchhoff, Kissel, Kleybolte, Klimenko, Knowles,
  Koch, Koehlenbeck, Koley, Kondrashov, Kontos, Korobko, Korth, Kowalska,
  Kozak, Krämer, Kringel, Krishnan, Kr{\'{o}}lak, Kuehn, Kumar, Kumar, Kumar,
  Kuo, Kutynia, Kwang, Lackey, Lai, Landry, Lang, Lange, Lantz, Lanza, Larson,
  Lartaux-Vollard, Lasky, Laxen, Lazzarini, Lazzaro, Leaci, Leavey, Lee, Lee,
  Lee, Lee, Lee, Lehmann, Lenon, Leonardi, Leroy, Letendre, Levin, Li, Linker,
  Littenberg, Liu, Lo, Lockerbie, London, Lord, Lorenzini, Loriette, Lormand,
  Losurdo, Lough, Lousto, Lovelace, Lück, Lumaca, Lundgren, Lynch, Ma, Macas,
  Macfoy, Machenschalk, MacInnis, Macleod, Hernandez, Maga{\~{n}}a-Sandoval,
  Zertuche, Magee, Majorana, Maksimovic, Man, Mandic, Mangano, Mansell, Manske,
  Mantovani, Marchesoni, Marion, M{\'{a}}rka, M{\'{a}}rka, Markakis, Markosyan,
  Markowitz, Maros, Marquina, Marsh, Martelli, Martellini, Martin, Martin,
  Martynov, Mason, Massera, Masserot, Massinger, Masso-Reid, Mastrogiovanni,
  Matas, Matichard, Matone, Mavalvala, Mazumder, McCarthy, McClelland,
  McCormick, McCuller, McGuire, McIntyre, McIver, McManus, McNeill, McRae,
  McWilliams, Meacher, Meadors, Mehmet, Meidam, Mejuto-Villa, Melatos, Mendell,
  Mercer, Merilh, Merzougui, Meshkov, Messenger, Messick, Metzdorff, Meyers,
  Miao, Michel, Middleton, Mikhailov, Milano, Miller, Miller, Miller,
  Millhouse, Milovich-Goff, Minazzoli, Minenkov, Ming, Mishra, Mitra,
  Mitrofanov, Mitselmakher, Mittleman, Moffa, Moggi, Mogushi, Mohan, Mohapatra,
  Montani, Moore, Moraru, Moreno, Morriss, Mours, Mow-Lowry, Mueller, Muir,
  Mukherjee, Mukherjee, Mukherjee, Mukund, Mullavey, Munch, Mu{\~{n}}iz,
  Muratore, Murray, Napier, Nardecchia, Naticchioni, Nayak, Neilson, Nelemans,
  Nelson, Nery, Neunzert, Nevin, Newport, Newton, Ng, Nguyen, Nguyen, Nichols,
  Nielsen, Nissanke, Nitz, Noack, Nocera, Nolting, North, Nuttall, Oberling,
  O'Dea, Ogin, Oh, Oh, Ohme, Okada, Oliver, Oppermann, Oram, O'Reilly,
  Ormiston, Ortega, O'Shaughnessy, Ossokine, Ottaway, Overmier, Owen, Pace,
  Page, Page, Pai, Pai, Palamos, Palashov, Palomba, Pal-Singh, Pan, Pan, Pang,
  Pang, Pankow, Pannarale, Pant, Paoletti, Paoli, Papa, Parida, Parker,
  Pascucci, Pasqualetti, Passaquieti, Passuello, Patil, Patricelli, Pearlstone,
  Pedraza, Pedurand, Pekowsky, Pele, Penn, Perez, Perreca, Perri, Pfeiffer,
  Phelps, Piccinni, Pichot, Piergiovanni, Pierro, Pillant, Pinard, Pinto,
  Pirello, Pitkin, Poe, Poggiani, Popolizio, Porter, Post, Powell, Prasad,
  Pratt, Pratten, Predoi, Prestegard, Price, Prijatelj, Principe, Privitera,
  Prodi, Prokhorov, Puncken, Punturo, Puppo, Pürrer, Qi, Quetschke, Quintero,
  Quitzow-James, Raab, Rabeling, Radkins, Raffai, Raja, Rajan, Rajbhandari,
  Rakhmanov, Ramirez, Ramos-Buades, Rapagnani, Raymond, Razzano, Read,
  Regimbau, Rei, Reid, Reitze, Ren, Reyes, Ricci, Ricker, Rieger, Riles, Rizzo,
  Robertson, Robie, Robinet, Rocchi, Rolland, Rollins, Roma, Romano, Romel,
  Romie, Rosi{\'{n}}ska, Ross, Rowan, Rüdiger, Ruggi, Rutins, Ryan, Sachdev,
  Sadecki, Sadeghian, Sakellariadou, Salconi, Saleem, Salemi, Samajdar, Sammut,
  Sampson, Sanchez, Sanchez, Sanchis-Gual, Sandberg, Sanders, Sassolas,
  Sathyaprakash, Saulson, Sauter, Savage, Sawadsky, Schale, Scheel, Scheuer,
  Schmidt, Schmidt, Schnabel, Schofield, Schönbeck, Schreiber, Schuette,
  Schulte, Schutz, Schwalbe, Scott, Scott, Seidel, Sellers, Sengupta, Sentenac,
  Sequino, Sergeev, Shaddock, Shaffer, Shah, Shahriar, Shaner, Shao, Shapiro,
  Shawhan, Sheperd, Shoemaker, Shoemaker, Siellez, Siemens, Sieniawska, Sigg,
  Silva, Singer, Singh, Singhal, Sintes, Slagmolen, Smith, Smith, Smith,
  Somala, Son, Sonnenberg, Sorazu, Sorrentino, Souradeep, Spencer, Srivastava,
  Staats, Staley, Steinke, Steinlechner, Steinlechner, Steinmeyer, Stevenson,
  Stone, Stops, Strain, Stratta, Strigin, Strunk, Sturani, Stuver,
  Summerscales, Sun, Sunil, Suresh, Sutton, Swinkels, Szczepa{\'{n}}czyk,
  Tacca, Tait, Talbot, Talukder, Tanner, T{\'{a}}pai, Taracchini, Tasson,
  Taylor, Taylor, Tewari, Theeg, Thies, Thomas, Thomas, Thomas, Thorne, Thorne,
  Thrane, Tiwari, Tiwari, Tokmakov, Toland, Tonelli, Tornasi,
  Torres-Forn{\'{e}}, Torrie, Töyrä, Travasso, Traylor, Trinastic, Tringali,
  Trozzo, Tsang, Tse, Tso, Tsukada, Tsuna, Tuyenbayev, Ueno, Ugolini,
  Unnikrishnan, Urban, Usman, Vahlbruch, Vajente, Valdes, van Bakel, van
  Beuzekom, van~den Brand, Broeck, Vander-Hyde, van~der Schaaf, van Heijningen,
  van Veggel, Vardaro, Varma, Vass, Vas{\'{u}}th, Vecchio, Vedovato, Veitch,
  Veitch, Venkateswara, Venugopalan, Verkindt, Vetrano, Vicer{\'{e}}, Viets,
  Vinciguerra, Vine, Vinet, Vitale, Vo, Vocca, Vorvick, Vyatchanin, Wade, Wade,
  Wade, Walet, Walker, Wallace, Walsh, Wang, Wang, Wang, Wang, Wang, Ward,
  Warner, Was, Watchi, Weaver, Wei, Weinert, Weinstein, Weiss, Wen, Wessel,
  Wessels, Westerweck, Westphal, Wette, Whelan, Whitcomb, Whiting, Whittle,
  Wilken, Williams, Williams, Williamson, Willis, Willke, Wimmer, Winkler,
  Wipf, Wittel, Woan, Woehler, Wofford, Wong, Worden, Wright, Wu, Wysocki,
  Xiao, Yamamoto, Yancey, Yang, Yap, Yazback, Yu, Yu, Yvert, Zadro{\.{z}}ny,
  Zanolin, Zelenova, Zendri, Zevin, Zhang, Zhang, Zhang, Zhang, Zhao, Zhou,
  Zhou, Zhu, Zhu, Zimmerman, Zucker, Zweizig, Wilson-Hodge, Bissaldi,
  Blackburn, Briggs, Burns, Cleveland, Connaughton, Gibby, Giles, Goldstein,
  Hamburg, Jenke, Hui, Kippen, Kocevski, McBreen, Meegan, Paciesas, Poolakkil,
  Preece, Racusin, Roberts, Stanbro, Veres, von Kienlin, Savchenko, Ferrigno,
  Kuulkers, Bazzano, Bozzo, Brandt, Chenevez, Courvoisier, Diehl, Domingo,
  Hanlon, Jourdain, Laurent, Lebrun, Lutovinov, Martin-Carrillo, Mereghetti,
  Natalucci, Rodi, Roques, Sunyaev, Ubertini, Aartsen, Ackermann, Adams,
  Aguilar, Ahlers, Ahrens, Samarai, Altmann, Andeen, Anderson, Ansseau, Anton,
  Argüelles, Auffenberg, Axani, Bagherpour, Bai, Barron, Barwick, Baum, Bay,
  Beatty, Tjus, Bernardini, Besson, Binder, Bindig, Blaufuss, Blot, Bohm,
  Börner, Bos, Bose, Böser, Botner, Bourbeau, Bourbeau, Bradascio, Braun,
  Brayeur, Brenzke, Bretz, Bron, Brostean-Kaiser, Burgman, Carver, Casey,
  Casier, Cheung, Chirkin, Christov, Clark, Classen, Coenders, Collin, Conrad,
  Cowen, Cross, Day, de~Andr{\'{e}}, Clercq, DeLaunay, Dembinski, Ridder,
  Desiati, de~Vries, de~Wasseige, de~With, DeYoung, D{\'{\i}}az-V{\'{e}}lez,
  di~Lorenzo, Dujmovic, Dumm, Dunkman, Dvorak, Eberhardt, Ehrhardt, Eichmann,
  Eller, Evenson, Fahey, Fazely, Felde, Filimonov, Finley, Flis, Franckowiak,
  Friedman, Fuchs, Gaisser, Gallagher, Gerhardt, Ghorbani, Giang, Glauch,
  Glüsenkamp, Goldschmidt, Gonzalez, Grant, Griffith, Haack, Hallgren, Halzen,
  Hanson, Hebecker, Heereman, Helbing, Hellauer, Hickford, Hignight, Hill,
  Hoffman, Hoffmann, Hokanson-Fasig, Hoshina, Huang, Huber, Hultqvist,
  Hünnefeld, In, Ishihara, Jacobi, Japaridze, Jeong, Jero, Jones, Kalaczynski,
  Kang, Kappes, Karg, Karle, Kauer, Keivani, Kelley, Kheirandish, Kim, Kim,
  Kintscher, Kiryluk, Kittler, Klein, Kohnen, Koirala, Kolanoski, Köpke,
  Kopper, Kopper, Koschinsky, Koskinen, Kowalski, Krings, Kroll, Krückl,
  Kunnen, Kunwar, Kurahashi, Kuwabara, Kyriacou, Labare, Lanfranchi, Larson,
  Lauber, Lesiak-Bzdak, Leuermann, Liu, Lu, Lünemann, Luszczak, Madsen, Maggi,
  Mahn, Mancina, Maruyama, Mase, Maunu, McNally, Meagher, Medici, Meier, Menne,
  Merino, Meures, Miarecki, Micallef, Moment{\'{e}}, Montaruli, Moore, Moulai,
  Nahnhauer, Nakarmi, Naumann, Neer, Niederhausen, Nowicki, Nygren, Pollmann,
  Olivas, O'Murchadha, Palczewski, Pandya, Pankova, Peiffer, Pepper, de~los
  Heros, Pieloth, Pinat, Price, Przybylski, Raab, Rädel, Rameez, Rawlins, Rea,
  Reimann, Relethford, Relich, Resconi, Rhode, Richman, Robertson, Rongen,
  Rott, Ruhe, Ryckbosch, Rysewyk, Sälzer, Herrera, Sandrock, Sandroos,
  Santander, Sarkar, Sarkar, Satalecka, Schlunder, Schmidt, Schneider,
  Schoenen, Schöneberg, Schumacher, Seckel, Seunarine, Soedingrekso, Soldin,
  Song, Spiczak, Spiering, Stachurska, Stamatikos, Stanev, Stasik, Stettner,
  Steuer, Stezelberger, Stokstad, Stössl, Strotjohann, Stuttard, Sullivan,
  Sutherland, Taboada, Tatar, Tenholt, Ter-Antonyan, Terliuk,
  Te{\v{s}}i{\'{c}}, Tilav, Toale, Tobin, Toscano, Tosi, Tselengidou, Tung,
  Turcati, Turley, Ty, Unger, Usner, Vandenbroucke, Driessche, van Eijndhoven,
  Vanheule, van Santen, Vehring, Vogel, Vraeghe, Walck, Wallace, Wallraff,
  Wandler, Wandkowsky, Waza, Weaver, Weiss, Wendt, Werthebach, Whelan, Wiebe,
  Wiebusch, Wille, Williams, Wills, Wolf, Wood, Woolsey, Woschnagg, Xu, Xu, Xu,
  Yanez, Yodh, Yoshida, Yuan, Zoll, Balasubramanian, Mate, Bhalerao,
  Bhattacharya, Vibhute, Dewangan, Rao, Vadawale, Svinkin, Hurley, Aptekar,
  Frederiks, Golenetskii, Kozlova, Lysenko, Oleynik, Tsvetkova, Ulanov, Cline,
  Li, Xiong, Zhang, Lu, Song, Cao, Chang, Chen, Chen, Chen, Chen, Chen, Chen,
  Cui, Cui, Deng, Dong, Du, Fu, Gao, Gao, Gao, Ge, Gu, Guan, Guo, Han, Hu,
  Huang, Huo, Jia, Jiang, Jiang, Jin, Jin, Li, Li, Li, Li, Li, Li, Li, Li, Li,
  Li, Li, Liang, Liao, Liu, Liu, Liu, Liu, Liu, Liu, Liu, Lu, Lu, Luo, Ma,
  Meng, Nang, Nie, Ou, Qu, Sai, Sun, Tan, Tao, Tao, Tuo, Wang, Wang, Wang,
  Wang, Wang, Wen, Wu, Wu, Xiao, Xu, Xu, Yan, Yang, Yang, Yang, Zhang, Zhang,
  Zhang, Zhang, Zhang, Zhang, Zhang, Zhang, Zhang, Zhang, Zhang, Zhang, Zhang,
  Zhang, Zhang, Zhang, Zhang, Zhang, Zhao, Zhao, Zhao, Zheng, Zhu, Zhu, Zou,
  Albert, Andr{\'{e}}, Anghinolfi, Ardid, Aubert, Aublin, Avgitas, Baret,
  Barrios-Mart{\'{\i}}, Basa, Belhorma, Bertin, Biagi, Bormuth, Bourret,
  Bouwhuis, Br{\^{a}}nza{\c{s}}, Bruijn, Brunner, Busto, Capone, Caramete,
  Carr, Celli, Moursli, Chiarusi, Circella, Coelho, Coleiro, Coniglione,
  Costantini, Coyle, Creusot, D{\'{\i}}az, Deschamps, Bonis, Distefano, Palma,
  Domi, Donzaud, Dornic, Drouhin, Eberl, Bojaddaini, Khayati, Elsässer,
  Enzenhöfer, Ettahiri, Fassi, Felis, Fusco, Gay, Giordano, Glotin,
  Gr{\'{e}}goire, Ruiz, Graf, Hallmann, van Haren, Heijboer, Hello,
  Hern{\'{a}}ndez-Rey, Hössl, Hofestädt, Hugon, Illuminati, James, de~Jong,
  Jongen, Kadler, Kalekin, Katz, Kiessling, Kouchner, Kreter, Kreykenbohm,
  Kulikovskiy, Lachaud, Lahmann, Lef{\`{e}}vre, Leonora, Lotze, Loucatos,
  Marcelin, Margiotta, Marinelli, Mart{\'{\i}}nez-Mora, Mele, Melis, Michael,
  Migliozzi, Moussa, Navas, Nezri, Organokov, P{\u{a}}v{\u{a}}la{\c{s}},
  Pellegrino, Perrina, Piattelli, Popa, Pradier, Quinn, Racca, Riccobene,
  S{\'{a}}nchez-Losa, Salda{\~{n}}a, Salvadori, Samtleben, Sanguineti,
  Sapienza, Sieger, Spurio, Stolarczyk, Taiuti, Tayalati, Trovato, Turpin,
  Tönnis, Vallage, Elewyck, Versari, Vivolo, Vizzoca, Wilms, Zornoza,
  Z{\'{u}}{\~{n}}iga, Beardmore, Breeveld, Burrows, Cenko, Cusumano,
  D'A{\`{\i}}, de~Pasquale, Emery, Evans, Giommi, Gronwall, Kennea, Krimm,
  Kuin, Lien, Marshall, Melandri, Nousek, Oates, Osborne, Pagani, Page, Palmer,
  Perri, Siegel, Sbarufatti, Tagliaferri, Tohuvavohu, Tavani, Verrecchia,
  Bulgarelli, Evangelista, Pacciani, Feroci, Pittori, Giuliani, Monte,
  Donnarumma, Argan, Trois, Ursi, Cardillo, Piano, Longo, Lucarelli,
  Munar-Adrover, Fuschino, Labanti, Marisaldi, Minervini, Fioretti,
  Parmiggiani, Gianotti, Trifoglio, Persio, Antonelli, Barbiellini, Caraveo,
  Cattaneo, Costa, Colafrancesco, D'Amico, Ferrari, Morselli, Paoletti,
  Picozza, Pilia, Rappoldi, Soffitta, Vercellone, Foley, Coulter, Kilpatrick,
  Drout, Piro, Shappee, Siebert, Simon, Ulloa, Kasen, Madore, Murguia-Berthier,
  Pan, Prochaska, Ramirez-Ruiz, Rest, Rojas-Bravo, Berger, Soares-Santos,
  Annis, Alexander, Allam, Balbinot, Blanchard, Brout, Butler, Chornock, Cook,
  Cowperthwaite, Diehl, Drlica-Wagner, Drout, Durret, Eftekhari, Finley, Fong,
  Frieman, Fryer, Garc{\'{\i}}a-Bellido, Gruendl, Hartley, Herner, Kessler,
  Lin, Lopes, Louren{\c{c}}o, Margutti, Marshall, Matheson, Medina, Metzger,
  Mu{\~{n}}oz, Muir, Nicholl, Nugent, Palmese, Paz-Chinch{\'{o}}n, Quataert,
  Sako, Sauseda, Schlegel, Scolnic, Secco, Smith, Sobreira, Villar, Vivas,
  Wester, Williams, Yanny, Zenteno, Zhang, Abbott, Banerji, Bechtol,
  Benoit-L{\'{e}}vy, Bertin, Brooks, Buckley-Geer, Burke, Capozzi, Rosell,
  Kind, Castander, Crocce, Cunha, D'Andrea, da~Costa, Davis, DePoy, Desai,
  Dietrich, Eifler, Fernandez, Flaugher, Fosalba, Gaztanaga, Gerdes,
  Giannantonio, Goldstein, Gruen, Gschwend, Gutierrez, Honscheid, James,
  Jeltema, Johnson, Johnson, Kent, Krause, Kron, Kuehn, Lahav, Lima, Maia,
  March, Martini, McMahon, Menanteau, Miller, Miquel, Mohr, Nichol, Ogando,
  Plazas, Romer, Roodman, Rykoff, Sanchez, Scarpine, Schindler, Schubnell,
  Sevilla-Noarbe, Sheldon, Smith, Smith, Stebbins, Suchyta, Swanson, Tarle,
  Thomas, Troxel, Tucker, Vikram, Walker, Wechsler, Weller, Carlin, Gill, Li,
  Marriner, Neilsen, Haislip, Kouprianov, Reichart, Sand, Tartaglia, Valenti,
  Yang, Benetti, Brocato, Campana, Cappellaro, Covino, D'Avanzo, D'Elia,
  Getman, Ghirlanda, Ghisellini, Limatola, Nicastro, Palazzi, Pian,
  Piranomonte, Possenti, Rossi, Salafia, Tomasella, Amati, Antonelli,
  Bernardini, Bufano, Capaccioli, Casella, Dadina, Cesare, Paola, Giuffrida,
  Giunta, Israel, Lisi, Maiorano, Mapelli, Masetti, Pescalli, Pulone,
  Salvaterra, Schipani, Spera, Stamerra, Stella, Testa, Turatto, Vergani,
  Aresu, Bachetti, Buffa, Burgay, Buttu, Caria, Carretti, Casasola, Castangia,
  Carboni, Casu, Concu, Corongiu, Deiana, Egron, Fara, Gaudiomonte, Gusai,
  Ladu, Loru, Leurini, Marongiu, Melis, Melis, Migoni, Milia, Navarrini,
  Orlati, Ortu, Palmas, Pellizzoni, Perrodin, Pisanu, Poppi, Righini, Saba,
  Serra, Serrau, Stagni, Surcis, Vacca, Vargiu, Hunt, Jin, Klose, Kouveliotou,
  Mazzali, M{\o}ller, Nava, Piran, Selsing, Vergani, Wiersema, Toma, Higgins,
  Mundell, di~Serego~Alighieri, G{\'{o}}tz, Gao, Gomboc, Kaper, Kobayashi,
  Kopac, Mao, Starling, Steele, van~der Horst, Acero, Atwood, Baldini,
  Barbiellini, Bastieri, Berenji, Bellazzini, Bissaldi, Blandford, Bloom,
  Bonino, Bottacini, Bregeon, Buehler, Buson, Cameron, Caputo, Caraveo,
  Cavazzuti, Chekhtman, Cheung, Chiang, Ciprini, Cohen-Tanugi, Cominsky,
  Costantin, Cuoco, D{\textquotesingle}Ammando, de~Palma, Digel, Lalla, Mauro,
  Venere, Dubois, Fegan, Focke, Franckowiak, Fukazawa, Funk, Fusco, Gargano,
  Gasparrini, Giglietto, Giordano, Giroletti, Glanzman, Green, Grondin,
  Guillemot, Guiriec, Harding, Horan, J{\'{o}}hannesson, Kamae, Kensei, Kuss,
  Mura, Latronico, Lemoine-Goumard, Longo, Loparco, Lovellette, Lubrano,
  Magill, Maldera, Manfreda, Mazziotta, McEnery, Meyer, Michelson, Mirabal,
  Monzani, Moretti, Morselli, Moskalenko, Negro, Nuss, Ojha, Omodei, Orienti,
  Orlando, Palatiello, Paliya, Paneque, Pesce-Rollins, Piron, Porter, Principe,
  Rain{\`{o}}, Rando, Razzano, Razzaque, Reimer, Reimer, Reposeur, Rochester,
  Parkinson, Sgr{\`{o}}, Siskind, Spada, Spandre, Suson, Takahashi, Tanaka,
  Thayer, Thayer, Thompson, Tibaldo, Torres, Torresi, Troja, Venters, Vianello,
  Zaharijas, Allison, Bannister, Dobie, Kaplan, Lenc, Lynch, Murphy, Sadler,
  Hotan, James, Oslowski, Raja, Shannon, Whiting, Arcavi, Howell, McCully,
  Hosseinzadeh, Hiramatsu, Poznanski, Barnes, Zaltzman, Vasylyev, Maoz, Cooke,
  Bailes, Wolf, Deller, Lidman, Wang, Gendre, Andreoni, Ackley, Pritchard,
  Bessell, Chang, Möller, Onken, Scalzo, Ridden-Harper, Sharp, Tucker,
  Farrell, Elmer, Johnston, Krishnan, Keane, Green, Jameson, Hu, Ma, Sun, Wu,
  Wang, Shang, Hu, Ashley, Yuan, Li, Tao, Zhu, Zhang, Suntzeff, Zhou, Yang,
  Orange, Morris, Cucchiara, Giblin, Klotz, Staff, Thierry, Schmidt, Tanvir,
  Levan, Cano, de~Ugarte-Postigo, Gonz{\'{a}}lez-Fern{\'{a}}ndez, Greiner,
  Hjorth, Irwin, Krühler, Mandel, Milvang-Jensen, O{\textquotesingle}Brien,
  Rol, Rosetti, Rosswog, Rowlinson, Steeghs, Thöne, Ulaczyk, Watson, Bruun,
  Cutter, Jaimes, Fujii, Fruchter, Gompertz, Jakobsson, Hodosan,
  J{\`{e}}rgensen, Kangas, Kann, Rabus, Schr{\o}der, Stanway, Wijers, Lipunov,
  Gorbovskoy, Kornilov, Tyurina, Balanutsa, Kuznetsov, Vlasenko, Podesta,
  Lopez, Podesta, Levato, Saffe, Mallamaci, Budnev, Gress, Kuvshinov, Gorbunov,
  Vladimirov, Zimnukhov, Gabovich, Yurkov, Sergienko, Rebolo, Serra-Ricart,
  Tlatov, Ishmuhametova, Abe, Aoki, Aoki, Asakura, Baar, Barway, Bond, Doi,
  Finet, Fujiyoshi, Furusawa, Honda, Itoh, Kanda, Kawabata, Kawabata, Kim,
  Koshida, Kuroda, Lee, Liu, Matsubayashi, Miyazaki, Morihana, Morokuma,
  Motohara, Murata, Nagai, Nagashima, Nagayama, Nakaoka, Nakata, Ohsawa,
  Ohshima, Ohta, Okita, Saito, Saito, Sako, Sekiguchi, Sumi, Tajitsu,
  Takahashi, Takayama, Tamura, Tanaka, Tanaka, Terai, Tominaga, Tristram,
  Uemura, Utsumi, Yamaguchi, Yasuda, Yoshida, Zenko, Adams, Anupama, Bally,
  Barway, Bellm, Blagorodnova, Cannella, Chandra, Chatterjee, Clarke, Cobb,
  Cook, Copperwheat, De, Emery, Feindt, Foster, Fox, Frail, Fremling,
  Frohmaier, Garcia, Ghosh, Giacintucci, Goobar, Gottlieb, Grefenstette,
  Hallinan, Harrison, Heida, Helou, Ho, Horesh, Hotokezaka, Ip, Itoh, Jacobs,
  Jencson, Kasen, Kasliwal, Kassim, Kim, Kiran, Kuin, Kulkarni, Kupfer, Lau,
  Madsen, Mazzali, Miller, Miyasaka, Mooley, Myers, Nakar, Ngeow, Nugent, Ofek,
  Palliyaguru, Pavana, Perley, Peters, Pike, Piran, Qi, Quimby, Rana, Rosswog,
  Rusu, Sadler, Sistine, Sollerman, Xu, Yan, Yatsu, Yu, Zhang, Zhao, Chambers,
  Huber, Schultz, Bulger, Flewelling, Magnier, Lowe, Wainscoat, Waters,
  Willman, Ebisawa, Hanyu, Harita, Hashimoto, Hidaka, Hori, Ishikawa, Isobe,
  Iwakiri, Kawai, Kawai, Kawamuro, Kawase, Kitaoka, Makishima, Matsuoka,
  Mihara, Morita, Morita, Nakahira, Nakajima, Nakamura, Negoro, Oda, Sakamaki,
  Sasaki, Serino, Shidatsu, Shimomukai, Sugawara, Sugita, Sugizaki, Tachibana,
  Takao, Tanimoto, Tomida, Tsuboi, Tsunemi, Ueda, Ueno, Yamada, Yamaoka,
  Yamauchi, Yatabe, Yoneyama, Yoshii, Coward, Crisp, Macpherson, Andreoni,
  Laugier, Noysena, Klotz, Gendre, Thierry, Turpin, Im, Choi, Kim, Yoon, Lim,
  Lee, Lee, Kim, Ko, Joe, Kwon, Kim, Lim, Choi, Fynbo, Malesani, Xu, Smartt,
  Jerkstrand, Kankare, Sim, Fraser, Inserra, Maguire, Leloudas, Magee,
  Shingles, Smith, Young, Kotak, Gal-Yam, Lyman, Homan, Agliozzo, Anderson,
  Angus, Ashall, Barbarino, Bauer, Berton, Botticella, Bulla, Cannizzaro,
  Cartier, Cikota, Clark, Cia, Valle, Dennefeld, Dessart, Dimitriadis,
  Elias-Rosa, Firth, Flörs, Frohmaier, Galbany, Gonz{\'{a}}lez-Gait{\'{a}}n,
  Gromadzki, Guti{\'{e}}rrez, Hamanowicz, Harmanen, Heintz, Hernandez, Hodgkin,
  Hook, Izzo, James, Jonker, Kerzendorf, Kostrzewa-Rutkowska, Kromer,
  Kuncarayakti, Lawrence, Manulis, Mattila, McBrien, Müller, Nordin,
  O{\textquotesingle}Neill, Onori, Palmerio, Pastorello, Patat, Pignata,
  Podsiadlowski, Razza, Reynolds, Roy, Ruiter, Rybicki, Salmon, Pumo, Prentice,
  Seitenzahl, Smith, Sollerman, Sullivan, Szegedi, Taddia, Taubenberger,
  Terreran, Soelen, Vos, Walton, Wright, Wyrzykowski, Yaron, Chen, Krühler,
  Schady, Wiseman, Greiner, Rau, Schweyer, Klose, Guelbenzu, Palliyaguru,
  Shara, Williams, Vaisanen, Potter, Colmenero, Crawford, Buckley, Mao,
  D{\'{\i}}az, Macri, Lambas, de~Oliveira, Castell{\'{o}}n, Ribeiro,
  S{\'{a}}nchez, Schoenell, Abramo, Akras, Alcaniz, Artola, Beroiz, Bonoli,
  Cabral, Camuccio, Chavushyan, Coelho, Colazo, Costa-Duarte, Larenas, Romero,
  Dultzin, Fern{\'{a}}ndez, Garc{\'{\i}}a, Girardini, Gon{\c{c}}alves,
  Gon{\c{c}}alves, Gurovich, Jim{\'{e}}nez-Teja, Kanaan, Lares, de~Oliveira,
  L{\'{o}}pez-Cruz, Melia, Molino, Padilla, Pe{\~{n}}uela, Placco,
  Qui{\~{n}}ones, Rivera, Renzi, Riguccini, R{\'{\i}}os-L{\'{o}}pez, Rodriguez,
  Sampedro, Schneiter, Sodr{\'{e}}, Starck, Torres-Flores, Tornatore,
  Zadro{\.{z}}ny, Castillo, Castro-Tirado, Tello, Hu, Zhang, Cunniffe,
  Castell{\'{o}}n, Hiriart, Caballero-Garc{\'{\i}}a, Jel{\'{\i}}nek,
  Kub{\'{a}}nek, del Pulgar, Park, Jeong, Cer{\'{o}}n, Pandey, Yock, Querel,
  Fan, Wang, Beardsley, Brown, Crosse, Emrich, Franzen, Gaensler, Horsley,
  Johnston-Hollitt, Kenney, Morales, Pallot, Sokolowski, Steele, Tingay, Trott,
  Walker, Wayth, Williams, Wu, Yoshida, Sakamoto, Kawakubo, Yamaoka, Takahashi,
  Asaoka, Ozawa, Torii, Shimizu, Tamura, Ishizaki, Cherry, Ricciarini,
  Penacchioni, Marrocchesi, Pozanenko, Volnova, Mazaeva, Minaev, Krugov,
  Kusakin, Reva, Moskvitin, Rumyantsev, Inasaridze, Klunko, Tungalag, Schmalz,
  Burhonov, Abdalla, Abramowski, Aharonian, Benkhali, Angüner, Arakawa,
  Arrieta, Aubert, Backes, Balzer, Barnard, Becherini, Tjus, Berge, Bernhard,
  Bernlöhr, Blackwell, Böttcher, Boisson, Bolmont, Bonnefoy, Bordas, Bregeon,
  Brun, Brun, Bryan, Büchele, Bulik, Capasso, Caroff, Carosi, Casanova,
  Cerruti, Chakraborty, Chaves, Chen, Chevalier, Colafrancesco, Condon, Conrad,
  Davids, Decock, Deil, Devin, deWilt, Dirson, Djannati-Ataï, Donath, Drury,
  Dutson, Dyks, Edwards, Egberts, Emery, Ernenwein, Eschbach, Farnier, Fegan,
  Fernandes, Fiasson, Fontaine, Funk, Füssling, Gabici, Gallant, Garrigoux,
  Gat{\'{e}}, Giavitto, Giebels, Glawion, Glicenstein, Gottschall, Grondin,
  Hahn, Haupt, Hawkes, Heinzelmann, Henri, Hermann, Hinton, Hofmann, Hoischen,
  Holch, Holler, Horns, Ivascenko, Iwasaki, Jacholkowska, Jamrozy, Jankowsky,
  Jankowsky, Jingo, Jouvin, Jung-Richardt, Kastendieck, Katarzy{\'{n}}ski,
  Katsuragawa, Kerszberg, Khangulyan, Kh{\'{e}}lifi, King, Klepser, Klochkov,
  Klu{\'{z}}niak, Komin, Kosack, Krakau, Kraus, Krüger, Laffon, Lamanna, Lau,
  Lees, Lefaucheur, Lemi{\`{e}}re, Lemoine-Goumard, Lenain, Leser, Lohse,
  Lorentz, Liu, Lypova, Malyshev, Marandon, Marcowith, Mariaud, Marx, Maurin,
  Maxted, Mayer, Meintjes, Meyer, Mitchell, Moderski, Mohamed, Mohrmann,
  Mor{\aa}, Moulin, Murach, Nakashima, de~Naurois, Ndiyavala, Niederwanger,
  Niemiec, Oakes, O{\textquotesingle}Brien, Odaka, Ohm, Ostrowski, Oya,
  Padovani, Panter, Parsons, Pekeur, Pelletier, Perennes, Petrucci, Peyaud,
  Piel, Pita, Poireau, Poon, Prokhorov, Prokoph, Pühlhofer, Punch,
  Quirrenbach, Raab, Rauth, Reimer, Reimer, Renaud, de~los Reyes, Rieger,
  Rinchiuso, Romoli, Rowell, Rudak, Rulten, Sahakian, Saito, Sanchez,
  Santangelo, Sasaki, Schlickeiser, Schüssler, Schulz, Schwanke, Schwemmer,
  Seglar-Arroyo, Settimo, Seyffert, Shafi, Shilon, Shiningayamwe, Simoni, Sol,
  Spanier, Spir-Jacob, Stawarz, Steenkamp, Stegmann, Steppa, Sushch, Takahashi,
  Tavernet, Tavernier, Taylor, Terrier, Tibaldo, Tiziani, Tluczykont, Trichard,
  Tsirou, Tsuji, Tuffs, Uchiyama, van~der Walt, van Eldik, van Rensburg, van
  Soelen, Vasileiadis, Veh, Venter, Viana, Vincent, Vink, Voisin, Völk,
  Vuillaume, Wadiasingh, Wagner, Wagner, Wagner, White, Wierzcholska, Willmann,
  Wörnlein, Wouters, Yang, Zaborov, Zacharias, Zanin, Zdziarski, Zech, Zefi,
  Ziegler, Zorn, {\.{Z}}ywucka, Fender, Broderick, Rowlinson, Wijers, Stewart,
  ter Veen, Shulevski, Kavic, Simonetti, League, Tsai, Obenberger, Nathaniel,
  Taylor, Dowell, Liebling, Estes, Lippert, Sharma, Vincent, Farella,
  Abeysekara, Albert, Alfaro, Alvarez, Arceo, Arteaga-Vel{\'{a}}zquez, Rojas,
  Solares, Barber, Gonzalez, Becerril, Belmont-Moreno, BenZvi, Berley, Bernal,
  Braun, Brisbois, Caballero-Mora, Capistr{\'{a}}n, Carrami{\~{n}}ana,
  Casanova, Castillo, Cotti, Cotzomi, de~Le{\'{o}}n, Le{\'{o}}n, la~Fuente,
  Hernandez, Dichiara, Dingus, DuVernois, D{\'{\i}}az-V{\'{e}}lez, Ellsworth,
  Engel, Enr{\'{\i}}quez-Rivera, Fiorino, Fleischhack, Fraija,
  Garc{\'{\i}}a-Gonz{\'{a}}lez, Garfias, Gerhardt, Mu{\~{n}}oz, Gonz{\'{a}}lez,
  Goodman, Hampel-Arias, Harding, Hernandez, Hernandez-Almada, Hona,
  Hüntemeyer, Iriarte, Jardin-Blicq, Joshi, Kaufmann, Kieda, Lara, Lauer,
  Lennarz, Vargas, Linnemann, Longinotti, Raya, Luna-Garc{\'{\i}}a,
  L{\'{o}}pez-Coto, Malone, Marinelli, Martinez, Martinez-Castellanos,
  Mart{\'{\i}}nez-Castro, Mart{\'{\i}}nez-Huerta, Matthews, Miranda-Romagnoli,
  Moreno, Mostaf{\'{a}}, Nellen, Newbold, Nisa, Noriega-Papaqui, Pelayo, Pretz,
  P{\'{e}}rez-P{\'{e}}rez, Ren, Rho, Rivi{\`{e}}re, Rosa-Gonz{\'{a}}lez,
  Rosenberg, Ruiz-Velasco, Salazar, Greus, Sandoval, Schneider, Schoorlemmer,
  Sinnis, Smith, Springer, Surajbali, Tibolla, Tollefson, Torres, Ukwatta,
  Weisgarber, Westerhoff, Wisher, Wood, Yapici, Yodh, Younk, Zhou,
  {\'{A}}lvarez, Aab, Abreu, Aglietta, Albuquerque, Albury, Allekotte, Almela,
  Castillo, Alvarez-Mu{\~{n}}iz, Anastasi, Anchordoqui, Andrada, Andringa,
  Aramo, Arsene, Asorey, Assis, Avila, Badescu, Balaceanu, Barbato, Luz,
  Becker, Bellido, Berat, Bertaina, Bertou, Biermann, Biteau, Blaess, Blanco,
  Blazek, Bleve, Boh{\'{a}}{\v{c}}ov{\'{a}}, Bonifazi, Borodai, Botti, Brack,
  Brancus, Bretz, Bridgeman, Briechle, Buchholz, Bueno, Buitink, Buscemi,
  Caballero-Mora, Caccianiga, Cancio, Canfora, Caruso, Castellina, Catalani,
  Cataldi, Cazon, Chavez, Chinellato, Chudoba, Clay, Cerutti, Colalillo,
  Coleman, Collica, Coluccia, Concei{\c{c}}{\~{a}}o, Consolati, Contreras,
  Cooper, Coutu, Covault, Cronin, D{\textquotesingle}Amico, Daniel, Dasso,
  Daumiller, Dawson, Day, de~Almeida, de~Jong, Mauro, de~Mello~Neto, Mitri,
  de~Oliveira, de~Souza, Debatin, Deligny, Castro, Diogo, Dobrigkeit,
  D{\textquotesingle}Olivo, Dorosti, Anjos, Dova, Dundovic, Ebr, Engel,
  Erdmann, Erfani, Escobar, Espadanal, Etchegoyen, Falcke, Farmer, Farrar,
  Fauth, Fazzini, Feldbusch, Fenu, Fick, Figueira, Filip{\v{c}}i{\v{c}},
  Freire, Fujii, Fuster, Gaïor, Garc{\'{\i}}a, Gat{\'{e}}, Gemmeke,
  Gherghel-Lascu, Ghia, Giaccari, Giammarchi, Giller, G{\l}as, Glaser, Golup,
  Berisso, Vitale, Gonz{\'{a}}lez, Gorgi, Gottowik, Grillo, Grubb, Guarino,
  Guedes, Halliday, Hampel, Hansen, Harari, Harrison, Harvey, Haungs, Hebbeker,
  Heck, Heimann, Herve, Hill, Hojvat, Holt, Homola, Hörandel, Horvath,
  Hrabovsk{\'{y}}, Huege, Hulsman, Insolia, Isar, Jandt, Johnsen, Josebachuili,
  Jurysek, Kääpä, Kampert, Keilhauer, Kemmerich, Kemp, Kieckhafer, Klages,
  Kleifges, Kleinfeller, Krause, Krohm, Kuempel, Mezek, Kunka, Awad, Lago,
  LaHurd, Lang, Lauscher, Legumina, de~Oliveira, Letessier-Selvon, Lhenry-Yvon,
  Link, Presti, Lopes, L{\'{o}}pez, Casado, Lorek, Luce, Lucero, Malacari,
  Mallamaci, Mandat, Mantsch, Mariazzi, Maris, Marsella, Martello, Martinez,
  Bravo, Meza, Mathes, Mathys, Matthews, Matthiae, Mayotte, Mazur, Medina,
  Medina-Tanco, Melo, Menshikov, Merenda, Michal, Micheletti, Middendorf,
  Miramonti, Mitrica, Mockler, Mollerach, Montanet, Morello, Morlino, Müller,
  Müller, Muller, Müller, Mussa, Naranjo, Nguyen, Niculescu-Oglinzanu,
  Niechciol, Niemietz, Niggemann, Nitz, Nosek, Novotny, No{\v{z}}ka,
  N{\'{u}}{\~{n}}ez, Oikonomou, Olinto, Palatka, Pallotta, Papenbreer, Parente,
  Parra, Paul, Pech, Pedreira, Pȩkala, Pe{\~{n}}a-Rodriguez, Pereira, Perlin,
  Perrone, Peters, Petrera, Phuntsok, Pierog, Pimenta, Pirronello, Platino,
  Plum, Poh, Porowski, Prado, Privitera, Prouza, Quel, Querchfeld, Quinn,
  Ramos-Pollan, Rautenberg, Ravignani, Ridky, Riehn, Risse, Ristori, Rizi,
  de~Carvalho, Fernandez, Rojo, Roncoroni, Roth, Roulet, Rovero, Ruehl, Saffi,
  Saftoiu, Salamida, Salazar, Saleh, Salina, S{\'{a}}nchez, Sanchez-Lucas,
  Santos, Santos, Sarazin, Sarmento, Sarmiento-Cano, Sato, Schauer, Scherini,
  Schieler, Schimp, Schmidt, Scholten, Schov{\'{a}}nek, Schröder, Schröder,
  Schulz, Schumacher, Sciutto, Segreto, Shadkam, Shellard, Sigl, Silli,
  {\v{S}}m{\'{\i}}da, Snow, Sommers, Sonntag, Soriano, Squartini, Stanca,
  Stani{\v{c}}, Stasielak, Stassi, Stolpovskiy, Strafella, Streich, Suarez,
  Suarez-Dur{\'{a}}n, Sudholz, Suomijärvi, Supanitsky, {\v{S}}up{\'{\i}}k,
  Swain, Szadkowski, Taboada, Taborda, Timmermans, Peixoto, Tomankova,
  Tom{\'{e}}, Elipe, Travnicek, Trini, Tueros, Ulrich, Unger, Urban, Galicia,
  Vali{\~{n}}o, Valore, van Aar, van Bodegom, van~den Berg, van Vliet, Varela,
  C{\'{a}}rdenas, V{\'{a}}zquez, Veberi{\v{c}}, Ventura, Quispe, Verzi, Vicha,
  Villase{\~{n}}or, Vorobiov, Wahlberg, Wainberg, Walz, Watson, Weber, Weindl,
  Wiede{\'{n}}ski, Wiencke, Wilczy{\'{n}}ski, Wirtz, Wittkowski, Wundheiler,
  Yang, Yushkov, Zas, Zavrtanik, Zavrtanik, Zepeda, Zimmermann, Ziolkowski,
  Zong, Zuccarello, Kim, Schulze, Bauer, Corral-Santana, de~Gregorio-Monsalvo,
  Gonz{\'{a}}lez-L{\'{o}}pez, Hartmann, Ishwara-Chandra, Mart{\'{\i}}n, Mehner,
  Misra, Micha{\l}owski, Resmi, Paragi, Agudo, An, Beswick, Casadio, Frey,
  Jonker, Kettenis, Marcote, Moldon, Szomoru, van Langevelde, Yang, Cwiek,
  Cwiok, Czyrkowski, Dabrowski, Kasprowicz, Mankiewicz, Nawrocki, Opiela,
  Piotrowski, Wrochna, Zaremba, {\.{Z}}arnecki, Haggard, Nynka, Ruan, Bland,
  Booler, Devillepoix, de~Gois, Hancock, Howie, Paxman, Sansom, Towner, Tonry,
  Coughlin, Stubbs, Denneau, Heinze, Stalder, Weiland, Eatough, Kramer, Kraus,
  Troja, Piro, Gonz{\'{a}}lez, Butler, Fox, Khandrika, Kutyrev, Lee, Ricci,
  Jr., S{\'{a}}nchez-Ram{\'{\i}}rez, Veilleux, Watson, Wieringa, Burgess, van
  Eerten, Fontes, Fryer, Korobkin, Wollaeger, Camilo, Foley, Goedhart,
  Makhathini, Oozeer, Smirnov, Fender, Woudt, , , , , , , , , , , , , , , , , ,
  , , , , , , , , , , , , , , , , , , , , , , , , , , , , , , , \&
  and}]{Abbott2017}
Abbott, B.~P., Abbott, R., Abbott, T.~D., {et~al.} 2017, The Astrophysical
  Journal, 848, L12, \dodoi{10.3847/2041-8213/aa91c9}

\bibitem[{Aguilera-Dena {et~al.}(2018)Aguilera-Dena, Langer, Moriya, \&
  Schootemeijer}]{AguileraDena2018}
Aguilera-Dena, D.~R., Langer, N., Moriya, T.~J., \& Schootemeijer, A. 2018, The
  Astrophysical Journal, 858, 115, \dodoi{10.3847/1538-4357/aabfc1}

\bibitem[{Aloy {et~al.}(1999)Aloy, Mueller, Ibanez, Marti, \&
  MacFadyen}]{Aloy1999}
Aloy, M.~A., Mueller, E., Ibanez, J.~M., Marti, J.~M., \& MacFadyen, A. 1999,
  Relativistic Jets from Collapsars.
\newblock \doarXiv{astro-ph/9910466}

\bibitem[{{Amati}(2008)}]{Amati2008MG}
{Amati}, L. 2008, in The Eleventh Marcel Grossmann Meeting On Recent
  Developments in Theoretical and Experimental General Relativity, Gravitation
  and Relativistic Field Theories, 1965--1967,
  \dodoi{10.1142/9789812834300\_0299}

\bibitem[{Amati {et~al.}(2002)Amati, Frontera, Tavani, in~’t Zand, Antonelli,
  Costa, Feroci, Guidorzi, Heise, Masetti, Montanari, Nicastro, Palazzi, Pian,
  Piro, \& Soffitta}]{Amati2002}
Amati, L., Frontera, F., Tavani, M., {et~al.} 2002, Astronomy \& Astrophysics,
  390, 81–89, \dodoi{10.1051/0004-6361:20020722}

\bibitem[{Arnett(1982)}]{Arnett1982}
Arnett, W.~D. 1982, The Astrophysical Journal, 253, 785, \dodoi{10.1086/159681}

\bibitem[{Ashall {et~al.}(2019)Ashall, Mazzali, Pian, Woosley, Palazzi,
  Prentice, Kobayashi, Holmbo, Levan, Perley, Stritzinger, Bufano, Filippenko,
  Melandri, Oates, Rossi, Selsing, Zheng, Castro-Tirado, Chincarini,
  D’Avanzo, De Pasquale, Emery, Fruchter, Hurley, Moller, Nomoto, Tanaka, \&
  Valeev}]{Ashall2019}
Ashall, C., Mazzali, P.~A., Pian, E., {et~al.} 2019, Monthly Notices of the
  Royal Astronomical Society, 487, 5824–5839, \dodoi{10.1093/mnras/stz1588}

\bibitem[{Barthelmy(2011)}]{Barthelmy2011}
Barthelmy, S. 2011, GRB 171205A: Swift-BAT refined analysis,  NASA.
\newblock \url{https://gcn.gsfc.nasa.gov/gcn3/22184.gcn3}

\bibitem[{Bartoli {et~al.}(2017)Bartoli, Bernardini, Bi, Cao, Catalanotti,
  Chen, Chen, Cui, Dai, D’Amone, Danzeng, Mitri, Piazzoli, Girolamo,
  Sciascio, Feng, Feng, Feng, Gao, Gou, Guo, He, Hu, Hu, Iacovacci, Iuppa, Jia,
  Labaci, Li, Liu, Liu, Liu, Lu, Ma, Ma, Mancarella, Mari, Marsella,
  Mastroianni, Montini, Ning, Perrone, Pistilli, Salvini, Santonico, Shen,
  Sheng, Shi, Surdo, Tan, Vallania, Vernetto, Vigorito, Wang, Wu, Wu, Xue,
  Yang, Yang, Yao, Yuan, Zha, Zhang, Zhang, Zhang, Zhang, Zhao, Zhaxici,
  Zhaxisang, Zhou, Zhu, \& Zhu}]{Bartoli2017}
Bartoli, B., Bernardini, P., Bi, X.~J., {et~al.} 2017, The Astrophysical
  Journal, 842, 31, \dodoi{10.3847/1538-4357/aa74bc}

\bibitem[{{Berger} {et~al.}(2011){Berger}, {Chornock}, {Holmes}, {Foley},
  {Cucchiara}, {Wolf}, {Podsiadlowski}, {Fox}, \& {Roth}}]{2011ApJ...743..204B}
{Berger}, E., {Chornock}, R., {Holmes}, T.~R., {et~al.} 2011, The Astrophysical
  Journal, 743, 204, \dodoi{10.1088/0004-637X/743/2/204}

\bibitem[{Berger {et~al.}(2011)Berger, Chornock, Holmes, Foley, Cucchiara,
  Wolf, Podsiadlowski, Fox, \& Roth}]{Berger2011}
Berger, E., Chornock, R., Holmes, T.~R., {et~al.} 2011, The Astrophysical
  Journal, 743, 204, \dodoi{10.1088/0004-637x/743/2/204}

\bibitem[{Bianco {et~al.}(2014)Bianco, Modjaz, Hicken, Friedman, Kirshner,
  Bloom, Challis, Marion, Wood-Vasey, \& Rest}]{Bianco2014}
Bianco, F.~B., Modjaz, M., Hicken, M., {et~al.} 2014, The Astrophysical Journal
  Supplement Series, 213, 19, \dodoi{10.1088/0067-0049/213/2/19}

\bibitem[{Biermann \& Cassinelli(1993)}]{Biermann1993}
Biermann, P.~L., \& Cassinelli, J.~P. 1993, Cosmic Rays, II. Evidence for a
  magnetic rotator Wolf Rayet star origin.
\newblock \doarXiv{astro-ph/9305003}

\bibitem[{Boggs \& Donaldson(1989)}]{Boggs1989}
Boggs, P.~T., \& Donaldson, J.~R. 1989, in Orthogonal distance regression

\bibitem[{Bosnjak {et~al.}(2006)Bosnjak, Celotti, Ghirlanda, Della~Valle, \&
  Pian}]{Bosnjak2006}
Bosnjak, Z., Celotti, A., Ghirlanda, G., Della~Valle, M., \& Pian, E. 2006,
  Astronomy \& Astrophysics, 447, 121, \dodoi{10.1051/0004-6361:20052803}

\bibitem[{Bromberg {et~al.}(2013)Bromberg, Nakar, Piran, \&
  Sari}]{Bromberg2013}
Bromberg, O., Nakar, E., Piran, T., \& Sari, R. 2013, The Astrophysical
  Journal, 764, 179, \dodoi{10.1088/0004-637x/764/2/179}

\bibitem[{Butler {et~al.}(2005)Butler, Sakamoto, Suzuki, Kawai, Lamb, Graziani,
  Donaghy, Dullighan, Vanderspek, Crew, Ford, Ricker, Atteia, Yoshida,
  Shirasaki, Tamagawa, Torii, Matsuoka, Fenimore, Galassi, Doty, Villasenor,
  Prigozhin, Jernigan, Barraud, Boer, Dezalay, Olive, Hurley, Levine, Martel,
  Morgan, Woosley, Cline, Braga, Manchanda, \& Pizzichini}]{Butler2005}
Butler, N., Sakamoto, T., Suzuki, M., {et~al.} 2005, The Astrophysical Journal,
  621, 884–893, \dodoi{10.1086/427746}

\bibitem[{Campana {et~al.}(2007)Campana, Guidorzi, Tagliaferri, Chincarini,
  Moretti, Rizzuto, \& Romano}]{Campana2007}
Campana, S., Guidorzi, C., Tagliaferri, G., {et~al.} 2007, Astron. Astrophys.,
  472, 395, \dodoi{10.1051/0004-6361:20066984}

\bibitem[{Campana {et~al.}(2006)Campana, Mangano, Blustin, Brown, Burrows,
  Chincarini, Cummings, Cusumano, Valle, Malesani, \& et~al.}]{Campana2006}
Campana, S., Mangano, V., Blustin, A.~J., {et~al.} 2006, Nature, 442,
  1008–1010, \dodoi{10.1038/nature04892}

\bibitem[{Campana {et~al.}(2008)}]{Campana2008}
Campana, S., {et~al.} 2008, Astrophys. J. Lett., 683, L9,
  \dodoi{10.1086/591421}

\bibitem[{Cannizzo \& Gehrels(2009)}]{Cannizzo2009}
Cannizzo, J.~K., \& Gehrels, N. 2009, The Astrophysical Journal, 700,
  1047–1058, \dodoi{10.1088/0004-637x/700/2/1047}

\bibitem[{Cannizzo {et~al.}(2011)Cannizzo, Troja, \& Gehrels}]{Cannizzo2011}
Cannizzo, J.~K., Troja, E., \& Gehrels, N. 2011, The Astrophysical Journal,
  734, 35, \dodoi{10.1088/0004-637x/734/1/35}

\bibitem[{Cano(2011)}]{Cano2011}
Cano, Z. 2011, PhD thesis.
\newblock \url{https://arxiv.org/ftp/arxiv/papers/1208/1208.0307.pdf}

\bibitem[{Cano(2013)}]{Cano2013}
---. 2013, Monthly Notices of the Royal Astronomical Society, 434, 1098,
  \dodoi{10.1093/mnras/stt1048}

\bibitem[{{Cano}(2014)}]{Cano2014a}
{Cano}, Z. 2014, \apj, 794, 121, \dodoi{10.1088/0004-637X/794/2/121}

\bibitem[{Cano {et~al.}(2017{\natexlab{a}})Cano, Wang, Dai, \& Wu}]{Cano2017a}
Cano, Z., Wang, S.-Q., Dai, Z.-G., \& Wu, X.-F. 2017{\natexlab{a}}, Advances in
  Astronomy, 2017, 1, \dodoi{10.1155/2017/8929054}

\bibitem[{Cano {et~al.}(2014)Cano, de~Ugarte~Postigo, Pozanenko, Butler,
  Thöne, Guidorzi, Krühler, Gorosabel, Jakobsson, Leloudas, Malesani, Hjorth,
  Melandri, Mundell, Wiersema, D’Avanzo, Schulze, Gomboc, Johansson, Zheng,
  Kann, Knust, Varela, Akerlof, Bloom, Burkhonov, Cooke, de~Diego, Dhungana,
  Farina, Ferrante, Flewelling, Fox, Fynbo, Gehrels, Georgiev, González,
  Greiner, Güver, Hartoog, Hatch, Jelinek, Kehoe, Klose, Klunko, Kopač,
  Kutyrev, Krugly, Lee, Levan, Linkov, Matkin, Minikulov, Molotov, Prochaska,
  Richer, Román-Zúñiga, Rumyantsev, Sánchez-Ramírez, Steele, Tanvir,
  Volnova, Watson, Xu, \& Yuan}]{Cano2014b}
Cano, Z., de~Ugarte~Postigo, A., Pozanenko, A., {et~al.} 2014, Astronomy \&
  Astrophysics, 568, A19, \dodoi{10.1051/0004-6361/201423920}

\bibitem[{Cano {et~al.}(2017{\natexlab{b}})Cano, Izzo, de~Ugarte~Postigo,
  Thöne, Krühler, Heintz, Malesani, Geier, Fuentes, Chen, Covino, D’Elia,
  Fynbo, Goldoni, Gomboc, Hjorth, Jakobsson, Kann, Milvang-Jensen, Pugliese,
  Sánchez-Ramírez, Schulze, Sollerman, Tanvir, \& Wiersema}]{Cano2017b}
Cano, Z., Izzo, L., de~Ugarte~Postigo, A., {et~al.} 2017{\natexlab{b}},
  Astronomy \& Astrophysics, 605, A107, \dodoi{10.1051/0004-6361/201731005}

\bibitem[{{Cao} {et~al.}(2022{\natexlab{a}}){Cao}, {Dainotti}, \&
  {Ratra}}]{Cao2022a}
{Cao}, S., {Dainotti}, M., \& {Ratra}, B. 2022{\natexlab{a}}.
\newblock \doarXiv{2201.05245}

\bibitem[{{Cao} {et~al.}(2022{\natexlab{b}}){Cao}, {Khadka}, \&
  {Ratra}}]{Cao2022b}
{Cao}, S., {Khadka}, N., \& {Ratra}, B. 2022{\natexlab{b}}, Monthly Notices of
  the Royal Astronomical Society, 510, 2928, \dodoi{10.1093/mnras/stab3559}

\bibitem[{Chen {et~al.}(2017)Chen, Moriya, Woosley, Sukhbold, Whalen, Suwa, \&
  Bromm}]{Chen2017}
Chen, K.-J., Moriya, T.~J., Woosley, S., {et~al.} 2017, The Astrophysical
  Journal, 839, 85, \dodoi{10.3847/1538-4357/aa68a4}

\bibitem[{Corsi {et~al.}(2017)Corsi, Cenko, Kasliwal, Quimby, Kulkarni, Frail,
  Goldstein, Blagorodnova, Connaughton, Perley, Singer, Copperwheat, Fremling,
  Kupfer, Piascik, Steele, Taddia, Vedantham, Kutyrev, Palliyaguru, Roberts,
  Sollerman, Troja, \& Veilleux}]{Corsi2017}
Corsi, A., Cenko, S.~B., Kasliwal, M.~M., {et~al.} 2017, The Astrophysical
  Journal, 847, 54, \dodoi{10.3847/1538-4357/aa85e5}

\bibitem[{Dado {et~al.}(2002)Dado, Dar, \& De~Rújula}]{Dado2002}
Dado, S., Dar, A., \& De~Rújula, A. 2002, Astronomy \& Astrophysics, 393, L25,
  \dodoi{10.1051/0004-6361:20021167}

\bibitem[{Dainotti \& Del~Vecchio(2017)}]{Dainotti2017a}
Dainotti, M., \& Del~Vecchio, R. 2017, New Astronomy Reviews, 77, 23,
  \dodoi{10.1016/j.newar.2017.04.001}

\bibitem[{Dainotti {et~al.}(2021{\natexlab{a}})Dainotti, Levine, Fraija, \&
  Chandra}]{DainottiLevine2021}
Dainotti, M., Levine, D., Fraija, N., \& Chandra, P. 2021{\natexlab{a}},
  Galaxies, 9, \dodoi{10.3390/galaxies9040095}

\bibitem[{Dainotti {et~al.}(2011)Dainotti, Ostrowski, \&
  Willingale}]{Dainotti2011b}
Dainotti, M., Ostrowski, M., \& Willingale, R. 2011, \mnras, 418, 2202,
  \dodoi{10.1111/j.1365-2966.2011.19433.x}

\bibitem[{Dainotti {et~al.}(2015)Dainotti, Petrosian, Willingale, O'Brien,
  Ostrowski, \& Nagataki}]{Dainotti2015}
Dainotti, M., Petrosian, V., Willingale, R., {et~al.} 2015, Monthly Notices of
  the Royal Astronomical Society, 451, 3898, \dodoi{10.1093/mnras/stv1229}

\bibitem[{Dainotti \& Amati(2018)}]{Dainotti2018}
Dainotti, M.~G., \& Amati, L. 2018, Publications of the Astronomical Society of
  the Pacific, 130, 051001, \dodoi{10.1088/1538-3873/aaa8d7}

\bibitem[{{Dainotti} {et~al.}(2007){Dainotti}, {Bernardini}, {Bianco}, {Caito},
  {Guida}, \& {Ruffini}}]{Dainotti2007}
{Dainotti}, M.~G., {Bernardini}, M.~G., {Bianco}, C.~L., {et~al.} 2007, \aap,
  471, L29, \dodoi{10.1051/0004-6361:20078068}

\bibitem[{Dainotti {et~al.}(2008)Dainotti, Cardone, \&
  Capozziello}]{Dainotti2008}
Dainotti, M.~G., Cardone, V.~F., \& Capozziello, S. 2008, Monthly Notices of
  the Royal Astronomical Society: Letters, 391, L79,
  \dodoi{10.1111/j.1745-3933.2008.00560.x}

\bibitem[{Dainotti {et~al.}(2021{\natexlab{b}})Dainotti, De~Simone, Schiavone,
  Montani, Rinaldi, \& Lambiase}]{DainottiDeSimone2021}
Dainotti, M.~G., De~Simone, B., Schiavone, T., {et~al.} 2021{\natexlab{b}}, The
  Astrophysical Journal, 912, 150, \dodoi{10.3847/1538-4357/abeb73}

\bibitem[{Dainotti {et~al.}(2022{\natexlab{a}})Dainotti, De~Simone, Schiavone,
  Montani, Rinaldi, Lambiase, Bogdan, \& Ugale}]{DainottiDeSimone2022}
Dainotti, M.~G., De~Simone, B.~D., Schiavone, T., {et~al.} 2022{\natexlab{a}},
  Galaxies, 10, 24, \dodoi{10.3390/galaxies10010024}

\bibitem[{Dainotti {et~al.}(2017{\natexlab{a}})Dainotti, Hernandez, Postnikov,
  Nagataki, O'brien, Willingale, \& Striegel}]{Dainotti2017b}
Dainotti, M.~G., Hernandez, X., Postnikov, S., {et~al.} 2017{\natexlab{a}},
  848, 88, \dodoi{10.3847/1538-4357/aa8a6b}

\bibitem[{Dainotti {et~al.}(2020{\natexlab{a}})Dainotti, Lenart, Sarracino,
  Nagataki, Capozziello, \& Fraija}]{Dainotti2020a}
Dainotti, M.~G., Lenart, A.~L., Sarracino, G., {et~al.} 2020{\natexlab{a}}, The
  Astrophysical Journal, 904, 97, \dodoi{10.3847/1538-4357/abbe8a}

\bibitem[{Dainotti {et~al.}(2017{\natexlab{b}})Dainotti, Nagataki, Maeda,
  Postnikov, \& Pian}]{Dainotti2017plateau}
Dainotti, M.~G., Nagataki, S., Maeda, K., Postnikov, S., \& Pian, E.
  2017{\natexlab{b}}, Astronomy \& Astrophysics, 600, A98,
  \dodoi{10.1051/0004-6361/201628384}

\bibitem[{Dainotti {et~al.}(2022{\natexlab{b}})Dainotti, Nielson, Sarracino,
  Rinaldi, Nagataki, Capozziello, Gnedin, \& Bargiacchi}]{DainottiNielson2022}
Dainotti, M.~G., Nielson, V., Sarracino, G., {et~al.} 2022{\natexlab{b}},
  Optical and X-ray GRB Fundamental Planes as Cosmological Distance Indicators,
   arXiv, \dodoi{10.48550/ARXIV.2203.15538}

\bibitem[{Dainotti {et~al.}(2013)Dainotti, Petrosian, Singal, \&
  Ostrowski}]{Dainotti2013}
Dainotti, M.~G., Petrosian, V., Singal, J., \& Ostrowski, M. 2013, The
  Astrophysical Journal, 774, 157, \dodoi{10.1088/0004-637x/774/2/157}

\bibitem[{Dainotti {et~al.}(2016)Dainotti, Postnikov, Hernandez, \&
  Ostrowski}]{Dainotti2016}
Dainotti, M.~G., Postnikov, S., Hernandez, X., \& Ostrowski, M. 2016, 825, L20,
  \dodoi{10.3847/2041-8205/825/2/l20}

\bibitem[{Dainotti {et~al.}(2020{\natexlab{b}})Dainotti, Livermore, Kann, Li,
  Oates, Yi, Zhang, Gendre, Cenko, \& Fraija}]{DainottiLivermore2020}
Dainotti, M.~G., Livermore, S., Kann, D.~A., {et~al.} 2020{\natexlab{b}}, The
  Astrophysical Journal Letters, 905, L26, \dodoi{10.3847/2041-8213/abcda9}

\bibitem[{{Dall'Osso} {et~al.}(2011){Dall'Osso}, {Stratta}, {Guetta}, {Covino},
  {De Cesare}, \& {Stella}}]{DallOsso2011}
{Dall'Osso}, S., {Stratta}, G., {Guetta}, D., {et~al.} 2011, \aap, 526, A121,
  \dodoi{10.1051/0004-6361/201014168}

\bibitem[{Della~Valle {et~al.}(2006)Della~Valle, Malesani, Bloom, Benetti,
  Chincarini, D'Avanzo, Foley, Covino, Melandri, Piranomonte, Tagliaferri,
  Stella, Gilmozzi, Antonelli, Campana, Chen, Filliatre, Fiore, Fugazza,
  Gehrels, Hurley, Mirabel, Pellizza, Piro, \& Prochaska}]{DellaValle2006}
Della~Valle, M., Malesani, D., Bloom, J.~S., {et~al.} 2006, The Astrophysical
  Journal, 642, L103, \dodoi{10.1086/504636}

\bibitem[{Demianski {et~al.}(2017)Demianski, Piedipalumbo, Sawant, \&
  Amati}]{Demianski2017}
Demianski, M., Piedipalumbo, E., Sawant, D., \& Amati, L. 2017, Astronomy \&
  Astrophysics, 598, A112, \dodoi{10.1051/0004-6361/201628909}

\bibitem[{Dereli {et~al.}(2017)Dereli, Boër, Gendre, Amati, Dichiara, \&
  Orange}]{Dereli2017}
Dereli, H., Boër, M., Gendre, B., {et~al.} 2017, The Astrophysical Journal,
  850, 117, \dodoi{10.3847/1538-4357/aa947d}

\bibitem[{Duncan(2001)}]{Duncan2001}
Duncan, R.~C. 2001, AIP Conference Proceedings, \dodoi{10.1063/1.1419599}

\bibitem[{D’Elia {et~al.}(2018)D’Elia, Campana, D’Aì, De~Pasquale,
  Emery, Frederiks, Lien, Melandri, Page, Starling, Burrows, Breeveld, Oates,
  O’Brien, Osborne, Siegel, Tagliaferri, Brown, Cenko, Svinkin, Tohuvavohu,
  \& Tsvetkova}]{Delia2018}
D’Elia, V., Campana, S., D’Aì, A., {et~al.} 2018, Astronomy \&
  Astrophysics, 619, A66, \dodoi{10.1051/0004-6361/201833847}

\bibitem[{Efron \& Petrosian(1992)}]{EfronPetrosian1992}
Efron, B., \& Petrosian, V. 1992, Astrophysical Journal, 399,
  \dodoi{10.1086/171931}

\bibitem[{Eisenberg {et~al.}(2022)Eisenberg, Gottlieb, \&
  Nakar}]{Eisenberg2022}
Eisenberg, M., Gottlieb, O., \& Nakar, E. 2022, Observational signatures of
  stellar explosions driven by relativistic jets.
\newblock \doarXiv{2201.08432}

\bibitem[{{Evans} {et~al.}(2009){Evans}, {Beardmore}, {Page}, {Osborne},
  {O'Brien}, {Willingale}, {Starling}, {Burrows}, {Godet}, {Vetere}, {Racusin},
  {Goad}, {Wiersema}, {Angelini}, {Capalbi}, {Chincarini}, {Gehrels}, {Kennea},
  {Margutti}, {Morris}, {Mountford}, {Pagani}, {Perri}, {Romano}, \&
  {Tanvir}}]{Evans2009}
{Evans}, P.~A., {Beardmore}, A.~P., {Page}, K.~L., {et~al.} 2009, \mnras, 397,
  1177, \dodoi{10.1111/j.1365-2966.2009.14913.x}

\bibitem[{Ferrero {et~al.}(2006)Ferrero, Kann, Zeh, Klose, Pian, Palazzi,
  Masetti, Hartmann, Sollerman, Deng, Filippenko, Greiner, Hughes, Mazzali, Li,
  Rol, Smith, \& Tanvir}]{Ferrero2006}
Ferrero, P., Kann, D.~A., Zeh, A., {et~al.} 2006, Astronomy \& Astrophysics,
  457, 857, \dodoi{10.1051/0004-6361:20065530}

\bibitem[{Frail {et~al.}(2001)Frail, Kulkarni, Sari, Djorgovski, Bloom, Galama,
  Reichart, Berger, Harrison, Price, Yost, Diercks, Goodrich, \&
  Chaffee}]{Frail2001}
Frail, D.~A., Kulkarni, S.~R., Sari, R., {et~al.} 2001, The Astrophysical
  Journal, 562, L55, \dodoi{10.1086/338119}

\bibitem[{Fynbo {et~al.}(2004)Fynbo, Sollerman, Hjorth, Grundahl, Gorosabel,
  Weidinger, Moller, Jensen, Vreeswijk, Fransson, \& et~al.}]{Fynbo2004}
Fynbo, J. P.~U., Sollerman, J., Hjorth, J., {et~al.} 2004, The Astrophysical
  Journal, 609, 962–971, \dodoi{10.1086/421260}

\bibitem[{Fynbo {et~al.}(2006)Fynbo, Watson, Thöne, Sollerman, Bloom, Davis,
  Hjorth, Jakobsson, Jørgensen, Graham, Fruchter, Bersier, Kewley, Cassan,
  Cerón, Foley, Gorosabel, Hinse, Horne, Jensen, Klose, Kocevski, Marquette,
  Perley, Ramirez-Ruiz, Stritzinger, Vreeswijk, Wijers, Woller, Xu, \&
  Zub}]{Fynbo2006}
Fynbo, J. P.~U., Watson, D., Thöne, C.~C., {et~al.} 2006, Nature, 444,
  1047–1049, \dodoi{10.1038/nature05375}

\bibitem[{Galama {et~al.}(1998)Galama, Vreeswijk, van Paradijs, Kouveliotou,
  Augusteijn, Böhnhardt, Brewer, Doublier, Gonzalez, Leibundgut, Lidman,
  Hainaut, Patat, Heise, in't Zand, Hurley, Groot, Strom, Mazzali, Iwamoto,
  Nomoto, Umeda, Nakamura, Young, Suzuki, Shigeyama, Koshut, Kippen, Robinson,
  de~Wildt, Wijers, Tanvir, Greiner, Pian, Palazzi, Frontera, Masetti,
  Nicastro, Feroci, Costa, Piro, Peterson, Tinney, Boyle, Cannon, Stathakis,
  Sadler, Begam, \& Ianna}]{Galama1998}
Galama, T.~J., Vreeswijk, P.~M., van Paradijs, J., {et~al.} 1998, Nature, 395,
  670, \dodoi{10.1038/27150}

\bibitem[{Garnavich {et~al.}(2003)Garnavich, Stanek, Wyrzykowski, Infante,
  Bendek, Bersier, Holland, Jha, Matheson, Kirshner, Krisciunas, Phillips, \&
  Carlberg}]{Garnavich2003}
Garnavich, P.~M., Stanek, K.~Z., Wyrzykowski, L., {et~al.} 2003, The
  Astrophysical Journal, 582, 924, \dodoi{10.1086/344785}

\bibitem[{Gehrels(2004)}]{Gehrels2004}
Gehrels, N. 2004, AIP Conference Proceedings, \dodoi{10.1063/1.1810924}

\bibitem[{Gendre {et~al.}(2007)Gendre, Galli, \& Piro}]{Gendre2007}
Gendre, B., Galli, A., \& Piro, L. 2007, Astronomy \& Astrophysics, 465,
  L13–L16, \dodoi{10.1051/0004-6361:20066896}

\bibitem[{Gendre {et~al.}(2019)Gendre, Joyce, Orange, Stratta, Atteia, \&
  Boër}]{Gendre2019}
Gendre, B., Joyce, Q.~T., Orange, N.~B., {et~al.} 2019, Monthly Notices of the
  Royal Astronomical Society, 486, 2471, \dodoi{10.1093/mnras/stz1036}

\bibitem[{Germany {et~al.}(2000)Germany, Reiss, Sadler, Schmidt, \&
  Stubbs}]{Germany2000}
Germany, L.~M., Reiss, D.~J., Sadler, E.~M., Schmidt, B.~P., \& Stubbs, C.~W.
  2000, The Astrophysical Journal, 533, 320, \dodoi{10.1086/308639}

\bibitem[{Ghirlanda {et~al.}(2004)Ghirlanda, Ghisellini, \&
  Lazzati}]{Ghirlanda2004}
Ghirlanda, G., Ghisellini, G., \& Lazzati, D. 2004, The Astrophysical Journal,
  616, 331–338, \dodoi{10.1086/424913}

\bibitem[{{Ghirlanda} {et~al.}(2010){Ghirlanda}, {Nava}, \&
  {Ghisellini}}]{Ghirlanda2010}
{Ghirlanda}, G., {Nava}, L., \& {Ghisellini}, G. 2010, \aap, 511, A43,
  \dodoi{10.1051/0004-6361/200913134}

\bibitem[{Ghisellini {et~al.}(2006)Ghisellini, Ghirlanda, Mereghetti, Bosnjak,
  Tavecchio, \& Firmani}]{Ghisellini2006}
Ghisellini, G., Ghirlanda, G., Mereghetti, S., {et~al.} 2006, Monthly Notices
  of the Royal Astronomical Society, 372, 1699,
  \dodoi{10.1111/j.1365-2966.2006.10972.x}

\bibitem[{Gompertz {et~al.}(2020)Gompertz, Levan, \& Tanvir}]{Gompertz2020}
Gompertz, B.~P., Levan, A.~J., \& Tanvir, N.~R. 2020, The Astrophysical
  Journal, 895, 58, \dodoi{10.3847/1538-4357/ab8d24}

\bibitem[{Granot {et~al.}(1999)Granot, Piran, \& Sari}]{Granot1999}
Granot, J., Piran, T., \& Sari, R. 1999, The Astrophysical Journal, 513,
  679–689, \dodoi{10.1086/306884}

\bibitem[{Granot \& Sari(2002)}]{GranotSari2002}
Granot, J., \& Sari, R. 2002, The Astrophysical Journal, 568, 820–829,
  \dodoi{10.1086/338966}

\bibitem[{Granot \& van~der Horst(2014)}]{Granot2014}
Granot, J., \& van~der Horst, A.~J. 2014, Publications of the Astronomical
  Society of Australia, 31, e008, \dodoi{10.1017/pasa.2013.44}

\bibitem[{Guessoum {et~al.}(2017)Guessoum, Alarayani, Al-Qassimi, AlShamsi,
  Sherif, Hamidani, Zitouni, \& Azzam}]{Guessoum2017}
Guessoum, N., Alarayani, O., Al-Qassimi, K., {et~al.} 2017, Journal of Physics:
  Conference Series, 869, 012080, \dodoi{10.1088/1742-6596/869/1/012080}

\bibitem[{Guetta \& Valle(2007)}]{Guetta2007}
Guetta, D., \& Valle, M.~D. 2007, The Astrophysical Journal, 657, L73,
  \dodoi{10.1086/511417}

\bibitem[{{Guillochon} {et~al.}(2017){Guillochon}, {Parrent}, {Kelley}, \&
  {Margutti}}]{TheOpenSupernovaCatalog}
{Guillochon}, J., {Parrent}, J., {Kelley}, L.~Z., \& {Margutti}, R. 2017, \apj,
  835, 64, \dodoi{10.3847/1538-4357/835/1/64}

\bibitem[{Götz {et~al.}(2014)Götz, Laurent, Antier, Covino, D’Avanzo,
  D’Elia, \& Melandri}]{Gotz2014}
Götz, D., Laurent, P., Antier, S., {et~al.} 2014, Monthly Notices of the Royal
  Astronomical Society, 444, 2776–2782, \dodoi{10.1093/mnras/stu1634}

\bibitem[{Heise(2003)}]{Heise2003}
Heise, J. 2003, AIP Conference Proceedings, \dodoi{10.1063/1.1579346}

\bibitem[{Hjorth(2013)}]{Hjort2013}
Hjorth, J. 2013, Philosophical Transactions of the Royal Society A:
  Mathematical, Physical and Engineering Sciences, 371, 20120275,
  \dodoi{10.1098/rsta.2012.0275}

\bibitem[{Hjorth \& Bloom(2012)}]{Hjorth2012}
Hjorth, J., \& Bloom, J.~S. 2012, The GRB–supernova connection,  Cambridge
  University Press.
\newblock
  \url{https://www.cambridge.org/core/books/gammaray-bursts/grbsupernova-connection/55BADFF3931C3239FE76146FDB15E145}

\bibitem[{Hu {et~al.}(2021)Hu, Castro-Tirado, Kumar, Gupta, Valeev, Pandey,
  Kann, Castellón, Agudo, Aryan, Caballero-García, Guziy, Martin-Carrillo,
  Oates, Pian, Sánchez-Ramírez, Sokolov, \& Zhang}]{Hu2021}
Hu, Y.-D., Castro-Tirado, A.~J., Kumar, A., {et~al.} 2021, Astronomy \&
  Astrophysics, 646, A50, \dodoi{10.1051/0004-6361/202039349}

\bibitem[{Hudec {et~al.}(1999)Hudec, Hudcová, \& Hroch}]{Hudec1999}
Hudec, R., Hudcová, V., \& Hroch, F. 1999, Astronomy and Astrophysics
  Supplement Series, 138, 475, \dodoi{10.1051/aas:1999316}

\bibitem[{Ito {et~al.}(2019)Ito, Matsumoto, Nagataki, Warren, Barkov, \&
  Yonetoku}]{Ito2019}
Ito, H., Matsumoto, J., Nagataki, S., {et~al.} 2019, Nature Communications, 10,
  \dodoi{10.1038/s41467-019-09281-z}

\bibitem[{Izzo {et~al.}(2019)Izzo, De~Ugarte~Postigo, Maeda, Thöne, Kann,
  Della~Valle, Carracedo, Micha\l{}owski, Schady, Schmidl, Selsing, Starling,
  Suzuki, Bensch, Bolmer, Campana, Cano, Covino, Fynbo, Hartmann, Heintz,
  Hjorth, Japelj, Kamiński, Kaper, Kouveliotou, Krużyński, Kwiatkowski,
  Leloudas, Levan, Malesani, Micha\l{}owski, Piranomonte, Pugliese, Rossi,
  Sánchez-Ramírez, Schulze, Steeghs, Tanvir, Ulaczyk, Vergani, \&
  Wiersema}]{Izzo2019}
Izzo, L., De~Ugarte~Postigo, A., Maeda, K., {et~al.} 2019, Jet cocoon
  signatures in the early spectra of a gamma-ray burst/supernova.
\newblock \url{https://arxiv.org/pdf/1901.05500.pdf}

\bibitem[{Kaneko {et~al.}(2015)Kaneko, Bostanc{\i}, Gö{\u{g} }ü{\c{s}}, \&
  Lin}]{Kaneko2015}
Kaneko, Y., Bostanc{\i}, Z.~F., Gö{\u{g} }ü{\c{s}}, E., \& Lin, L. 2015,
  Monthly Notices of the Royal Astronomical Society, 452, 824,
  \dodoi{10.1093/mnras/stv1286}

\bibitem[{Kann {et~al.}(2019)Kann, Schady, Olivares~E., Klose, Rossi, Perley,
  Krühler, Greiner, Nicuesa~Guelbenzu, Elliott, Knust, Filgas, Pian, Mazzali,
  Fynbo, Leloudas, Afonso, Delvaux, Graham, Rau, Schmidl, Schulze, Tanga,
  Updike, \& Varela}]{Kann2019}
Kann, D.~A., Schady, P., Olivares~E., F., {et~al.} 2019, Astronomy \&
  Astrophysics, 624, A143, \dodoi{10.1051/0004-6361/201629162}

\bibitem[{Kendall(1948)}]{kendall1948rank}
Kendall, M. 1948, Rank Correlation Methods (C. Griffin).
\newblock \url{https://books.google.it/books?id=hiBMAAAAMAAJ}

\bibitem[{Khatami \& Kasen(2019)}]{Khatami2019}
Khatami, D.~K., \& Kasen, D.~N. 2019, The Astrophysical Journal, 878, 56,
  \dodoi{10.3847/1538-4357/ab1f09}

\bibitem[{Klose {et~al.}(2019)Klose, Schmidl, Kann, Nicuesa~Guelbenzu, Schulze,
  Greiner, Olivares~E., Krühler, Schady, Afonso, Filgas, Fynbo, Rau, Rossi,
  Takats, Tanga, Updike, \& Varela}]{Klose2019}
Klose, S., Schmidl, S., Kann, D.~A., {et~al.} 2019, Astronomy \& Astrophysics,
  622, A138, \dodoi{10.1051/0004-6361/201832728}

\bibitem[{Klotz {et~al.}(2008)Klotz, Gendre, Stratta, Galli, Corsi, Preger,
  Cutini, Pélangeon, Atteia, Boër, \& et~al.}]{Klotz2008}
Klotz, A., Gendre, B., Stratta, G., {et~al.} 2008, Astronomy \& Astrophysics,
  483, 847–855, \dodoi{10.1051/0004-6361:20078677}

\bibitem[{Kocevski \& Liang(2006)}]{Kocevski2006}
Kocevski, D., \& Liang, E. 2006, The Astrophysical Journal, 642, 371,
  \dodoi{10.1086/500816}

\bibitem[{Kong {et~al.}(2009)Kong, Huang, Cheng, \& Lu}]{Kong2009}
Kong, S., Huang, Y., Cheng, K., \& Lu, T. 2009, Science in China Series G:
  Physics, Mechanics and Astronomy, 52, 2047–2053,
  \dodoi{10.1007/s11433-009-0275-y}

\bibitem[{Kouveliotou {et~al.}(1993)Kouveliotou, Meegan, Fishman, Bhat, Briggs,
  Koshut, Paciesas, \& Pendleton}]{Kouveliotou1993}
Kouveliotou, C., Meegan, C.~A., Fishman, G.~J., {et~al.} 1993, The
  Astrophysical Journal Letters, 413, L101–L104, \dodoi{10.1086/186969}

\bibitem[{{Kumar} {et~al.}(2008{\natexlab{a}}){Kumar}, {Narayan}, \&
  {Johnson}}]{Kumar2008a}
{Kumar}, P., {Narayan}, R., \& {Johnson}, J.~L. 2008{\natexlab{a}}, Science,
  321, 376–379, \dodoi{10.1126/science.1159003}

\bibitem[{{Kumar} {et~al.}(2008{\natexlab{b}}){Kumar}, {Narayan}, \&
  {Johnson}}]{Kumar2008b}
---. 2008{\natexlab{b}}, \mnras, 388, 1729,
  \dodoi{10.1111/j.1365-2966.2008.13493.x}

\bibitem[{Kumar \& Zhang(2015)}]{Kumar2015}
Kumar, P., \& Zhang, B. 2015, Physics Reports, 561, 1–109,
  \dodoi{10.1016/j.physrep.2014.09.008}

\bibitem[{Levan {et~al.}(2014)Levan, Tanvir, Fruchter, Hjorth, Pian, Mazzali,
  Hounsell, Perley, Cano, Graham, Cenko, Fynbo, Kouveliotou,
  Pe{\textquotesingle}er, Misra, \& Wiersema}]{Levan2014}
Levan, A.~J., Tanvir, N.~R., Fruchter, A.~S., {et~al.} 2014, The Astrophysical
  Journal, 792, 115, \dodoi{10.1088/0004-637x/792/2/115}

\bibitem[{Li(2006{\natexlab{a}})}]{LiXinLi2006}
Li, L.-X. 2006{\natexlab{a}}, Monthly Notices of the Royal Astronomical
  Society, 372, 1357, \dodoi{10.1111/j.1365-2966.2006.10943.x}

\bibitem[{Li(2006{\natexlab{b}})}]{Li2006}
---. 2006{\natexlab{b}}, Monthly Notices of the Royal Astronomical Society,
  372, 1357, \dodoi{10.1111/j.1365-2966.2006.10943.x}

\bibitem[{{Li}(2008{\natexlab{a}})}]{Li080109}
{Li}, L.-X. 2008{\natexlab{a}}, \mnras, 388, 603,
  \dodoi{10.1111/j.1365-2966.2008.13461.x}

\bibitem[{{Li}(2008{\natexlab{b}})}]{Li2008}
---. 2008{\natexlab{b}}, {The GRB-Supernova Connection},
  \dodoi{10.1063/1.3027928}

\bibitem[{Li {et~al.}(2002)Li, Filippenko, Van~Dyk, Hu, Qiu, Modjaz, \&
  Leonard}]{Li2002}
Li, W., Filippenko, A., Van~Dyk, S., {et~al.} 2002, A Hubble Space Telescope
  snapshot survey of nearby supernovae.
\newblock \url{https://arxiv.org/pdf/astro-ph/0201228.pdf}

\bibitem[{Liang {et~al.}(2008)Liang, Racusin, Zhang, Zhang, \&
  Burrows}]{Liang2008}
Liang, E., Racusin, J.~L., Zhang, B., Zhang, B., \& Burrows, D.~N. 2008, The
  Astrophysical Journal, 675, 528–552, \dodoi{10.1086/524701}

\bibitem[{Liang \& Zhang(2005)}]{LiangZhang2005}
Liang, E., \& Zhang, B. 2005, 633, 611, \dodoi{10.1086/491594}

\bibitem[{Liang {et~al.}(2007)Liang, Zhang, Virgili, \& Dai}]{Liang2007}
Liang, E., Zhang, B., Virgili, F., \& Dai, Z.~G. 2007, The Astrophysical
  Journal, 662, 1111, \dodoi{10.1086/517959}

\bibitem[{Lin {et~al.}(2020)Lin, Lu, Lin, \& Wang}]{Lin2020}
Lin, J., Lu, R.-J., Lin, D.-B., \& Wang, X.-G. 2020, The Astrophysical Journal,
  895, 46, \dodoi{10.3847/1538-4357/ab88a7}

\bibitem[{Lloyd \& Petrosian(2000)}]{Lloyd2000}
Lloyd, N.~M., \& Petrosian, V. 2000, The Astrophysical Journal, 543, 722,
  \dodoi{10.1086/317125}

\bibitem[{Lloyd-Ronning {et~al.}(2020)Lloyd-Ronning, Hurtado, Aykutalp,
  Johnson, \& Ceccobello}]{Lloyd2020}
Lloyd-Ronning, N., Hurtado, V.~U., Aykutalp, A., Johnson, J., \& Ceccobello, C.
  2020, Monthly Notices of the Royal Astronomical Society, 494, 4371,
  \dodoi{10.1093/mnras/staa1057}

\bibitem[{Lloyd-Ronning {et~al.}(2019)Lloyd-Ronning, Aykutalp, \&
  Johnson}]{Lloyd2019}
Lloyd-Ronning, N.~M., Aykutalp, A., \& Johnson, J.~L. 2019, Monthly Notices of
  the Royal Astronomical Society, 488, 5823, \dodoi{10.1093/mnras/stz2155}

\bibitem[{Lyman {et~al.}(2016)Lyman, Bersier, James, Mazzali, Eldridge, Fraser,
  \& Pian}]{Lyman2016}
Lyman, J.~D., Bersier, D., James, P.~A., {et~al.} 2016, Monthly Notices of the
  Royal Astronomical Society, 457, 328, \dodoi{10.1093/mnras/stv2983}

\bibitem[{Lü {et~al.}(2018)Lü, Lan, Zhang, Liang, Kann, Du, \& Shen}]{Lu2018}
Lü, H.-J., Lan, L., Zhang, B., {et~al.} 2018, The Astrophysical Journal, 862,
  130, \dodoi{10.3847/1538-4357/aacd03}

\bibitem[{Madjaz {et~al.}(1999)Madjaz, Li, Garnavich, Jha, Challis, \&
  Kirshner}]{Madjaz1999}
Madjaz, M., Li, W.~D., Garnavich, P., {et~al.} 1999, IAUC 7268: 1999eb; 1999ec;
  1999dj.
\newblock \url{http://www.cbat.eps.harvard.edu/iauc/07200/07268.html#Item2}

\bibitem[{{Malmquist}(1922)}]{Malmquist}
{Malmquist}, K.~G. 1922, Meddelanden fran Lunds Astronomiska Observatorium
  Serie I, 100, 1

\bibitem[{Martone {et~al.}(2017)Martone, Izzo, Della~Valle, Amati, Longo, \&
  Götz}]{Martone2017}
Martone, R., Izzo, L., Della~Valle, M., {et~al.} 2017, Astronomy \&
  Astrophysics, 608, A52, \dodoi{10.1051/0004-6361/201730704}

\bibitem[{{Mazets} {et~al.}(1981){Mazets}, {Golenetskii}, {Ilinskii}, {Panov},
  {Aptekar}, {Gurian}, {Proskura}, {Sokolov}, {Sokolova}, \&
  {Kharitonova}}]{Mazets1981}
{Mazets}, E.~P., {Golenetskii}, S.~V., {Ilinskii}, V.~N., {et~al.} 1981, \apss,
  80, 3, \dodoi{10.1007/BF00649140}

\bibitem[{Mazzali(2011)}]{Mazzali2011}
Mazzali, P.~A. 2011, Proceedings of the International Astronomical Union, 7,
  75, \dodoi{10.1017/s1743921312012720}

\bibitem[{Mazzali {et~al.}(2008)Mazzali, Valenti, Della~Valle, Chincarini,
  Sauer, Benetti, Pian, Piran, D'Elia, Elias-Rosa, Margutti, Pasotti,
  Antonelli, Bufano, Campana, Cappellaro, Covino, D'Avanzo, Fiore, Fugazza,
  Gilmozzi, Hunter, Maguire, Maiorano, Marziani, Masetti, Mirabel, Navasardyan,
  Nomoto, Palazzi, Pastorello, Panagia, Pellizza, Sari, Smartt, Tagliaferri,
  Tanaka, Taubenberger, Tominaga, Trundle, \& Turatto}]{Mazzali2008}
Mazzali, P.~A., Valenti, S., Della~Valle, M., {et~al.} 2008, Science, 321,
  1185, \dodoi{10.1126/science.1158088}

\bibitem[{{Melandri} {et~al.}(2014){Melandri}, {Pian}, {D'Elia}, {D'Avanzo},
  {Della Valle}, {Mazzali}, {Tagliaferri}, {Cano}, {Levan}, {M{\o}oller},
  {Amati}, {Bernardini}, {Bersier}, {Bufano}, {Campana}, {Castro-Tirado},
  {Covino}, {Ghirlanda}, {Hurley}, {Malesani}, {Masetti}, {Palazzi},
  {Piranomonte}, {Rossi}, {Salvaterra}, {Starling}, {Tanaka}, {Tanvir}, \&
  {Vergani}}]{Melandri2014}
{Melandri}, A., {Pian}, E., {D'Elia}, V., {et~al.} 2014, \aap, 567, A29,
  \dodoi{10.1051/0004-6361/201423572}

\bibitem[{Melandri {et~al.}(2019)Melandri, Malesani, Izzo, Japelj, Vergani,
  Schady, Sagués~Carracedo, de~Ugarte~Postigo, Anderson, Barbarino, Bolmer,
  Breeveld, Calissendorff, Campana, Cano, Carini, Covino, D’Avanzo, D’Elia,
  Della~Valle, De~Pasquale, Fynbo, Gromadzki, Hammer, Hartmann, Heintz,
  Inserra, Jakobsson, Kann, Kotilainen, Maguire, Masetti, Nicholl, Olivares~E,
  Pugliese, Rossi, Salvaterra, Sollerman, Stone, Tagliaferri, Tomasella,
  Thöne, Xu, \& Young}]{Melandri2019}
Melandri, A., Malesani, D.~B., Izzo, L., {et~al.} 2019, Monthly Notices of the
  Royal Astronomical Society, 490, 5366, \dodoi{10.1093/mnras/stz2900}

\bibitem[{Meszaros(2000)}]{Meszaros2000}
Meszaros, P. 2000, AIP Conference Proceedings, \dodoi{10.1063/1.1361591}

\bibitem[{{M{\'e}sz{\'a}ros}(2001)}]{Meszaros2001}
{M{\'e}sz{\'a}ros}, P. 2001, Progress of Theoretical Physics Supplement, 143,
  33, \dodoi{10.1143/PTPS.143.33}

\bibitem[{Minaev \& Pozanenko(2019)}]{Minaev2019}
Minaev, P.~Y., \& Pozanenko, A.~S. 2019, Monthly Notices of the Royal
  Astronomical Society, 492, 1919, \dodoi{10.1093/mnras/stz3611}

\bibitem[{Modjaz {et~al.}(2009)Modjaz, Li, Butler, Chornock, Perley, Blondin,
  Bloom, Filippenko, Kirshner, Kocevski, Poznanski, Hicken, Foley,
  Stringfellow, Berlind, Barrado~y Navascues, Blake, Bouy, Brown, Challis,
  Chen, de~Vries, Dufour, Falco, Friedman, Ganeshalingam, Garnavich, Holden,
  Illingworth, Lee, Liebert, Marion, Olivier, Prochaska, Silverman, Smith,
  Starr, Steele, Stockton, Williams, \& Wood-Vasey}]{Modjaz2009}
Modjaz, M., Li, W., Butler, N., {et~al.} 2009, The Astrophysical Journal, 702,
  226, \dodoi{10.1088/0004-637x/702/1/226}

\bibitem[{Moriya {et~al.}(2020)Moriya, Marchant, \& Blinnikov}]{Moriya2020UL}
Moriya, T.~J., Marchant, P., \& Blinnikov, S.~I. 2020, Astronomy \&
  Astrophysics, 641, L10, \dodoi{10.1051/0004-6361/202038903}

\bibitem[{Nagataki(2009)}]{Nagataki2009jet}
Nagataki, S. 2009, The Astrophysical Journal, 704, 937–950,
  \dodoi{10.1088/0004-637x/704/2/937}

\bibitem[{Nakauchi {et~al.}(2013)Nakauchi, Kashiyama, Suwa, \&
  Nakamura}]{Nakauchi2013}
Nakauchi, D., Kashiyama, K., Suwa, Y., \& Nakamura, T. 2013, The Astrophysical
  Journal, 778, 67, \dodoi{10.1088/0004-637x/778/1/67}

\bibitem[{Nicholl(2021)}]{Nicholl2021}
Nicholl, M. 2021, Astronomy \& Geophysics, 62, 5.34,
  \dodoi{10.1093/astrogeo/atab092}

\bibitem[{Nicolas {et~al.}(2021)Nicolas, Rigault, Copin, Graziani, Aldering,
  Briday, Kim, Nordin, Perlmutter, \& Smith}]{Nicolas2021}
Nicolas, N., Rigault, M., Copin, Y., {et~al.} 2021, Astronomy \& Astrophysics,
  649, A74, \dodoi{10.1051/0004-6361/202038447}

\bibitem[{Nomoto {et~al.}(2010)Nomoto, Moriya, Tominaga, \&
  Suzuki}]{Nomoto2010}
Nomoto, K., Moriya, T., Tominaga, N., \& Suzuki, T. 2010, AIP Conference
  Proceedings, 1279, 60, \dodoi{10.1063/1.3509354}

\bibitem[{{Nomoto} {et~al.}(2008){Nomoto}, {Tominaga}, {Tanaka}, {Maeda}, \&
  {Umeda}}]{Nomoto2008}
{Nomoto}, K., {Tominaga}, N., {Tanaka}, M., {Maeda}, K., \& {Umeda}, H. 2008,
  in Massive Stars as Cosmic Engines, ed. F.~{Bresolin}, P.~A. {Crowther}, \&
  J.~{Puls}, Vol. 250, 463--470, \dodoi{10.1017/S1743921308020838}

\bibitem[{Nomoto {et~al.}(2000)Nomoto, Mazzali, Nakamura, Iwamoto, Maeda,
  Suzuki, Turatto, Danziger, \& Patat}]{Nomoto:2000vi}
Nomoto, K., Mazzali, P.~A., Nakamura, T., {et~al.} 2000, AIP Conf. Proc., 526,
  622, \dodoi{10.1063/1.1361611}

\bibitem[{{Oates} {et~al.}(2012){Oates}, {Page}, {De Pasquale}, {Schady},
  {Breeveld}, {Holland}, {Kuin}, \& {Marshall}}]{Oates2012}
{Oates}, S.~R., {Page}, M.~J., {De Pasquale}, M., {et~al.} 2012, \mnras, 426,
  L86, \dodoi{10.1111/j.1745-3933.2012.01331.x}

\bibitem[{Ofek {et~al.}(2007)Ofek, Cenko, Gal-Yam, Fox, Nakar, Rau, Frail,
  Kulkarni, Price, Schmidt, Soderberg, Peterson, Berger, Sharon, Shemmer,
  Penprase, Chevalier, Brown, Burrows, Gehrels, Harrison, Holland, Mangano,
  McCarthy, Moon, Nousek, Persson, Piran, \& Sari}]{Ofek2007}
Ofek, E.~O., Cenko, S.~B., Gal-Yam, A., {et~al.} 2007, The Astrophysical
  Journal, 662, 1129, \dodoi{10.1086/518082}

\bibitem[{Pearson(1901)}]{Pearson}
Pearson, K. 1901, The London, Edinburgh, and Dublin Philosophical Magazine and
  Journal of Science, 2, 559, \dodoi{10.1080/14786440109462720}

\bibitem[{Perley {et~al.}(2010)Perley, Bloom, Klein, Covino, Minezaki,
  Woźniak, Vestrand, Williams, Milne, Butler, Updike, Krühler, Afonso,
  Antonelli, Cowie, Ferrero, Greiner, Hartmann, Kakazu, Küpcü~Yoldaş,
  Morgan, Price, Prochaska, \& Yoshii}]{Perley2010}
Perley, D.~A., Bloom, J.~S., Klein, C.~R., {et~al.} 2010, Monthly Notices of
  the Royal Astronomical Society, 406, 2473,
  \dodoi{10.1111/j.1365-2966.2010.16772.x}

\bibitem[{Perna {et~al.}(2018)Perna, Lazzati, \& Cantiello}]{Perna2018}
Perna, R., Lazzati, D., \& Cantiello, M. 2018, The Astrophysical Journal, 859,
  48, \dodoi{10.3847/1538-4357/aabcc1}

\bibitem[{Petrosian {et~al.}(2009)Petrosian, Bouvier, \& Ryde}]{Petrosian2009}
Petrosian, V., Bouvier, A., \& Ryde, F. 2009, Gamma-ray Bursts as Cosmological
  Tools.
\newblock \url{https://arxiv.org/pdf/0909.5051.pdf}

\bibitem[{{Phillips}(1993)}]{Phillips1993}
{Phillips}, M.~M. 1993, The Astrophysical Journal Letters, 413, L105,
  \dodoi{10.1086/186970}

\bibitem[{{Pian}(2005)}]{Pian2005}
{Pian}, E. 2005, in Astronomical Society of the Pacific Conference Series, Vol.
  342, 1604-2004: Supernovae as Cosmological Lighthouses, ed. M.~{Turatto},
  S.~{Benetti}, L.~{Zampieri}, \& W.~{Shea}, 315

\bibitem[{Pian {et~al.}(2004)Pian, Giommi, Amati, Costa, Danziger, Feroci,
  Fiocchi, Frontera, Kouveliotou, Masetti, \& et~al.}]{Pian2004}
Pian, E., Giommi, P., Amati, L., {et~al.} 2004, Advances in Space Research, 34,
  2711–2714, \dodoi{10.1016/j.asr.2003.04.072}

\bibitem[{Pian {et~al.}(2006)Pian, Mazzali, Masetti, Ferrero, Klose, Palazzi,
  Ramirez-Ruiz, Woosley, Kouveliotou, Deng, Filippenko, Foley, Fynbo, Kann, Li,
  Hjorth, Nomoto, Patat, Sauer, Sollerman, Vreeswijk, Guenther, Levan,
  O’Brien, Tanvir, Wijers, Dumas, Hainaut, Wong, Baade, Wang, Amati,
  Cappellaro, Castro-Tirado, Ellison, Frontera, Fruchter, Greiner, Kawabata,
  Ledoux, Maeda, Møller, Nicastro, Rol, \& Starling}]{Pian2006}
Pian, E., Mazzali, P.~A., Masetti, N., {et~al.} 2006, Nature, 442, 1011–1013,
  \dodoi{10.1038/nature05082}

\bibitem[{Pian {et~al.}(2017)Pian, D’Avanzo, Benetti, Branchesi, Brocato,
  Campana, Cappellaro, Covino, D’Elia, Fynbo, \& et~al.}]{Pian2017}
Pian, E., D’Avanzo, P., Benetti, S., {et~al.} 2017, Nature, 551, 67–70,
  \dodoi{10.1038/nature24298}

\bibitem[{Piran(1999)}]{Piran1999}
Piran, T. 1999, Physics Reports, 314, 575–667,
  \dodoi{10.1016/s0370-1573(98)00127-6}

\bibitem[{{Piran}(2000)}]{Piran2000}
{Piran}, T. 2000, \physrep, 333, 529, \dodoi{10.1016/S0370-1573(00)00036-3}

\bibitem[{{Piro} {et~al.}(2014){Piro}, {Troja}, {Gendre}, {Ghisellini},
  {Ricci}, {Bannister}, {Fiore}, {Kidd}, {Piranomonte}, \&
  {Wieringa}}]{Piro2014}
{Piro}, L., {Troja}, E., {Gendre}, B., {et~al.} 2014, \apjl, 790, L15,
  \dodoi{10.1088/2041-8205/790/2/L15}

\bibitem[{Prentice {et~al.}(2016)Prentice, Mazzali, Pian, Gal-Yam, Kulkarni,
  Rubin, Corsi, Fremling, Sollerman, Yaron, Arcavi, Zheng, Kasliwal,
  Filippenko, Cenko, Cao, \& Nugent}]{Prentice2016}
Prentice, S.~J., Mazzali, P.~A., Pian, E., {et~al.} 2016, Monthly Notices of
  the Royal Astronomical Society, 458, 2973, \dodoi{10.1093/mnras/stw299}

\bibitem[{Price {et~al.}(2003)Price, Kulkarni, Schmidt, Galama, Bloom, Berger,
  Frail, Djorgovski, Fox, Henden, Klose, Harrison, Reichart, Sari, Yost,
  Axelrod, McCarthy, Holtzman, Halpern, Kimble, Wheeler, Chevalier, Hurley,
  Ricker, Costa, Frontera, \& Piro}]{Price2003}
Price, P.~A., Kulkarni, S.~R., Schmidt, B.~P., {et~al.} 2003, The Astrophysical
  Journal, 584, 931, \dodoi{10.1086/345734}

\bibitem[{Qin \& Chen(2013)}]{Qin2013}
Qin, Y.-P., \& Chen, Z.-F. 2013, Monthly Notices of the Royal Astronomical
  Society, 430, 163, \dodoi{10.1093/mnras/sts547}

\bibitem[{Rastinejad {et~al.}(2022)Rastinejad, Gompertz, Levan, Fong, Nicholl,
  Lamb, Malesani, Nugent, Oates, Tanvir, Postigo, Kilpatrick, Moore, Metzger,
  Ravasio, Rossi, Schroeder, Jencson, Sand, Smith, Fernández, Berger,
  Blanchard, Chornock, Cobb, De~Pasquale, Fynbo, Izzo, Kann, Laskar, Marini,
  Paterson, Escorial, Sears, \& Thöne}]{Rastinejad2022KN}
Rastinejad, J.~C., Gompertz, B.~P., Levan, A.~J., {et~al.} 2022, A Kilonova
  Following a Long-Duration Gamma-Ray Burst at 350 Mpc,  arXiv,
  \dodoi{10.48550/ARXIV.2204.10864}

\bibitem[{{Rea} {et~al.}(2015){Rea}, {Gull{\'o}n}, {Pons}, {Perna}, {Dainotti},
  {Miralles}, \& {Torres}}]{Rea2015}
{Rea}, N., {Gull{\'o}n}, M., {Pons}, J.~A., {et~al.} 2015, \apj, 813, 92,
  \dodoi{10.1088/0004-637X/813/2/92}

\bibitem[{Rhodes {et~al.}(2021)Rhodes, Fender, Williams, \&
  Mooley}]{Rhodes2021}
Rhodes, L., Fender, R., Williams, D. R.~A., \& Mooley, K. 2021, Monthly Notices
  of the Royal Astronomical Society, 503, 2966, \dodoi{10.1093/mnras/stab640}

\bibitem[{Rigon {et~al.}(2003)Rigon, Turatto, Benetti, Pastorello, Cappellaro,
  Aretxaga, Vega, Chavushyan, Patat, Danziger, \& Salvo}]{Rigon2003}
Rigon, L., Turatto, M., Benetti, S., {et~al.} 2003, Monthly Notices of the
  Royal Astronomical Society, 340, 191,
  \dodoi{10.1046/j.1365-8711.2003.06282.x}

\bibitem[{Rossi {et~al.}(2021)Rossi, Rothberg, Palazzi, Kann, D'avanzo, Klose,
  Perego, Pian, Savaglio, Stratta, Agapito, Covino, Cusano, D'Elia,
  De~Pasquale, Della~Valle, Guidorzi, Kuhn, Izzo, Loffredo, Masetti, Melandri,
  Nicuesa~Guelbenzu, Paris, Plantet, Rossi, Salvaterra, \& Veillet}]{Rossi2021}
Rossi, A., Rothberg, B., Palazzi, E., {et~al.} 2021, submitted to ApJ, 1,
  \dodoi{arXiv:2105.03829}

\bibitem[{Rowlinson {et~al.}(2014)Rowlinson, Gompertz, Dainotti, O'Brien,
  Wijers, \& van~der Horst}]{Rowlinson2014}
Rowlinson, A., Gompertz, B.~P., Dainotti, M., {et~al.} 2014, Monthly Notices of
  the Royal Astronomical Society, 443, 1779, \dodoi{10.1093/mnras/stu1277}

\bibitem[{{Rowlinson} {et~al.}(2013){Rowlinson}, {O'Brien}, {Metzger},
  {Tanvir}, \& {Levan}}]{Rowlinson2013}
{Rowlinson}, A., {O'Brien}, P.~T., {Metzger}, B.~D., {Tanvir}, N.~R., \&
  {Levan}, A.~J. 2013, \mnras, 430, 1061, \dodoi{10.1093/mnras/sts683}

\bibitem[{Rueda \& Ruffini(2012)}]{Rueda2012}
Rueda, J.~A., \& Ruffini, R. 2012, The Astrophysical Journal, 758, L7,
  \dodoi{10.1088/2041-8205/758/1/l7}

\bibitem[{Ruffini {et~al.}(2001)Ruffini, Bianco, Fraschetti, Xue, \&
  Chardonnet}]{Ruffini2001}
Ruffini, R., Bianco, C.~L., Fraschetti, F., Xue, S.-S., \& Chardonnet, P. 2001,
  The Astrophysical Journal, 555, L117, \dodoi{10.1086/323177}

\bibitem[{Ruffini {et~al.}(2016)Ruffini, Rueda, Muccino, Aimuratov, Becerra,
  Bianco, Kovacevic, Moradi, Oliveira, Pisani, \& Wang}]{Ruffini2016}
Ruffini, R., Rueda, J.~A., Muccino, M., {et~al.} 2016, The Astrophysical
  Journal, 832, 136, \dodoi{10.3847/0004-637x/832/2/136}

\bibitem[{Ruffini {et~al.}(2021)Ruffini, Moradi, Aimuratov, Bianco, Karlica,
  Rueda, Sahakyan, Wang, Li, Xue, \& team}]{Ruffini2021}
Ruffini, R., Moradi, R., Aimuratov, Y., {et~al.} 2021, GRB 210210A as a BdHN
  II,  NASA.
\newblock \url{https://gcn.gsfc.nasa.gov/gcn/gcn3/29481.gcn3}

\bibitem[{Sakamoto(2004)}]{Sakamoto2004}
Sakamoto, T. 2004, AIP Conference Proceedings, \dodoi{10.1063/1.1810811}

\bibitem[{Sakamoto {et~al.}(2008)Sakamoto, Barthelmy, Barbier, Cummings,
  Fenimore, Gehrels, Hullinger, Krimm, Markwardt, Palmer, Parsons, Sato,
  Stamatikos, Tueller, Ukwatta, \& Zhang}]{Sakamoto2008Swift}
Sakamoto, T., Barthelmy, S.~D., Barbier, L., {et~al.} 2008, The Astrophysical
  Journal Supplement Series, 175, 179–190, \dodoi{10.1086/523646}

\bibitem[{Schady(2017)}]{Schady2017}
Schady, P. 2017, Gamma-ray bursts and their use as cosmic probes.
\newblock \doarXiv{1707.05214}

\bibitem[{Schaerer {et~al.}(1994)Schaerer, de~Koter, \& Schmutz}]{Schaerer1994}
Schaerer, D., de~Koter, A., \& Schmutz, W. 1994, Combined Stellar Structure and
  Atmosphere Models: Exploratory Results for Wolf-Rayet Stars.
\newblock \doarXiv{astro-ph/9410046}

\bibitem[{Schmidt(2005)}]{Schmidt2005}
Schmidt, M. 2005, Il Nuovo Cimento, 28 (2005), 347,
  \dodoi{10.1393/ncc/i2005-10057-9}

\bibitem[{{Schulze} {et~al.}(2014){Schulze}, {Malesani}, {Cucchiara}, {Tanvir},
  {Kr{\"u}hler}, {de Ugarte Postigo}, {Leloudas}, {Lyman}, {Bersier},
  {Wiersema}, {Perley}, {Schady}, {Gorosabel}, {Anderson}, {Castro-Tirado},
  {Cenko}, {De Cia}, {Ellerbroek}, {Fynbo}, {Greiner}, {Hjorth}, {Kann},
  {Kaper}, {Klose}, {Levan}, {Mart{\'\i}n}, {O'Brien}, {Page}, {Pignata},
  {Rapaport}, {S{\'a}nchez-Ram{\'\i}rez}, {Sollerman}, {Smith}, {Sparre},
  {Th{\"o}ne}, {Watson}, {Xu}, {Bauer}, {Bayliss}, {Bj{\"o}rnsson}, {Bremer},
  {Cano}, {Covino}, {D'Elia}, {Frail}, {Geier}, {Goldoni}, {Hartoog},
  {Jakobsson}, {Korhonen}, {Lee}, {Milvang-Jensen}, {Nardini}, {Nicuesa
  Guelbenzu}, {Oguri}, {Pandey}, {Petitpas}, {Rossi}, {Sandberg}, {Schmidl},
  {Tagliaferri}, {Tilanus}, {Winters}, {Wright}, \& {Wuyts}}]{Schulze2014}
{Schulze}, S., {Malesani}, D., {Cucchiara}, A., {et~al.} 2014, \aap, 566, A102,
  \dodoi{10.1051/0004-6361/201423387}

\bibitem[{Shahmoradi \& Nemiroff(2015)}]{Shahmoradi2015}
Shahmoradi, A., \& Nemiroff, R.~J. 2015, Monthly Notices of the Royal
  Astronomical Society, 451, 126, \dodoi{10.1093/mnras/stv714}

\bibitem[{Singal {et~al.}(2011)Singal, Petrosian, Lawrence, \&
  Stawarz}]{Singal2011}
Singal, J., Petrosian, V., Lawrence, A., \& Stawarz, L. 2011, The Astrophysical
  Journal, 743, 104, \dodoi{10.1088/0004-637x/743/2/104}

\bibitem[{Soderberg {et~al.}(2006{\natexlab{a}})Soderberg, Nakar, Berger, \&
  Kulkarni}]{Soderberg2006latetime}
Soderberg, A.~M., Nakar, E., Berger, E., \& Kulkarni, S.~R. 2006{\natexlab{a}},
  The Astrophysical Journal, 638, 930–937, \dodoi{10.1086/499121}

\bibitem[{Soderberg {et~al.}(2006{\natexlab{b}})Soderberg, Kulkarni, Nakar,
  Berger, Cameron, Fox, Frail, Gal-Yam, Sari, Cenko, \& et~al.}]{Soderberg2006}
Soderberg, A.~M., Kulkarni, S.~R., Nakar, E., {et~al.} 2006{\natexlab{b}},
  Nature, 442, 1014–1017, \dodoi{10.1038/nature05087}

\bibitem[{Soderberg {et~al.}(2008)Soderberg, Berger, Page, Schady, Parrent,
  Pooley, Wang, Ofek, Cucchiara, Rau, Waxman, Simon, Bock, Milne, Page,
  Barentine, Barthelmy, Beardmore, Bietenholz, Brown, Burrows, Burrows,
  Byrngelson, Cenko, Chandra, Cummings, Fox, Gal-Yam, Gehrels, Immler,
  Kasliwal, Kong, Krimm, Kulkarni, Maccarone, Mészáros, Nakar, O’Brien,
  Overzier, de~Pasquale, Racusin, Rea, \& York}]{Soderberg2008}
Soderberg, A.~M., Berger, E., Page, K.~L., {et~al.} 2008, Nature, 453, 469,
  \dodoi{10.1038/nature06997}

\bibitem[{{Sollerman} {et~al.}(2006){Sollerman}, {Jaunsen}, {Fynbo}, {Hjorth},
  {Jakobsson}, {Stritzinger}, {F{\'e}ron}, {Laursen}, {Ovaldsen}, {Selj},
  {Th{\"o}ne}, {Xu}, {Davis}, {Gorosabel}, {Watson}, {Duro}, {Ilyin}, {Jensen},
  {Lysfjord}, {Marquart}, {Nielsen}, {N{\"a}r{\"a}nen}, {Schwarz}, {Walch},
  {Wold}, \& {{\"O}stlin}}]{Sollerman2006}
{Sollerman}, J., {Jaunsen}, A.~O., {Fynbo}, J.~P.~U., {et~al.} 2006, \aap, 454,
  503, \dodoi{10.1051/0004-6361:20065226}

\bibitem[{Spearman(1904)}]{Spearman}
Spearman, C. 1904, The American Journal of Psychology, 15, 72.
\newblock \url{http://www.jstor.org/stable/1412159}

\bibitem[{Starling {et~al.}(2010)Starling, Wiersema, Levan, Sakamoto, Bersier,
  Goldoni, Oates, Rowlinson, Campana, Sollerman, Tanvir, Malesani, Fynbo,
  Covino, D'Avanzo, O'Brien, Page, Osborne, Vergani, Barthelmy, Burrows, Cano,
  Curran, De~Pasquale, D'Elia, Evans, Flores, Fruchter, Garnavich, Gehrels,
  Gorosabel, Hjorth, Holland, van~der Horst, Hurkett, Jakobsson, Kamble,
  Kouveliotou, Kuin, Kaper, Mazzali, Nugent, Pian, Stamatikos, Thöne, \&
  Woosley}]{Starling2010}
Starling, R. L.~C., Wiersema, K., Levan, A.~J., {et~al.} 2010, Monthly Notices
  of the Royal Astronomical Society, 411, 2792,
  \dodoi{10.1111/j.1365-2966.2010.17879.x}

\bibitem[{Stratta {et~al.}(2018)Stratta, Dainotti, Dall'Osso, Hernandez, \&
  De~Cesare}]{Stratta:2018xza}
Stratta, G., Dainotti, M.~G., Dall'Osso, S., Hernandez, X., \& De~Cesare, G.
  2018, Astrophys. J., 869, 155, \dodoi{10.3847/1538-4357/aadd8f}

\bibitem[{Stritzinger {et~al.}(2002)Stritzinger, Hamuy, Suntzeff, Smith,
  Phillips, Maza, Strolger, Antezana, González, Wischnjewsky, \&
  et~al.}]{Stritzinger2002}
Stritzinger, M., Hamuy, M., Suntzeff, N.~B., {et~al.} 2002, The Astronomical
  Journal, 124, 2100–2117, \dodoi{10.1086/342544}

\bibitem[{Sultana {et~al.}(2013)Sultana, Kazanas, \&
  Mastichiadis}]{Sultana2013}
Sultana, J., Kazanas, D., \& Mastichiadis, A. 2013, The Astrophysical Journal,
  779, 16, \dodoi{10.1088/0004-637x/779/1/16}

\bibitem[{Suzuki {et~al.}(2019)Suzuki, Maeda, \& Shigeyama}]{Suzuki2019}
Suzuki, A., Maeda, K., \& Shigeyama, T. 2019, The Astrophysical Journal, 870,
  38, \dodoi{10.3847/1538-4357/aaef85}

\bibitem[{Tanaka {et~al.}(2012)Tanaka, Moriya, Yoshida, \& Nomoto}]{Tanaka2012}
Tanaka, M., Moriya, T.~J., Yoshida, N., \& Nomoto, K. 2012, Monthly Notices of
  the Royal Astronomical Society, 422, 2675–2684,
  \dodoi{10.1111/j.1365-2966.2012.20833.x}

\bibitem[{Terlevich \& al(199)}]{Terlevich199}
Terlevich, R., \& al, e. 199, IAUC 7269: 1999eb, GRB 991002; 1999ec; 1999do,
  www.cbat.eps.harvard.edu.
\newblock \url{http://www.cbat.eps.harvard.edu/iauc/07200/07269.html#Item1}

\bibitem[{Toffano {et~al.}(2021)Toffano, Ghirlanda, Nava, Ghisellini, Ravasio,
  \& Oganesyan}]{Toffano2021}
Toffano, M., Ghirlanda, G., Nava, L., {et~al.} 2021, Astronomy \& Astrophysics,
  652, A123, \dodoi{10.1051/0004-6361/202141032}

\bibitem[{Tominaga {et~al.}(2004)Tominaga, Deng, Mazzali, Maeda, Nomoto, Pian,
  Hjorth, \& Fynbo}]{Tominaga2004}
Tominaga, N., Deng, J., Mazzali, P.~A., {et~al.} 2004, The Astrophysical
  Journal, 612, L105–L108, \dodoi{10.1086/424841}

\bibitem[{Tominaga {et~al.}(2007)Tominaga, Maeda, Umeda, Nomoto, Tanaka,
  Iwamoto, Suzuki, \& Mazzali}]{Tominaga2007}
Tominaga, N., Maeda, K., Umeda, H., {et~al.} 2007, The Astrophysical Journal,
  657, L77, \dodoi{10.1086/513193}

\bibitem[{Troja {et~al.}(2017)Troja, Piro, van Eerten, Wollaeger, Im, Fox,
  Butler, Cenko, Sakamoto, Fryer, \& et~al.}]{Troja2017}
Troja, E., Piro, L., van Eerten, H., {et~al.} 2017, Nature, 551, 71–74,
  \dodoi{10.1038/nature24290}

\bibitem[{Tsutsui \& Shigeyama(2013)}]{Tsutsui2013b}
Tsutsui, R., \& Shigeyama, T. 2013, Publications of the Astronomical Society of
  Japan, 65, \dodoi{10.1093/pasj/65.3.L3}

\bibitem[{Tsutsui {et~al.}(2013)Tsutsui, Yonetoku, Nakamura, Takahashi, \&
  Morihara}]{Tsutsui2013a}
Tsutsui, R., Yonetoku, D., Nakamura, T., Takahashi, K., \& Morihara, Y. 2013,
  Monthly Notices of the Royal Astronomical Society, 431, 1398–1404,
  \dodoi{10.1093/mnras/stt262}

\bibitem[{Tsvetkova {et~al.}(2021)Tsvetkova, Frederiks, Svinkin, Aptekar,
  Cline, Golenetskii, Hurley, Lysenko, Ridnaia, \& Ulanov}]{Tsvetkova2021}
Tsvetkova, A., Frederiks, D., Svinkin, D., {et~al.} 2021, The Astrophysical
  Journal, 908, 83, \dodoi{10.3847/1538-4357/abd569}

\bibitem[{Turatto {et~al.}(2000)Turatto, Suzuki, Mazzali, Benetti, Cappellaro,
  Danziger, Nomoto, Nakamura, Young, \& Patat}]{Turatto2000}
Turatto, M., Suzuki, T., Mazzali, P.~A., {et~al.} 2000, The Astrophysical
  Journal, 534, L57, \dodoi{10.1086/312653}

\bibitem[{Umeda {et~al.}(2005)Umeda, Tominaga, Maeda, \& Nomoto}]{Umeda2005}
Umeda, H., Tominaga, N., Maeda, K., \& Nomoto, K. 2005, The Astrophysical
  Journal, 633, L17–L20, \dodoi{10.1086/498136}

\bibitem[{Usov(1994)}]{Usov1994}
Usov, V.~V. 1994, Hydrodynamics and High-Energy Physics of WR Colliding Winds.
\newblock \doarXiv{astro-ph/9405067}

\bibitem[{{van Eerten}(2014{\natexlab{a}})}]{VanEerten2014a}
{van Eerten}, H. 2014{\natexlab{a}}, \mnras, 442, 3495,
  \dodoi{10.1093/mnras/stu1025}

\bibitem[{{van Eerten}(2014{\natexlab{b}})}]{VanEerten2014b}
{van Eerten}, H.~J. 2014{\natexlab{b}}, \mnras, 445, 2414,
  \dodoi{10.1093/mnras/stu1921}

\bibitem[{Virgili {et~al.}(2009)Virgili, Liang, \& Zhang}]{Virgili2009}
Virgili, F.~J., Liang, E.-W., \& Zhang, B. 2009, Monthly Notices of the Royal
  Astronomical Society, 392, 91–103, \dodoi{10.1111/j.1365-2966.2008.14063.x}

\bibitem[{Virgili {et~al.}(2013)Virgili, Mundell, Pal’shin, Guidorzi,
  Margutti, Melandri, Harrison, Kobayashi, Chornock, Henden, Updike, Cenko,
  Tanvir, Steele, Cucchiara, Gomboc, Levan, Cano, Mottram, Clay, Bersier,
  Kopač, Japelj, Filippenko, Li, Svinkin, Golenetskii, Hartmann, Milne,
  Williams, O’Brien, Fox, \& Berger}]{Virgili2013}
Virgili, F.~J., Mundell, C.~G., Pal’shin, V., {et~al.} 2013, The
  Astrophysical Journal, 778, 54, \dodoi{10.1088/0004-637x/778/1/54}

\bibitem[{von Kienlin {et~al.}(2020)von Kienlin, Meegan, Paciesas, Bhat,
  Bissaldi, Briggs, Burns, Cleveland, Gibby, Giles, Goldstein, Hamburg, Hui,
  Kocevski, Mailyan, Malacaria, Poolakkil, Preece, Roberts, Veres, \&
  Wilson-Hodge}]{vonKienlin2020}
von Kienlin, A., Meegan, C.~A., Paciesas, W.~S., {et~al.} 2020, The
  Astrophysical Journal, 893, 46, \dodoi{10.3847/1538-4357/ab7a18}

\bibitem[{{Wang} \& {Wheeler}(1998)}]{Wang1998}
{Wang}, L., \& {Wheeler}, J.~C. 1998, \apjl, 504, L87, \dodoi{10.1086/311580}

\bibitem[{Wang {et~al.}(2019)Wang, Rueda, Ruffini, Becerra, Bianco, Becerra,
  Li, \& Karlica}]{Wang:2018slh}
Wang, Y., Rueda, J.~A., Ruffini, R., {et~al.} 2019, Astrophys. J., 874, 39,
  \dodoi{10.3847/1538-4357/ab04f8}

\bibitem[{Waxman {et~al.}(2007)Waxman, Meszaros, \& Campana}]{Waxman2007}
Waxman, E., Meszaros, P., \& Campana, S. 2007, American Astronomical Society,
  667, 351, \dodoi{10.1086/520715}

\bibitem[{Willingale {et~al.}(2007)Willingale, O'Brien, Osborne, Godet, Page,
  Goad, Burrows, Zhang, Rol, Gehrels, \& Chincarini}]{Willingale2007}
Willingale, R., O'Brien, P.~T., Osborne, J.~P., {et~al.} 2007, The
  Astrophysical Journal, 662, 1093, \dodoi{10.1086/517989}

\bibitem[{Woosley \& Bloom(2006)}]{Woosley2006}
Woosley, S., \& Bloom, J. 2006, Annual Review of Astronomy and Astrophysics,
  44, 507–556, \dodoi{10.1146/annurev.astro.43.072103.150558}

\bibitem[{Xu {et~al.}(2021)Xu, Tang, Geng, Wang, Wang, Kuerban, \&
  Huang}]{Xu2021}
Xu, F., Tang, C.-H., Geng, J.-J., {et~al.} 2021, The Astrophysical Journal,
  920, 135, \dodoi{10.3847/1538-4357/ac158a}

\bibitem[{Yamazaki(2008)}]{Yamazaki2008}
Yamazaki, R. 2008, The Astrophysical Journal, 690, L118–L121,
  \dodoi{10.1088/0004-637x/690/2/l118}

\bibitem[{Yonetoku {et~al.}(2004)Yonetoku, Murakami, Nakamura, Yamazaki, Inoue,
  \& Ioka}]{Yonetoku2004}
Yonetoku, D., Murakami, T., Nakamura, T., {et~al.} 2004, 609, 935,
  \dodoi{10.1086/421285}

\bibitem[{Zhang {et~al.}(2018{\natexlab{a}})Zhang, Zhang, Sun, Lei, Gao, Li,
  Shao, Zhao, Hu, Lü, Wu, Fan, Wang, Castro-Tirado, Zhang, Yu, Cao, \&
  Liang}]{Zhang2018NAT}
Zhang, B.-B., Zhang, B., Sun, H., {et~al.} 2018{\natexlab{a}}, Nature
  Communications, 9, \dodoi{10.1038/s41467-018-02847-3}

\bibitem[{Zhang {et~al.}(2021)Zhang, Liu, Peng, Li, Lü, Yang, Yang, Yang,
  Meng, Zou, Ye, Wang, Mao, Zhao, Bai, Castro-Tirado, Hu, Dai, Liang, \&
  Zhang}]{Zhang2021}
Zhang, B.-B., Liu, Z.-K., Peng, Z.-K., {et~al.} 2021, Nature Astronomy,
  \dodoi{10.21203/rs.3.rs-131126/v1}

\bibitem[{Zhang {et~al.}(2018{\natexlab{b}})Zhang, Murase, Kimura, Horiuchi, \&
  Mészáros}]{Zhang2018}
Zhang, B.~T., Murase, K., Kimura, S.~S., Horiuchi, S., \& Mészáros, P.
  2018{\natexlab{b}}, Physical Review D, 97, \dodoi{10.1103/physrevd.97.083010}

\bibitem[{Zhang \& Woosley(2002)}]{Zhang2002jet}
Zhang, W., \& Woosley, S.~E. 2002, Relativistic Jets from Collapsars: Gamma-Ray
  Bursts.
\newblock \doarXiv{astro-ph/0209482}

\bibitem[{Zhang {et~al.}(2022)Zhang, Yu, \& Liu}]{Zhang2022-SN2006aj}
Zhang, Z.-D., Yu, Y.-W., \& Liu, L.-D. 2022, The effects of a magnetar engine
  on the gamma-ray burst-associated supernovae: Application to double-peaked SN
  2006aj,  arXiv, \dodoi{10.48550/ARXIV.2204.11092}

\bibitem[{Zhao {et~al.}(2020)Zhao, Zhang, Zhang, Liang, Luan, Zhou, Yi, Wang,
  \& Zhang}]{Zhao2020}
Zhao, W., Zhang, J.-C., Zhang, Q.-X., {et~al.} 2020, The Astrophysical Journal,
  900, 112, \dodoi{10.3847/1538-4357/aba43a}

\end{thebibliography}
\bibliographystyle{aasjournal}



\newpage

\newpage
\begin{longrotatetable}
\begin{deluxetable*}{llllllllllllllllllll}
\tablecaption{Observable characteristics of GRB and SNe\label{chartable04}}
\tablewidth{600pt}
\tabletypesize{\scriptsize}
\tablehead{
\colhead{GRB ID} & \colhead{SN ID} & 
\colhead{GRB Type} & \colhead{SN Type} & 
\colhead{z} & \colhead{$T_{90}$} & 
\colhead{$E_{\gamma,\rm iso}$} & \colhead{$E^{*}_{\rm p}$} & 
\colhead{$L_{\gamma,\rm iso}$} & \colhead{Grade} & 
\colhead{$L_{\rm p,bol}$} \\
\colhead{} & \colhead{} &
\colhead{} & \colhead{} & 
\colhead{} & \colhead{(s)} &
\colhead{($10^{52}erg$)} & \colhead{(keV)} & 
\colhead{($10^{50} erg/s$)} & \colhead{} &
\colhead{($10^{42} erg/s$)}
}

\startdata
910423 & 1991aa & \nodata & Ib & 0.011 & $208.6\pm 1.1$ & \nodata & \nodata & \nodata & \nodata & \nodata \\
951107C & 1995bc & \nodata & II & 0.0477 & $43.52\pm 5.961$ & \nodata & \nodata & \nodata & \nodata & \nodata \\
960221 & 1996N & \nodata & Ib & \nodata & $31.328\pm 4.868$ & \nodata & \nodata & \nodata & \nodata & \nodata \\
960925 & 1996at & \nodata & Ic-Ib/c & 0.09 & $1.792\pm 0.453$ & \nodata & \nodata & \nodata & \nodata & \nodata \\
961218 & 1997B & \nodata & Ic-Ib/c & 0.01 & $8.768\pm 0.932$ & \nodata & \nodata & \nodata & \nodata & \nodata \\
970228 & \nodata & GRB & \nodata & 0.695 & $56$ & $1.6 ^{+0.12}_{-0.12}$ & $195 ^{+64}_{-64}$ & 4.84 & C & \nodata \\
970508 & \nodata & \nodata & Ib/c & 0.835 & $23.104\pm 3.789$ & $0.546$ & $145 ^{+43}_{-43}$ & 4.34 & E & \nodata \\
970514 & 1997cy & INT & II & 0.063 & $1.28\pm 0.771$ & $0.0004$ & \nodata & 0.0332 & C & 18.6 \\
971013 & 1997dq & \nodata & Ib & 0.003 & $12.288\pm 5.514$ & \nodata & \nodata & \nodata & \nodata & \nodata \\
971115 & 1997ef & \nodata & Ic & 0.01169 & \nodata & \nodata & \nodata & \nodata & C & \nodata \\
971120 & 1997ei & \nodata & Ic & 0.0106 & $13.632\pm 2.241$ & \nodata & \nodata & \nodata & B & \nodata \\
971221 & 1997ey & \nodata & Ia & 0.58 & $1.02\pm 0.2$ & \nodata & \nodata & \nodata & E & \nodata \\
980326 & \nodata & GRB & \nodata & 1 & $9$ & $0.48 ^{+0.09}_{-0.09}$ & $71 ^{+36}_{-36}$ & 10.7 & D & \nodata \\
980425 & 1998bw & llGRB & Ic & 0.00867 & $18$ & $0.000086 ^{+0.000002}_{-0.000002}$ & $55 ^{+21}_{-21}$ & 0.000482 & A & 7.33 \\
980525 & 1998ce & \nodata & II & \nodata & $39.68\pm 2.433$ & \nodata & \nodata & \nodata & \nodata & \nodata \\
980703 & \nodata & \nodata & \nodata & 0.967 & $108.352\pm 2.625$ & $7.41683 ^{+0.714875}_{-0.714875}$ & $503 ^{+64}_{-64}$ & 28 & D & \nodata \\
980910 & 1999E & \nodata & II & 0.0261 & $0.72\pm 0.3$ & \nodata & \nodata & \nodata & B & 10 \\
990712 & \nodata & GRB & \nodata & 0.4331 & $19$ & $0.67 ^{+0.13}_{-0.13}$ & $93 ^{+15}_{-15}$ & 5.05 & C & \nodata \\
990902 & 1999dp & \nodata & II & 0.016 & $19.2\pm 5$ & \nodata & \nodata & \nodata & \nodata & \nodata \\
991002 & 1999eb & \nodata & \nodata & 0.018113 & $1.918\pm 1.995$ & \nodata & \nodata & \nodata & \nodata & \nodata \\
991021 & 1999ex & \nodata & Ic & 0.011 & \nodata & \nodata & \nodata & \nodata & \nodata & \nodata \\
991208 & \nodata & GRB & \nodata & 0.7063 & $60$ & $22.3 ^{+1.8}_{-1.8}$ & $313 ^{+31}_{-31}$ & 63.4 & E & \nodata \\
000114 & 2000C & \nodata & Ic & 0.012 & $0.579\pm 0.035$ & \nodata & \nodata & \nodata & \nodata & \nodata \\
000418 & \nodata & \nodata & \nodata & 1.1185 & $2.288\pm 0.92$ & $9.57 ^{+0.49}_{-0.49}$ & $284 ^{+21}_{-21}$ & 886 & D & \nodata \\
000911 & \nodata & GRB & \nodata & 1.0585 & $500$ & $67 ^{+14}_{-14}$ & $1859 ^{+371}_{-371}$ & 27.5 & E & \nodata \\
011121 & 2001ke & GRB & II & 0.362 & $47$ & $7.8 ^{+2.1}_{-2.1}$ & $1060 ^{+275}_{-275}$ & 22.6 & B & 5.9 \\
020405 & \nodata & GRB & \nodata & 0.68986 & $40$ & $10 ^{+0.9}_{-0.9}$ & $354 ^{+10}_{-10}$ & 42.2 & C & \nodata \\
020903 & \nodata & llGRB & \nodata & 0.2506 & $3.3$ & $0.0011 ^{+0.0006}_{-0.0006}$ & $3.37 ^{+1.79}_{-1.79}$ & 0.042 & B & \nodata \\
021211 & 2002lt & GRB & Ic & 1.004 & $2.8$ & $1.12 ^{+0.13}_{-0.13}$ & $108 ^{+26}_{-50}$ & 80.2 & B & \nodata \\
030329 & 2003dh & GRB & Ic-HN & 0.16867 & $22.76\pm 0.5$ & $1.5 ^{+0.3}_{-0.3}$ & $113.3 ^{+2.3}_{-2.3}$ & 7.7 & A & 10.1 \\
030723 & \nodata & XRF & \nodata & 0.38 & $28.3\pm 2.5$ & $0.021 ^{+0.087}_{-0.016}$ & $<0.023 $ & 0.0845 & D & \nodata \\
031203 & 2003lw & INT & Ic & 0.10536 & $37$ & $0.0086 ^{+0.004}_{-0.004}$ & $158 ^{+51}_{-51}$ & 0.0255 & A & 12.6 \\
040924 & \nodata & GRB & \nodata & 0.858 & $2.39\pm 0.24$ & $0.95 ^{+0.09}_{-0.09}$ & $102 ^{+35}_{-35}$ & 73.8 & C & \nodata \\
041006 & \nodata & GRB & \nodata & 0.716 & $18$ & $3 ^{+0.9}_{-0.9}$ & $98 ^{+20}_{-20}$ & 28.6 & C & \nodata \\
050416A & \nodata & GRB & \nodata & 0.6528 & $2.4\pm 0.2$ & $0.1 ^{+0.01}_{-0.01}$ & $25.1 ^{+4.2}_{-4.2}$ & 6.89 & D & \nodata \\
050525A & 2005nc & GRB & Ic & 0.606 & $8.84\pm 0.5$ & $2.5 ^{+0.43}_{-0.43}$ & $127 ^{+10}_{-10}$ & 45.4 & B & \nodata \\
050824 & \nodata & GRB & \nodata & 0.8281 & $25\pm 5$ & $0.1905 ^{+0.1495}_{-0.1495}$ & $21.5 ^{+10.5}_{-10.5}$ & 1.39 & E & \nodata \\
060218 & 2006aj & llGRB & Ic-BL & 0.03342 & $2100\pm 100$ & $0.0053 ^{+0.0003}_{-0.0003}$ & $4.9 ^{+0.3}_{-0.3}$ & 0.00026 & A & 6.47 \\
060729 & \nodata & GRB & \nodata & 0.5428 & $115\pm 10$ & $1.6 ^{+0.6}_{-0.6}$ & $77 ^{+38}_{-38}$ & 2.14 & D & \nodata \\
060904B & \nodata & GRB & \nodata & 0.7029 & $192\pm 5$ & $2.4 ^{+0.2}_{-0.2}$ & $163 ^{+31}_{-31}$ & 2.12 & C & \nodata \\
070419A & \nodata & INT & \nodata & 0.9705 & $116\pm 6$ & $0.16$ & $<69 $ & 27.1 & D & \nodata \\
071025 & \nodata & \nodata & \nodata & 4.88 & $109\pm 2$ & $65$ & $1023.3 ^{+204.99}_{-204.99}$ & 351 & E & \nodata \\
071112C & \nodata & \nodata & \nodata & 0.823 & $15\pm 2$ & $1.18 ^{+0.19}_{-0.19}$ & $740 ^{+326}_{-326}$ & 74.2 & C & 5.4 \\
080109 & 2008D & XRF & Ib & 0.006494 & $470\pm 30$ & $0.0000013$ & $0.1685 ^{+0.1315}_{-0.1315}$ & 0.000000278 & A & 1.66 \\
080319B & \nodata & GRB & \nodata & 0.9371 & $124.86$ & $114 ^{+9}_{-9}$ & $1261 ^{+65}_{-65}$ & 176 & C & \nodata \\
081007A & 2008hw & GRB & Ic & 0.5295 & $9.01$ & $0.15 ^{+0.04}_{-0.04}$ & $61 ^{+15}_{-15}$ & 2.54 & B & 14 \\
090618 & \nodata & GRB & \nodata & 0.54 & $113.2\pm 0.6$ & $25.7 ^{+5}_{-5}$ & $288 ^{+9.2}_{-9.2}$ & 34.9 & C & \nodata \\
091127 & 2009nz & GRB & Ic-BL & 0.49044 & $7.42$ & $1.5 ^{+0.2}_{-0.2}$ & $51 ^{+5}_{-5}$ & 30.1 & B & 12 \\
100316D & 2010bh & llGRB & Ic & 0.0592 & $1300$ & $>0.0059$ & $26 ^{+16}_{-16}$ & 0.00048 & A & 5.67 \\
100418A & \nodata & GRB & \nodata & 0.6239 & $8\pm 2$ & $0.099 ^{+0.063}_{-0.063}$ & $47.1 ^{+3.2}_{-3.2}$ & 2 & DE & \nodata \\
101219B & 2010ma & GRB & Ic & 0.55185 & $51$ & $0.42 ^{+0.05}_{-0.05}$ & $108 ^{+12}_{-12}$ & 1.27 & AB & 15 \\
101225A & \nodata & ULGRB & \nodata & 0.847 & $1088\pm 20$ & $1.2 ^{+0.3}_{-0.3}$ & $70 ^{+37}_{-37}$ & 0.0316 & D & \nodata \\
111209A & 2011kl & ULGRB & SLSN & 0.67702 & $10000$ & $58.2 ^{+7.3}_{-7.3}$ & $520 ^{+89}_{-89}$ & 0.976 & AB & 29.1 \\
111211A & \nodata & \nodata & \nodata & 0.478 & $15$ & $0.74$ & \nodata & 7.3 & BC & \nodata \\
111228A & \nodata & GRB & \nodata & 0.71627 & $101.2\pm 5.42$ & $4.2 ^{+0.6}_{-0.6}$ & $58.4 ^{+6.9}_{-6.9}$ & 7.12 & E & 6.76 \\
120422A & 2012bz & GRB & Ic & 0.28253 & $5.4\pm 1.4$ & $0.024 ^{+0.008}_{-0.008}$ & $<72 $ & 57 & A & 14.8 \\
120714B & 2012eb & INT & Ib/c & 0.3984 & $159\pm 34$ & $0.0594 ^{+0.0195}_{-0.0195}$ & $69 ^{+43}_{-43}$ & 0.0522 & B & 6.17 \\
120729A & \nodata & GRB & \nodata & 0.8 & $71.5\pm 17.5$ & $2.3 ^{+1.5}_{-1.5}$ & $559 ^{+57}_{-57}$ & 5.79 & DE & \nodata \\
130215A & 2013ez & GRB & Ic & 0.597 & $65.7\pm 10.8$ & $3.1 ^{+1.6}_{-1.6}$ & $248 ^{+101}_{-101}$ & 7.53 & B & \nodata \\
130427A & 2013cq & GRB & Ic & 0.3399 & $162.83\pm 1.36$ & $81 ^{+10}_{-10}$ & $1415 ^{+13}_{-13}$ & 6.65E+52 & B & 9.12 \\
130702A & 2013dx & INT & Ic & 0.145 & $58.881$ & $0.064 ^{+0.013}_{-0.013}$ & $17.2 ^{+5.7}_{-5.7}$ & 0.124 & A & 10.8 \\
130831A & 2013fu & GRB & Ib/c & 0.479 & $32.5\pm 2.5$ & $0.46 ^{+0.02}_{-0.02}$ & $79.9 ^{+10.4}_{-13.3}$ & 2.09 & AB & 6.92 \\
140206A & \nodata & \nodata & \nodata & 2.739 & $93.2\pm 13.5$ & $240 ^{+2}_{-2}$ & $1780 ^{+120}_{-120}$ & 963 & \nodata & \nodata \\
140606B & iPTF14bfu & GRB & Ic-BL & 0.384 & $22.78\pm 2.06$ & $0.347 ^{+0.02}_{-0.02}$ & $352 ^{+46}_{-37}$ & 2.1 & AB & \nodata \\
150818A & \nodata & INT & \nodata & 0.285 & $123.3\pm 31.3$ & $0.1 ^{+0.02}_{-0.02}$ & $128 ^{+13}_{-13}$ & 0.103 & B & \nodata \\
161219B & 2016jca & INT & Ic & 0.1475 & $6.94\pm 0.79$ & $0.0085 ^{+0.0043}_{-0.0043}$ & $71 ^{+19.3}_{-33.2}$ & 0.141 & B & 4.6 \\
161228B & iPTF17cw & \nodata & Ic-BL & 0.093 & $8.5\pm 0.5$ & $0.023 ^{+0.006}_{-0.006}$ & $214 ^{+26}_{-26}$ & \nodata & \nodata & 18.3 \\
171010A & 2017htp & \nodata & Ic-BL & 0.3285 & $107.266\pm 0.81$ & $18 ^{+0.55}_{-0.55}$ & $227 ^{+7}_{-7}$ & 22.3 & A & \nodata \\
171205A & 2017iuk & \nodata & Ic-BL & 0.0368 & $189.4\pm 35$ & $0.00218 ^{+0.00063}_{-0.0005}$ & $125 ^{+141}_{-141}$ & 0.001193 & B & \nodata \\
180720B & \nodata & \nodata & \nodata & 0.653 & $48.897\pm 0.362$ & $33.97 ^{+0.01}_{-0.01}$ & $472 ^{+15}_{-14}$ & 115 & \nodata & \nodata \\
180728A & 2018fip & XRF & Ic & 0.117 & $8.68\pm 0.3$ & $0.233 ^{+0.1}_{-0.1}$ & $108 ^{+8}_{-7}$ & 3 & B & \nodata \\
190114C & 2019jrj & \nodata & \nodata & 0.4245 & $361.5\pm 11.7$ & $27.03 ^{+0.24}_{-0.24}$ & $929.3 ^{+9.4}_{-9.4}$ & 10.7 & \nodata & \nodata \\
190829A & 2019oyw & \nodata & Ic-BL & 0.0785 & $58.2\pm 8.9$ & $0.018$ & $140.205 ^{+21.57}_{-21.57}$ & 0.0334 & B & \nodata \\
200826A & \nodata & GRB & \nodata & 0.748577 & $1.14\pm 0.13$ & $0.709 ^{+0.028}_{-0.028}$ & $210.33 ^{+6.87}_{-6.42}$ & 109 & C & \nodata \\
210210A & \nodata & GRB & \nodata & 0.715 & $6.6\pm 0.59$ & $0.17$ & $28.469 ^{+12.348}_{-18.3}$ & 4.42 & \nodata & \nodata \\
\enddata
\end{deluxetable*}
\end{longrotatetable}

\addtocounter{table}{-1} 
\begin{longrotatetable}
\begin{deluxetable*}{llllllllllllllllllll}
\tablecaption{\textit{(continued)}\label{chartable04}}
\tablewidth{600pt}
\tabletypesize{\scriptsize}
\tablehead{
\colhead{GRB ID} & \colhead{$t^{*}_{\rm p}$} & 
\colhead{$\Delta m_{\rm 15,bol}$} & 
\colhead{$E_{\rm K}$} & 
\colhead{$M_{\rm ej}$} & \colhead{$M_{\rm Ni}$} &
\colhead{$v_{\rm ph}$} & \colhead{$k_{\rm avg}$} 
& \colhead{$s_{\rm avg}$} \\
\colhead{} & \colhead{(days)} &
\colhead{(mag)} &
\colhead{($10^{52}erg$)} & \colhead{($M_{\rm \odot}$)} &
\colhead{($M_{\rm \odot}$)} & \colhead{($km/s$)} & 
\colhead{} & \colhead{} }

\startdata
910423 & $\cdots$ & $\cdots$ & 
$\cdots$ & $\cdots$ & $\cdots$ & $\cdots$ & $\cdots$ & $\cdots$ \\
951107C & $\cdots$ & $\cdots$ & 
$\cdots$ & $\cdots$ & $\cdots$ & $\cdots$ & $\cdots$ & $\cdots$ \\
960221 & $\cdots$ & $\cdots$ & 
$\cdots$ & $\cdots$ & $\cdots$ & $\cdots$ & $\cdots$ & $\cdots$ \\
960925 & $\cdots$ & $\cdots$ & 
$\cdots$ & $\cdots$ & $\cdots$ & $\cdots$ & $\cdots$ & $\cdots$ \\
961218 & $\cdots$ & $\cdots$ & 
$\cdots$ & $\cdots$ & $\cdots$ & $\cdots$ & $\cdots$ & $\cdots$ \\
970228 & $\cdots$ & $\cdots$ & 
$\cdots$ & $\cdots$ & $\cdots$ & $\cdots$ & $\cdots$ & $\cdots$ \\
970508 & $\cdots$ & $\cdots$ 
$\cdots$ & $\cdots$ & $\cdots$ & $\cdots$ & $\cdots$ & $\cdots$ \\
970514 & $\cdots$ & $\cdots$ & 
30 & 5 & 2.6 & $15000^{-}_{-}$ & $\cdots$ & $\cdots$ \\
971013 & $\cdots$ & $\cdots$ & 
$\cdots$ & $\cdots$ & $\cdots$ & $\cdots$ & $\cdots$ & $\cdots$ \\
971115 & $\cdots$ & $\cdots$ & 
$20^{+4}_{-4}$ & $7.6^{+2}_{-2}$ & $0.15^{+0.02}_{-0.02}$ & $\cdots$ & $\cdots$ & $\cdots$ \\
971120 & $\cdots$ & $\cdots$ & 
$\cdots$ & $\cdots$ & $\cdots$ & $13000^{-}_{-}$ & $\cdots$ & $\cdots$ \\
971221 & $\cdots$ & $\cdots$ & 
$\cdots$ & $\cdots$ & $\cdots$ & $\cdots$ & $\cdots$ & $\cdots$ \\
980326 & $\cdots$ & $\cdots$ & 
$\cdots$ & $\cdots$ & $\cdots$ & $\cdots$ & $\cdots$ & $\cdots$ \\
980425 & 15.16 & 0.8 & 
$25^{+5}_{-5}$ & $8^{+2}_{-2}$ & $0.45^{+0.15}_{-0.15}$ & $18000^{-}_{-}$ & 1 & 1 \\
980525 & $\cdots$ & $\cdots$ & 
$\cdots$ & $\cdots$ & $\cdots$ & $\cdots$ & $\cdots$ & $\cdots$ \\
980703 & $\cdots$ & $\cdots$ & 
$\cdots$ & $\cdots$ & $\cdots$ & $\cdots$ & $\cdots$ & $\cdots$ \\
980910 & $\cdots$ & $\cdots$ & 
$\cdots$ & $\cdots$ & $\cdots$ & $\cdots$ & $\cdots$ & $\cdots$ \\
990712 & $\cdots$ & $\cdots$ & 
$26.1^{+24.6}_{-15}$ & $6.6^{+3.5}_{-2.9}$ & $0.14^{+0.04}_{-0.04}$ & $\cdots$ & $0.36\pm 0.05$ & $0.94\pm 0.12$ \\
990902 & $\cdots$ & $\cdots$ & 
$\cdots$ & $\cdots$ & $\cdots$ & $\cdots$ & $\cdots$ & $\cdots$ \\
991002 & $\cdots$ & $\cdots$ & 
$\cdots$ & $\cdots$ & $\cdots$ & $\cdots$ & $\cdots$ & $\cdots$ \\
991021 & $\cdots$ & $\cdots$ & 
$1.3^{+0.8}_{-0.5}$ & $2.9^{+0.9}_{-0.7}$ & $0.15^{+0.04}_{-0.03}$ & $\cdots$ & $\cdots$ & $\cdots$ \\
991208 & $\cdots$ & $\cdots$ & 
$38.7^{+44.6}_{-26}$ & $9.7^{+6.8}_{-5.6}$ & $0.96^{+0.48}_{-0.48}$ & $\cdots$ & $2.11\pm 0.58$ & $1.1\pm 0.2$ \\
000114 & $\cdots$ & $\cdots$ & 
$\cdots$ & $\cdots$ & $\cdots$ & $\cdots$ & $\cdots$ & $\cdots$ \\
000418 & $\cdots$ & $\cdots$ & 
$\cdots$ & $\cdots$ & $\cdots$ & $\cdots$ & $\cdots$ & $\cdots$ \\
000911 & $\cdots$ & $\cdots$ & 
$\cdots$ & $\cdots$ & $\cdots$ & $\cdots$ & $0.85\pm 0.35$ & $1.4\pm 0.32$ \\
011121 & 17 & $\cdots$ & 
$17.7^{+8.8}_{-6.4}$ & $4.4^{+0.8}_{-0.8}$ & $0.35^{+0.01}_{-0.01}$ & $\cdots$ & $1.13\pm 0.23$ & $0.84\pm 0.17$ \\
020405 & $\cdots$ & $\cdots$ & 
$8.9^{+5.4}_{-3.8}$ & $2.2^{+0.6}_{-0.5}$ & $0.23^{+0.02}_{-0.02}$ & $\cdots$ & $0.82\pm 0.14$ & $0.62\pm 0.03$ \\
020903 & $\cdots$ & $\cdots$ & 
$28.9^{+32.2}_{-18.9}$ & $7.3^{+4.9}_{-4}$ & $0.25^{+0.13}_{-0.13}$ & $\cdots$ & $0.61\pm 0.19$ & $0.98\pm 0.02$ \\
021211 & $\cdots$ & $\cdots$ & 
$28.5^{+45}_{-13}$ & $7.2^{+7.4}_{-6}$ & $0.16^{+0.14}_{-0.14}$ & $\cdots$ & $0.4\pm 0.19$ & $0.98\pm 0.26$ \\
030329 & 12.75 & 0.7 & 
$35^{+15}_{-15}$ & $7.5^{+2.5}_{-2.5}$ & $0.5^{+0.1}_{-0.1}$ & $20000^{-}_{-}$ & $1.28\pm 0.28$ & $0.87\pm 0.18$ \\
030723 & $\cdots$ & $\cdots$ & 
$\cdots$ & $\cdots$ & $\cdots$ & $\cdots$ & $\cdots$ & $\cdots$ \\
031203 & 17.33 & 0.62 & 
$60^{+15}_{-15}$ & $13^{+4}_{-4}$ & $0.55^{+0.2}_{-0.2}$ & $18000^{-}_{-}$ & $1.65\pm 0.36$ & $1.1\pm 0.24$ \\
040924 & $\cdots$ & $\cdots$ & 
$\cdots$ & $\cdots$ & $\cdots$ & $\cdots$ & $0.203\pm 0.202$ & $1.371\pm 0.971$ \\
041006 & $\cdots$ & $\cdots$ & 
$76.4^{+39.8}_{-28.7}$ & $19.2^{+3.9}_{-3.6}$ & $0.69^{+0.07}_{-0.07}$ & $\cdots$ & $1.16\pm 0.06$ & $1.47\pm 0.04$ \\
050416A & $\cdots$ & $\cdots$ & 
$\cdots$ & $\cdots$ & $\cdots$ & $\cdots$ & $\cdots$ & $\cdots$ \\
050525A & $\cdots$ & $\cdots$ & 
$18.9^{+10.7}_{-7.5}$ & $4.8^{+1.1}_{-1}$ & $0.24^{+0.02}_{-0.02}$ & $\cdots$ & $0.69\pm 0.03$ & $0.83\pm 0.03$ \\
050824 & $\cdots$ & $\cdots$ & 
$5.7^{+9.3}_{-3.7}$ & $1.4^{+1.6}_{-0.6}$ & $0.26^{+0.17}_{-0.17}$ & $\cdots$ & $1.05\pm 0.42$ & $0.52\pm 0.14$ \\
060218 & 10.42 & 0.83 & $1.0^{+0.5}_{-0.5}$ & $2.0^{+0.5}_{-0.5}$ & $0.20^{+0.10}_{-0.10}$ & $20000^{-}_{-}$ & $0.58\pm0.13$ & $0.67\pm0.14$ \\
060729 & $\cdots$ & $\cdots$ & 
$24.4^{+14.3}_{-9.9}$ & $6.1^{+1.6}_{-1.4}$ & $0.36^{+0.05}_{-0.05}$ & $\cdots$ & $0.94\pm 0.1$ & $0.92\pm 0.04$ \\
060904B & $\cdots$ & $\cdots$ & 
$9.9^{+5.1}_{-3.7}$ & $2.5^{+0.5}_{-0.5}$ & $0.12^{+0.01}_{-0.01}$ & $\cdots$ & $0.42\pm 0.02$ & $0.65\pm 0.01$ \\
070419A & $\cdots$ & $\cdots$ & 
$\cdots$ & $\cdots$ & $\cdots$ & $\cdots$ & $\cdots$ & $\cdots$ \\
071025 & $\cdots$ & $\cdots$ & 
$\cdots$ & $\cdots$ & $\cdots$ & $\cdots$ & $\cdots$ & $\cdots$ \\
071112C & 11.9 & $\cdots$ & 
$\cdots$ & $\cdots$ & 0.14 & $\cdots$ & $0.6\pm 0.1$ & $0.75\pm 0.15$ \\
080109 & 19.30 & $\cdots$ & $7.0^{+1.0}_{-1.0}$ & $7.0^{+2.0}_{-2.0}$ & $0.09^{+0.02}_{-0.02}$ & $30000^{-}_{-}$ & $\cdots$ & $\cdots$ \\
080319B & $\cdots$ & $\cdots$ & 
$22.7^{+19.1}_{-11.9}$ & $5.7^{+2.6}_{-2.2}$ & $0.86^{+0.45}_{-0.45}$ & $\cdots$ & $2.3\pm 0.9$ & $0.89\pm 0.1$ \\
081007A & 12 & $\cdots$ & 
$19^{+15}_{-15}$ & $2.3^{+1}_{-1}$ & $0.39^{+0.08}_{-0.08}$ & $12600^{-}_{-}$ & $0.71\pm 0.1$ & $0.85\pm 0.11$ \\
090618 & $\cdots$ & $\cdots$ & 
$36.5^{+20}_{-14.2}$ & $9.2^{+2.1}_{-1.9}$ & $0.37^{+0.03}_{-0.03}$ & $\cdots$ & $1.11\pm 0.22$ & $0.98\pm 0.2$ \\
091127 & 15 & 0.5 & 
$13.5^{+0.4}_{-0.4}$ & $4.7^{+0.1}_{-0.1}$ & $0.33^{+0.01}_{-0.01}$ & $17000^{+2500}_{-2500}$ & $0.89\pm 0.01$ & $0.88\pm 0.01$ \\
100316D & 8.76 & 0.89 & 
$15.4^{+1.4}_{-1.4}$ & $2.5^{+0.2}_{-0.2}$ & $0.12^{+0.02}_{-0.02}$ & $35000^{-}_{-}$ & $0.53\pm 0.15$ & $0.53\pm 0.11$ \\
100418A & $\cdots$ & $\cdots$ & 
$\cdots$ & $\cdots$ & $\cdots$ & $\cdots$ & $\cdots$ & $\cdots$ \\
101219B & 11.8 & 0.99 & 
$10^{+6}_{-6}$ & $1.3^{+0.5}_{-0.5}$ & $0.43^{+0.03}_{-0.03}$ & $\cdots$ & $1.16\pm 0.63$ & $0.76\pm 0.1$ \\
101225A & $\cdots$ & $\cdots$ & 
$32^{+16}_{-16}$ & $8.1^{+1.5}_{-1.5}$ & $0.41^{+0.03}_{-0.03}$ & $19934^{-}_{-}$ & $0.96\pm 0.05$ & $1.02\pm 0.03$ \\
111209A & 14.80 & 0.78 & $55^{+35}_{-35}$ & $4.0^{+1.0}_{-1.0}$ & $1.0^{+0.1}_{-0.1}$ & $21000^{-}_{-}$ & $1.81\pm0.19$ & $1.08\pm0.11$ \\
111211A & $\cdots$ & $\cdots$ & 
$\cdots$ & $\cdots$ & $\cdots$ & $\cdots$ & $\cdots$ & $\cdots$ \\
111228A & 22.7 & $\cdots$ & 
$\cdots$ & $\cdots$ & 0.28 & $\cdots$ & $0.75\pm 0.13$ & $1.43\pm 0.14$ \\
120422A & 14.45 & 0.62 & 
$25.5^{+2.1}_{-2.1}$ & $6.1^{+0.5}_{-0.5}$ & $0.57^{+0.07}_{-0.07}$ & $20500^{-}_{-}$ & $1.13\pm 0.25$ & $0.93\pm 0.19$ \\
120714B & 13.6 & $\cdots$ & 
$\cdots$ & $\cdots$ & 0.18 & $42000^{+5500}_{-5500}$ & 0.69 & $0.86\pm 0.04$ \\
120729A & $\cdots$ & $\cdots$ & 
$\cdots$ & $\cdots$ & $\cdots$ & $\cdots$ & $1.02\pm 0.26$ & 1 \\
130215A & $\cdots$ & $\cdots$ & 
$\cdots$ & $\cdots$ & $\cdots$ & $\cdots$ & $0.675\pm 0.075$ & 1 \\
130427A & 13 & $\cdots$ & 
$\cdots$ & $\cdots$ & $0.56^{+0.21}_{-0.18}$ & $\cdots$ & $0.85\pm 0.03$ & $0.77\pm 0.03$ \\
130702A & 12.94 & 0.85 & 
$8.2^{+0.4}_{-0.4}$ & $3.1^{+0.1}_{-0.1}$ & $0.37^{+0.01}_{-0.01}$ & $21300^{-}_{-}$ & $0.98\pm 0.07$ & $0.78\pm 0.05$ \\
130831A & 11.9 & $\cdots$ & 
$18.7^{+9}_{-9}$ & $4.7^{+0.8}_{-0.8}$ & $0.3^{+0.07}_{-0.07}$ & $30300^{+6000}_{-6000}$ & $0.95\pm 0.19$ & $0.82\pm 0.19$ \\
140206A & $\cdots$ & $\cdots$ & 
$\cdots$ & $\cdots$ & $\cdots$ & $\cdots$ & $\cdots$ & $\cdots$ \\
140606B & $\cdots$ & $\cdots$ & 
$19^{+11}_{-11}$ & $4.8^{+1.9}_{-1.9}$ & $0.42^{+0.17}_{-0.17}$ & $19800^{-}_{-}$ & $1.04\pm 0.24$ & $0.81\pm 0.13$ \\
150818A & $\cdots$ & $\cdots$ & 
$\cdots$ & $\cdots$ & $\cdots$ & $\cdots$ & $\cdots$ & $\cdots$ \\
161219B & $\cdots$ & $\cdots$ & 
$51^{+8}_{-8}$ & $5.8^{+0.3}_{-0.3}$ & $0.22^{+0.08}_{-0.08}$ & $29700^{1500}_{1500}$ & $0.74\pm 0.09$ & $0.75\pm 0.07$ \\
161228B & 16.47 & $\cdots$ & 
4 & 2 & $0.644^{+0.065}_{-0.059}$ & $16900^{-}_{-}$ & $\cdots$ & $\cdots$ \\
171010A & $\cdots$ & $\cdots$ & 
$8.1^{+2.5}_{-2.5}$ & $4.1^{+0.7}_{-0.7}$ & $0.33^{+0.02}_{-0.02}$ & $14000^{+1000}_{-1000}$ & $\cdots$ & $\cdots$ \\
171205A & $\cdots$ & $\cdots$ & 
$4.9^{+0.9}_{-0.9}$ & $0.18^{+0.01}_{-0.01}$ & $22000^{-}_{-}$ & $\cdots$ & $\cdots$ \\
180720B & $\cdots$ & $\cdots$ & 
$\cdots$ & $\cdots$ & $\cdots$ & $\cdots$ & $\cdots$ & $\cdots$ \\
180728A & $\cdots$ & $\cdots$ & 
$\cdots$ & $\cdots$ & $\cdots$ & $\cdots$ & $\cdots$ & $\cdots$ \\
190114C & $\cdots$ & $\cdots$ & 
$\cdots$ & $\cdots$ & $\cdots$ & $\cdots$ & $\cdots$ & $\cdots$ \\
190829A & $\cdots$ & $\cdots$ & 
$13.55^{+5.08}_{-5.08}$ & $5.67^{+0.72}_{-0.72}$ & $0.5^{+0.1}_{-0.1}$ & $\cdots$ & $\cdots$ & $\cdots$ \\
200826A & $\cdots$ & $\cdots$ & 
$\cdots$ & $\cdots$ & $\cdots$ & $\cdots$ & $1.45\pm 0.27$ & 1 \\
210210A & $\cdots$ & $\cdots$ & 
$\cdots$ & $\cdots$ & $\cdots$ & $\cdots$ & $\cdots$ & $\cdots$ \\
\enddata
\end{deluxetable*}
\end{longrotatetable}

\addtocounter{table}{-1}
\begin{longrotatetable}
\begin{deluxetable*}{llllllllllllllll}
\tablecaption{\textit{(continued)}\label{chartable04}}
\tablewidth{600pt}
\tabletypesize{\scriptsize}
\tablehead{
\colhead{GRB ID} & \colhead{SN ID} & 
\colhead{$\log_{10}T^{*}_{\rm a,opt}$} & \colhead{$\log_{10}L_{\rm a,opt}$} & 
\colhead{$\log_{10}T^{*}_{\rm a,X}$} & \colhead{ $\log_{10}L_{\rm a,X}$} & 
\colhead{$\theta_{\rm jet}$} & 
\colhead{$T_{\rm jet}$} & \colhead{Sources} \\
\colhead{} & \colhead{} &
\colhead{(s)} & \colhead{(ergs$^{-1}$)} & 
\colhead{(s)} & \colhead{(ergs$^{-1}$)} &
\colhead{($^{\circ}$)} &\colhead{(days)} & \colhead{}}

\startdata
910423 & 1991aa & $\cdots$ & $\cdots$ & $\cdots$ & $\cdots$ & $\cdots$ & $\cdots$
& (3), (53)\\
951107C & 1995bc & $\cdots$ & $\cdots$ & $\cdots$ & $\cdots$ & $\cdots$ & $\cdots$
&\\
960221 & 1996N & $\cdots$ & $\cdots$ & $\cdots$ & $\cdots$ & $\cdots$ & $\cdots$
& (32)\\
960925 & 1996at & $\cdots$ & $\cdots$ & $\cdots$ & $\cdots$ & $\cdots$ & $\cdots$
& (54)\\ 
961218 & 1997B & $\cdots$ & $\cdots$ & $\cdots$ & $\cdots$ & $\cdots$ &$\cdots$
 & (32), (54)\\
970228 & $\cdots$ & $\cdots$ & $\cdots$ & $\cdots$ & $\cdots$ & $\cdots$ &$\cdots$
 & (1),(17)\\
970508 & $\cdots$ & $\cdots$ & $\cdots$ & $\cdots$ & $\cdots$ & $16.79\pm 3.30$ &$25.00\pm 5.00$ & (4), (7), (8), (54)\\
970514 & 1997cy & $\cdots$ & $\cdots$ & $\cdots$ & $\cdots$ & $\cdots$ &$\cdots$
 & (2), (18),(33)\\
971013 & 1997dq & $\cdots$ & $\cdots$ & $\cdots$ & $\cdots$ & $\cdots$ & $\cdots$
 & (33), (54) \\
971115 & 1997ef & $\cdots$ & $\cdots$ & $\cdots$ & $\cdots$ & $\cdots$ &$\cdots$
 & (33), \\
971120 & 1997ei & $\cdots$ & $\cdots$ & $\cdots$ & $\cdots$ & $\cdots$ &$\cdots$
 & (54) \\
971221 & 1997ey & $\cdots$ & $\cdots$ & $\cdots$ & $\cdots$ & $\cdots$ &$\cdots$
 & (4), (30)\\
980326 & $\cdots$ & $\cdots$ & $\cdots$ & $\cdots$ & $\cdots$ & $>5.08$ & 0.4 & (1), (5), (17) \\
980425 & 1998bw & $\cdots$ & $\cdots$ & $\cdots$ & $\cdots$ & $11.00\pm 3.00$ &$\cdots$
 & (13) \\
980525 & 1998ce & $\cdots$ & $\cdots$ & $\cdots$ & $\cdots$ & $\cdots$ & $\cdots$
 & (63) \\
980703 & $\cdots$ & $\cdots$ & $\cdots$ & $\cdots$ & $\cdots$ & $11.00\pm 0.80$ &$3.40\pm 0.50$
 & (6), (17), (49) \\
980910 & 1999E & $\cdots$ & $\cdots$ & $\cdots$ & $\cdots$ & $\cdots$ &$\cdots$
 & (4), (30),(33), (34),(35)\\
990712 & $\cdots$ & $\cdots$ & $\cdots$ & $\cdots$ & $\cdots$ & $>23.55$ & $11.57\pm 0.20$
& (1), (7), (8)\\
990902 & 1999dp & $\cdots$ & $\cdots$ & $\cdots$ & $\cdots$ & $\cdots$ &$\cdots$
 & (4) \\
991002 & 1999eb & $\cdots$ & $\cdots$ & $\cdots$ & $\cdots$ & $\cdots$ &$\cdots$
 & (4), (33), (36), (60), (61)\\
991021 & 1999ex & $\cdots$ & $\cdots$ & $\cdots$ & $\cdots$ & $\cdots$ & $\cdots$
& (37), (55)\\
991208 & $\cdots$ & $\cdots$ & $\cdots$ & $\cdots$ & $\cdots$ & $<4.53$ & 2.1 & (1) , (7)\\ 
000114 & 2000C & $\cdots$ & $\cdots$ & $\cdots$ & $\cdots$ & $\cdots$ & $\cdots$
& (4)\\
000418 & $\cdots$ & $\cdots$ & $\cdots$ & $\cdots$ & $\cdots$ & $11.35$ &$\cdots$
 & (6), (7), (17)\\
000911 & $\cdots$ & $\cdots$ & $\cdots$ & $\cdots$ & $\cdots$ & $<4.4$ &$\cdots$
 & (1) \\
011121 & 2001ke & $\cdots$ & $\cdots$ & $\cdots$ & $\cdots$ & $4.49\pm 0.16$ &$\cdots$
 & (1), (17),(30), (38) & \\
020405 & $\cdots$ & $\cdots$ & $\cdots$ & $\cdots$ & $\cdots$ & $6.40\pm 1.05$ & $1.67\pm 0.52$
 & (1), (17), (50)\\
020903 & $\cdots$ & $\cdots$ & $\cdots$ & $\cdots$ & $\cdots$ & $\cdots$ &$\cdots$
 & (1), (17) \\
021211 & 2002lt & $\cdots$ & $\cdots$ & $\cdots$ & $\cdots$ & $4.82\pm 0.68$ &$\cdots$
 & (1),(8), (17)\\
030329 & 2003dh & $5.43\pm 0.05$ & $44.11\pm 0.09$ & $\cdots$ & $\cdots$ & $3.80\pm 0.05$ & $0.47\pm 0.05$
& (1),(8), (10), (17)\\
030723 & $\cdots$ & $\cdots$ & $\cdots$ & $\cdots$ & $\cdots$ & $\cdots$& $\cdots$
& (9),(39), (57), (58), (59) \\
031203 & 2003lw & $\cdots$ & $\cdots$ & $\cdots$ & $\cdots$ & $9.00\pm 2.00$ & $\cdots$
& (1), (6), (57) \\
040924 & $\cdots$ & $3.23\pm 0.04$ & $45.26\pm 0.13$ & $\cdots$ & $\cdots$ & $>6.9$ &$\cdots$
 & (1), (8), (17)\\
041006 & $\cdots$ & $3.84\pm 0.03$ & $44.76\pm 0.07$ & $\cdots$ & $\cdots$ & $2.90\pm 0.40$ & $\cdots$ & (1), (8), (17)\\
050416A & $\cdots$ & $3.93\pm 0.06$ & $43.70\pm 0.08$ & $2.78\pm 0.07$ & $46.69\pm 0.05$ & $\cdots$ & $\cdots$ & (1)\\
050525A & 2005nc & $3.69\pm 0.04$ & $45.51\pm 0.04$ & $\cdots$ & $\cdots$ & $2.12\pm 0.46$ & $0.16\pm 0.09$ & (1), (30)\\
050824 & $\cdots$ & $3.39\pm 0.06$ & $44.87\pm 0.06$ & $4.82\pm 0.13$ & $45.24\pm 0.07$ & $\cdots$ & $\cdots$ & (1)\\
060218 & 2006aj & $\cdots$ & $\cdots$ & $5.06\pm 0.14$ & $42.61\pm 0.16$ & $12.60\pm 3.95$ & $\cdots$ & (1), (56)\\
060729 & $\cdots$ & $4.88\pm 0.03$ & $44.88\pm 0.03$ & $4.92\pm 0.01$ & $45.97\pm 0.04$ & $6.6$ & $26.23\pm 6.12$ & (1), (17), (51)\\
060904B & $\cdots$ & $3.66\pm 0.04$ & $45.25\pm 0.05$ & $3.64\pm 0.08$ & $46.36\pm 0.12$ & $\cdots$ & 0.02 & (1),(11)\\
070419A & $\cdots$ & $2.98\pm 0.07$ & $45.05\pm 0.14$ & $\cdots$ & $\cdots$ & $\cdots$ & 0.02
 & (13)\\
071025 & $\cdots$ & $\cdots$ & $\cdots$ & $\cdots$ & $\cdots$ & $\cdots$ & 
$\cdots$ & (10) \\
071112C & $\cdots$ & $\cdots$ & $\cdots$ & $\cdots$ & $\cdots$ & $\cdots$ & $\cdots$ & (6), (24)\\
080109 & 2008D & $\cdots$ & $\cdots$ & $\cdots$ & $\cdots$ & $>8.09$ & $\cdots$ & (1), (19),(41),(42),(43),(44)\\
080319B & $\cdots$ & $\cdots$ & $\cdots$ & $5.08\pm 0.09$ & $45.42\pm 0.10$ & $10.24\pm 0.71$ & $0.03\pm 0.01$ & (1), (62)\\
081007A & 2008hw & $4.65\pm 0.07$ & $43.98\pm 0.10$ & $3.40\pm 0.08$ & $46.44\pm 0.06$ & $>11.09$ & 11.57 & (1) \\
090618 & $\cdots$ & $3.88\pm 0.01$ & $45.17\pm 0.01$ & $3.49\pm 0.01$ & $47.40\pm 0.02$ & $\cdots$& $\cdots$ & (1)\\
091127 & 2009nz & $4.25\pm 0.01$ & $44.83\pm 0.04$ & $3.81\pm 0.02$ & $47.07\pm 0.02$ & $5.50\pm 1.50$ & $0.37\pm 0.10$ & ,(45),(46), (52) \\
100316D & 2010bh & $6.38\pm 0.06$ & $42.06\pm 0.11$ & $\cdots$ & $\cdots$ & $>5.6$ &$\cdots$
 & (1), (12), (24)\\
100418A & $\cdots$ & $4.65\pm 0.15$ & $44.19\pm 0.12$ & $5.33\pm 0.08$ & $44.69\pm 0.11$ & $\cdots$ &$\cdots$
 & (1), (6) \\
101219B & 2010ma & $3.68\pm 0.02$ & $44.97\pm 0.02$ & $4.23\pm 0.17$ & $45.08\pm 0.10$ & $>9.07$ & $\cdots$
& (1) \\
101225A & $\cdots$ & $5.26\pm 0.17$ & $42.43\pm 0.06$ & $\cdots$ & $\cdots$ & $\cdots$ &$\cdots$
 & (1), (13), (58) \\
111209A & 2011kl & $4.15\pm 0.08$ & $45.10\pm 0.08$ & $3.91\pm 0.01$ & $48.08\pm 0.02$ & $9.17\pm 1.50$ & $\cdots$
& (1),(29),(30)\\
111211A & $\cdots$ & $\cdots$ & $\cdots$ & $\cdots$ & $\cdots$ & $\cdots$ & $\cdots$
&(13)\\
111228A & $\cdots$ & $4.05\pm 0.01$ & $45.52\pm 0.01$ & $3.87\pm 0.04$ & $46.70\pm 0.03$ & $\cdots$ & $\cdots$
 & (1), (24)\\
120422A & 2012bz & $\cdots$ & $\cdots$ & $5.13\pm 0.22$ & $43.66\pm 0.12$ & $23.00\pm 7.00$ & $\cdots$
& (1) \\
120714B & 2012eb & $\cdots$ & $\cdots$ & $\cdots$ & $\cdots$ & $\cdots$ & $\cdots$
& (1), (24)\\
120729A & $\cdots$ & $\cdots$ & $\cdots$ & $\cdots$ & $\cdots$ & $\cdots$ & $0.06\pm 0.01$ & (1) \\
130215A & 2013ez & $\cdots$ & $\cdots$ & $\cdots$ & $\cdots$ & $>10.16$ & $\cdots$ & (1), (6)\\
130427A & 2013cq & $\cdots$ & $\cdots$ & $\cdots$ & $\cdots$ & $>7.5$ & $0.43\pm 0.05$
 &(1), (6), (24),(43) \\
130702A & 2013dx & $5.01\pm 0.01$ & $43.59\pm 0.02$ & $\cdots$ & $\cdots$ & $14.00\pm 4.00$ & $\cdots$
& (1), (6) \\
130831A & 2013fu & $3.70\pm 0.01$ & $45.52\pm 0.03$ & $3.15\pm 0.09$ & $47.05\pm 0.06$ & $>3.17$ &$\cdots$
 & (1), (24),(2)\\
140206A & $\cdots$ & $\cdots$ & $\cdots$ & $\cdots$ & $\cdots$ & $\cdots$ & $\cdots$
& (6), (20) \\
140606B & iPTF14bfu & $\cdots$ & $\cdots$ & $\cdots$ & $\cdots$ & $>11.5$ & $\cdots$
& (1) \\
150818A & $\cdots$ & $\cdots$ & $\cdots$ & $\cdots$ & $\cdots$ & $\cdots$ & $\cdots$
& (1) \\
161219B & 2016jca & $\cdots$ & $\cdots$ & $\cdots$ & $\cdots$ & $42\pm3$ & $\cdots$
& (1), (6), (30), (66), (67) \\
161228B & iPTF17cw & $\cdots$ & $\cdots$ & $\cdots$ & $\cdots$ & $\cdots$ & $\cdots$
 & (20), (21)\\
171010A & 2017htp & $\cdots$ & $\cdots$ & $\cdots$ & $\cdots$ & $\cdots$ &$6.48\pm 27.09$
 & (14), (25)\\
171205A & 2017iuk & $\cdots$ & $\cdots$ & $\cdots$ & $\cdots$ & $\cdots$ & $\cdots$
&(6), (15), (27), (47) \\
180720B & $\cdots$ & $\cdots$ & $\cdots$ & $\cdots$ & $\cdots$ & $\cdots$ & $1.08\pm 0.27$
 & (1), (6), (28) \\
180728A & 2018fip & $\cdots$ & $\cdots$ & $\cdots$ & $\cdots$ & $\cdots$ & $1.82\pm 1.08$
 &(1), (6),(47),(48), (64) \\
190114C & 2019jrj & $\cdots$ & $\cdots$ & $\cdots$ & $\cdots$ & $\cdots$ & $0.10\pm 0.04$
&(6),(25), (46)\\
190829A & 2019oyw & $\cdots$ & $\cdots$ & $\cdots$ & $\cdots$ & $>57.30$ & $\cdots$
 &(22),(23), (26)\\
200826A & $\cdots$ & $\cdots$ & $\cdots$ & $\cdots$ & $\cdots$ & $>16$ & $\cdots$
&(16),(18), (23),(31)\\
210210A & $\cdots$ & $\cdots$ & $\cdots$ & $\cdots$ & $\cdots$ & $\cdots$ & $\cdots$
 & (40)\\
\enddata
\end{deluxetable*}
\tablecomments{Sources. (1):\cite{Cano2017a}, (2):\cite{Cano2014b}, (3): \cite{Schmidt2005}, (4): \cite{Bosnjak2006}, (5): \cite{Qin2013}, (6): \cite{Minaev2019}, (7): \cite{Frail2001}, (8) \cite{Ghirlanda2004} , (9): \cite{Butler2005}, (10): \cite{Perley2010}, (11): \cite{Klotz2008}, (12): \cite{Starling2010}, (13): \cite{Dereli2017}, (14): \cite{vonKienlin2020}, (15): \cite{Bartoli2017}, (16): \cite{Rhodes2021}, (17): \cite{Demianski2017}, (18): \cite{Turatto2000}, (19): \cite{Li2008}, (20): \cite{Gotz2014}, (21): \cite{Corsi2017}, (22): \cite{Dado2002}, (23): \cite{Zhang2021}, (24): \cite{Klose2019}, (25): \cite{Melandri2019}, (26): \cite{Hu2021}, (27):\cite{Izzo2019}, (28):\cite{Toffano2021}, (29) : \cite{Kann2019} , (30) : \cite{Lu2018}
    ,(31) : \cite{Rossi2021}, (32) : \href{https://heasarc.gsfc.nasa.gov/w3browse/all/batsegrb.html}{BATSEGRB}, (33) : \cite{Nomoto:2000vi}, (34) : \cite{Germany2000},(35) : \cite{Rigon2003}, (36) : \cite{Li2002}, ,(37): \cite{Lyman2016},(38): \cite{Garnavich2003}, (39): \href{https://space.mit.edu/HETE/Bursts/Data/}{HETE}, (40) : \cite{Ruffini2021},(41) : \cite{Soderberg2008},(42) : \cite{Bianco2014}, (43) : \cite{Prentice2016},(44) : \cite{Mazzali2011},(45): \cite{2011ApJ...743..204B}, (46): \href{https://sne.space/}{OSC} , (47): \cite{Wang:2018slh}, (48) : \cite{Amati2002}, (49): \cite{Kong2009}, (50):\cite{Price2003}, (51):\cite{Liang2008}, (52):\cite{Zhao2020}, (53): \cite{Hudec1999}, (54):\cite{Wang1998}, (55):\cite{Stritzinger2002}, (56):\cite{Berger2011}, (57):\cite{Fynbo2004}, (58):\cite{Pian2005}, (59):\cite{Tominaga2004}, (60):\cite{Madjaz1999}, (61):\cite{Terlevich199}, (62):\href{GCN CIRCULAR 7627}{https://gcn.gsfc.nasa.gov/gcn3/7627.gcn3}, (63):\href{http://www.cbat.eps.harvard.edu/iauc/06900/06955.html}{IAUC 6955}, (64):\href{https://gcn.gsfc.nasa.gov/cgi-bin/swish-cgi.pl}{NASA/GSFC LHEA GCN Bacodine}, (65):\cite{Lin2020}, (66):\cite{Cano2017b}, (67):\cite{Ashall2019}. Other information come from The Open Supernova Catalog \href{The Open Supernova Catalog}{https://sne.space} \citep{TheOpenSupernovaCatalog} and the Transient Name Server \href{Transient Name Server}{https://www.wis-tns.org/}.}
\end{longrotatetable}

\end{document}